\documentclass[12pt,3p,times]{elsarticle}

\usepackage{fullpage}
\usepackage{amsmath,amsfonts,amssymb}
\usepackage{amsthm}
\usepackage{bm}
\usepackage{gensymb}
\usepackage{graphicx,epstopdf}
\usepackage[position=top,labelfont=normalfont,textfont=normalfont,singlelinecheck=off,justification=raggedright]{subfig}
\usepackage{floatrow}
\usepackage[dvipsnames]{xcolor}

\usepackage{setspace}
\usepackage{enumerate,etaremune}
\usepackage{booktabs}
\usepackage{tabularx,multirow}
\usepackage{framed}
\usepackage{hhline}
\usepackage{url}
\usepackage[unicode,bookmarks=false]{hyperref}
\hypersetup{
  colorlinks,
  citecolor=Blue,linkcolor=Blue,urlcolor=Blue
}

\usepackage[textsize=footnotesize,linecolor=red,backgroundcolor=red!25,bordercolor=red]
  {todonotes}

\usepackage{tikz}
\usetikzlibrary{shapes.geometric}
\usetikzlibrary{shapes,arrows}
\usetikzlibrary{plotmarks}

\usetikzlibrary{calc}

\usepackage{physics}

\usepackage{algorithm}
\usepackage{algorithmicx,algpseudocode}
\usepackage{cases}

\newcommand{\eg}{{\it e.g.}}
\newcommand{\ie}{{\it i.e.}}

\newcommand{\etal}{{\it et al.}}

\newcommand{\rn}[1]{\uppercase\expandafter{\romannumeral #1\relax}}

\newsavebox{\dotbox}

\theoremstyle{remark}

\definecolor{hku-blue}{RGB}{20,135,200}
\definecolor{hku-dark-blue}{RGB}{0,65,145}
\definecolor{hku-medium-blue}{RGB}{0,65,135}
\definecolor{hku-light-blue}{RGB}{95,190,235}
\definecolor{hku-dark-gray}{RGB}{124,124,124}

\newcommand{\reviewerone}[1]{{\color{black} #1}}
\newcommand{\reviewertwo}[1]{{\color{black} #1}}
\newcommand{\reviewerthr}[1]{{\color{black} #1}}

\newcolumntype{L}[1]{>{\raggedright\let\newline\\arraybackslash\hspace{0pt}}m{#1}}
\newcolumntype{C}[1]{>{\centering\let\newline\\arraybackslash\hspace{0pt}}m{#1}}
\newcolumntype{R}[1]{>{\raggedleft\let\newline\\arraybackslash\hspace{0pt}}m{#1}}

\allowdisplaybreaks

\biboptions{sort&compress,square,comma,numbers}
\AtBeginDocument{\hypersetup{citecolor=MidnightBlue,linkcolor=MidnightBlue,urlcolor=MidnightBlue}}

\usepackage{etoolbox}
\makeatletter
\patchcmd{\ps@pprintTitle}
{Preprint submitted to}
{~}
{}{}
\makeatother
\journal{~}

\begin{document}

\begin{frontmatter}

\title{Fully Coupled Implicit Hydro-Mechanical Multiphase Flow Simulation in Deformable Porous Media Using DEM}

\author[HKU]{Quanwei Dai}
\author[SDU]{Kang Duan}
\author[HKU]{Chung-Yee Kwok\corref{corr}}
\cortext[corr]{Corresponding Author}
\ead{fkwok8@hku.hk}

\address[HKU]{Department of Civil Engineering, The University of Hong Kong, Hong Kong, China}
\address[SDU]{School of Civil Engineering, Shandong University, Jinan, China}

\journal{~}

\begin{abstract}
Knowledge of the underlying mechanisms of multiphase flow dynamics in porous media is crucial for optimizing subsurface engineering applications like geological carbon sequestration. However, studying the micro-mechanisms of multiphase fluid--grain interactions in the laboratory is challenging due to the difficulty in obtaining mechanical data such as force and displacement. Traditional discrete element method models coupled with pore networks offer insights into these interactions but struggle with accurate pressure prediction during pore expansion from fracturing and efficient simulation during the slow drainage of compressible fluids. To address these limitations, we develop an advanced two-way coupled hydro-mechanical discrete element method model that accurately and efficiently captures fluid--fluid and fluid--grain interactions in deformable porous media. Our model integrates an unconditionally stable implicit finite volume approach, enabling significant timesteps for advancing fluids. A pressure-volume iteration scheme dynamically balances injection-induced pressure buildup with substantial pore structure deformation, while flow front-advancing criteria precisely locate the fluid--fluid interface and adaptively refine timesteps, particularly when capillary effects block potential flow paths. The model is validated against benchmark Hele-Shaw experiments in both rigid and deformable porous media, providing quantitative insights into the micro-mechanisms governing multiphase flow. For the first time, grain-scale inputs such as viscous and capillary pressures, energies, contact forces, and flow resistances are utilized to provide a detailed understanding of micro-scale fluid--fluid and fluid--grain flow patterns and their transitions.

\end{abstract}

 
\end{frontmatter}

\setcounter{page}{1}

\section{Introduction}
\label{sec:introduction}

Understanding the mechanisms of multiphase flow in deformable porous media---namely, immiscible two-phase flow interacts with pore deformation---is essential for the success of subsurface engineering practices such as radioactive waste disposal, enhanced oil recovery, and geological carbon sequestration (\eg~\cite{xu2013coupled,berg2013real,wang2020multiphase,szulczewski2012lifetime,al2016pore}). In geological CO$_2$ storage, for instance, effective residual trapping of CO$_2$ in deep-ground reservoirs is crucial for maximizing storage capacity and ensuring long-term stability \cite{krevor2015capillary,huang2015parallel}. The balance between viscous and capillary pressures governs the trapping mechanisms \cite{blunt1995pore,oren1998extending}, and optimizing injection flow rates and fluid viscosities can present unfavorable viscous fingering, thereby enhancing trapping capacity \cite{ferer2004crossover,wang2013experimental,zheng2017effect}. Nevertheless, rapid injection and high fluid viscosities can cause pore pressure buildup, altering geomechanical properties such as pore structure \cite{yu2019co2} and permeability \cite{rutqvist2002study}, potentially leading to CO$_2$ leakage \cite{white2014geomechanical,zoback2012earthquake}. Accurately describing the pore-scale physics of multiphase flow in deformable porous media is therefore crucial for balancing efficiency and stability in geological systems---a challenge that remains significant \cite{jain2009preferential,blunt2013pore,blunt2017multiphase,juanes2020multiphase}.

Existing studies have explored the micro-mechanisms of multiphase flow in porous media using both experimental and computational approaches. The Hele-Shaw test, a classic experimental setup, is widely employed to visualize pore-scale phenomena and approximate Darcy flows in laboratory settings (\eg~\cite{hele1898flow,saffman1958penetration,paterson1981radial}). 
In this experiment, an invading fluid displaces a defending fluid with varying viscosities between two closely packed transparent plates, facilitating the study of fluid--fluid dynamics (\eg~\cite{nittmann1985fractal,lenormand1988numerical,furuberg1996intermittent,zhao2016wettability,chen2017visualizing,irannezhad2023fluid}) and fluid--grain interactions (\eg~\cite{cheng2008towards,holtzman2012capillary,sandnes2011patterns,huang2012granular,meng2022fracturing,zadeh2023pore}) in rigid and deformable porous media. 
While most experimental research focuses on invasion morphology characteristics, such as finger width and length, it often overlooks mechanical features like force and displacement, which are crucial for understanding how grain motion influences hydrodynamic behavior. Although attempts have been made to measure grain motion through particle tracking \cite{chevalier2009morphodynamics,macminn2015fluid} and to visualize effective stress evolution via photo-poromechanics \cite{meng2022fracturing,meng2023crossover}, these efforts fall short of providing comprehensive insights into the micro-mechanisms of multiphase flow in porous media. 

On the other hand, computational approaches, including continuum and discrete methods, are widely used to elucidate the micro-mechanisms quantitatively. For pure fluid invasion in rigid porous media, continuum models utilize techniques such as the level-set method (\eg~\cite{sussman1994level,jettestuen2013level}), lattice Boltzmann method (\eg~\cite{porter2009lattice,chen2019inertial}), and smoothed particle hydrodynamics (\eg~\cite{tartakovsky2006pore,tartakovsky2016smoothed}) to generate diverse fluid--fluid displacement patterns. Advanced continuum approaches like the Darcy-Cahn-Hilliard and Darcy-Brinkman-Biot models investigate multiphase fluid-driven mechanics in deformable porous media (\eg~\cite{carrillo2021capillary,paulin2022fluid,guevel2023darcy}), where fluid migration is coupled with grain motion. 
However, these continuum models are computationally intensive and often struggle to capture invasion patterns accurately due to challenges in generating detailed pore geometries, resolving Navier-Stokes equations, and establishing locally distributed permeabilities and capillary pressures. 

In contrast, discrete methods such as the pore-network model (\eg~\cite{holtzman2015wettability,primkulov2019signatures,yang2019modeling}) simplify pore geometry and fluid flow by representing them through an interconnected network of grain-enclosed pores and capillary throats. Integrating the pore-network model with the discrete element method (DEM) enables hydro-mechanical (HM) coupling to predict fluid--grain interactions, offering fundamental insights into micro-mechanics \cite{jain2009preferential,zhang2013coupled}. 
Nonetheless, the need for small timesteps to stabilize explicit pressure solutions limits the application of existing HM-DEM coupled models in complex multiphase flow problems, particularly during slow drainage. 
Increasing timestep sizes is a direct way to enhance simulation efficiency. 
Zhang \etal~\cite{zhang2013coupled} addressed this by reducing the fluid bulk modulus of water from 2000 to 0.75 MPa, while Meng \etal~\cite{meng2020jamming} eliminated the timestep dependence by assuming incompressible fluids. Nonetheless, a recent study \cite{cuttle2023compression} highlights the crucial role of fluid compressibility in developing multiphase flow patterns, emphasizing the importance of using realistic values. 

To address the simulation efficiency issue while considering reasonable fluid compressibility, we have developed a fully coupled HM-DEM framework with an adaptive variable-timestep scheme that operates efficiently within capillarity and viscous instability in dynamic pore geometries. 
Specifically, we advance our original single-phase HM-DEM model \cite{duan2020initiation} by incorporating an implicit pressure solver that is unconditionally stable under any injection scenario.
Two numerical accuracy challenges for multiphase flow simulations with significant timesteps are (1) accounting for the viscous pressure response in injection-induced pore deformation and (2) locating the flow front interface and updating the mixing fluid properties in relevant pores. We address these challenges with a fully coupled pressure-volume iteration scheme and multiphase flow front advancing criteria, respectively. 
We validate the upgraded model with Hele-Shaw test observations in rigid and deformable porous media, demonstrating that it effectively reproduces comprehensive fluid-fluid and fluid-grain displacement patterns. 
Moreover, we utilize the model to gain insights into the micro-mechanisms of multiphase flow in porous media, proving it a reliable tool for addressing critical issues in geological systems and engineering practices, such as subsurface fluid migration and trapping mechanisms.

\section{Methodology and formulations}
\label{sec:methodology}
This section outlines our methodology for pore-scale numerical modeling of multiphase flow in porous media. To investigate the interplay among viscous pressures, capillary effects, and grain motion, we simplify the multiphase flow problem with the following assumptions:
\begin{itemize}\itemsep=0pt
  \item Invading and defending fluids remain immiscible throughout the injection process.
  \item Gravity and inertia are neglected for the lateral fluid invasion in Hele-Shaw tests. \reviewerthr{Gravity has minimal impact in the lateral direction, and inertia is negligible due to the low Reynolds numbers ($\text{Re}<60$), where viscous forces dominate \cite{huang2012granular}. For scenarios such as field-scale CO$_2$ plume migration (upward) \cite{rutqvist2002study} or high inflow rate Hele-Shaw tests ($\text{Re}>300$) \cite{chen2017visualizing}, where gravity or inertia significantly influence flow behavior, the model would need to be extended, though this is beyond the scope of the current study.}
  \item For Plane Poiseuille flow, both in-plane and out-of-plane capillary pressures are assumed to have identical interfacial curvature \cite{lenormand1988numerical}, easing curvature determination in multilayer porous media.
  \item Strong drainage and imbibition scenarios are simulated without considering film and corner flows. The front advancing mechanism is thus the “Burst” event described by Cieplak and Robbins \cite{cieplak1988dynamical}---\ie~advancing unstable fluid--fluid interfaces via pressure jumps. 
\end{itemize}

\begin{figure}[htbp]
  \centering
  \includegraphics[width=1.0\textwidth]{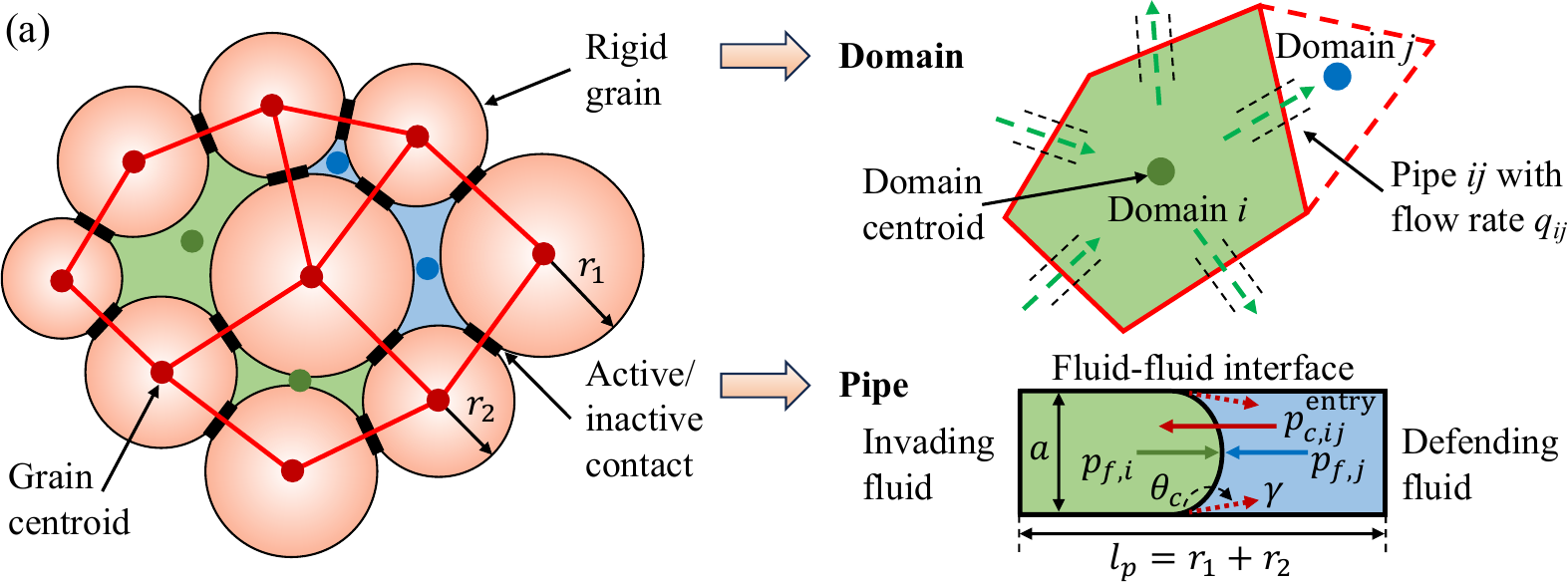}
  \includegraphics[width=1.0\textwidth]{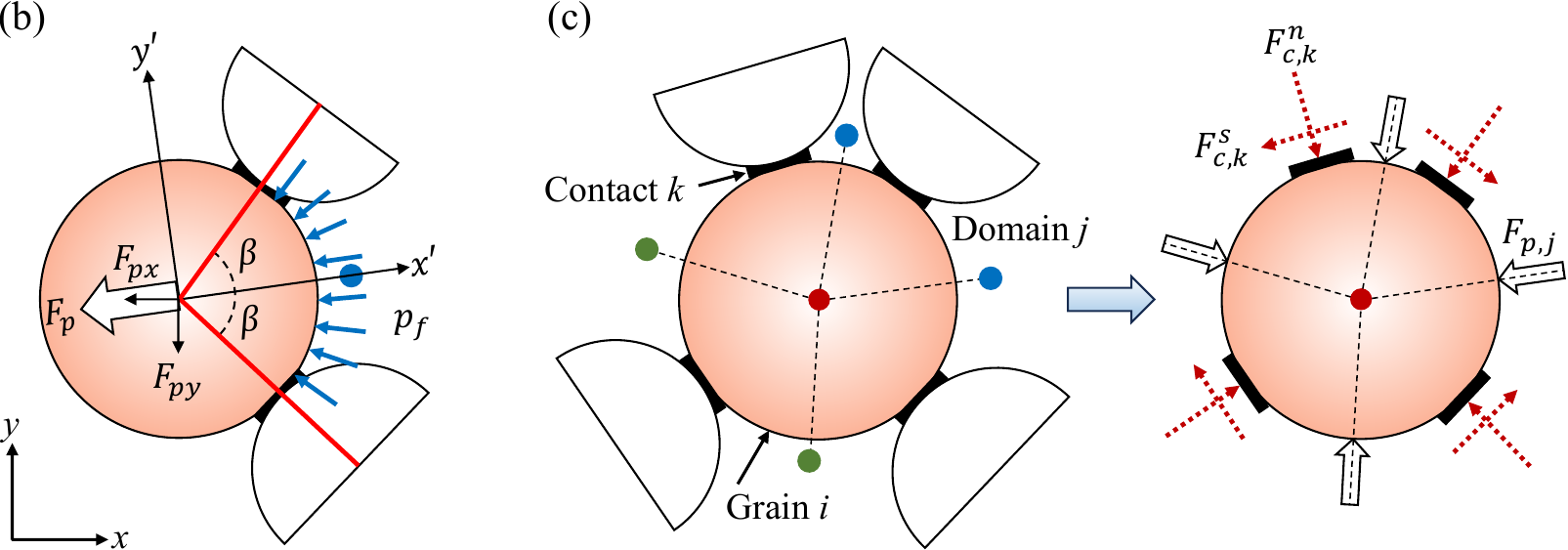}
  \caption{Schematic diagram of the fully coupled HM-DEM model. (a) Pore network, (b) calculation of applied pressure forces, and (c) grain-based force equilibrium.}
  \label{fig:pore-network-DEM}
\end{figure}

The HM-DEM model is described in two integral components: (1) an advanced fluid flow model with an implicit time integration scheme and (2) a DEM-based grain motion model that responds to injection-induced pressure buildup. To accurately capture the hydrodynamic responses of multiphase flow in deformable porous media, we generate a pore network and solid matrix using the fully coupled two-dimensional (2D) particle flow code---PFC2D \cite{itasca2008}, as shown in Figure~\ref{fig:pore-network-DEM}a. Each grain (represented as a disc in 2D) is modeled as an ideal spherical particle. The grain-enclosed polygons represent pore domains interconnected by grain--grain contact-identified pipes. The network of domains and pipes represents the invading and defending fluid phases. The fluid flow model calculates the pressure distribution induced by injection, providing pressure forces for the grain motion model. The grain motion model then balances the injection energy and updates the domain volume and pipe conductance for the fluid flow model. The iterative interaction between these models ensures a dynamic and accurate representation of multiphase flow in deformable porous media.

\subsection{Implicit fluid flow model}
The fluid flow model employs the Plane Poiseuille flow equation to achieve momentum balance in pipes, determining the volumetric flow rate from domain $i$ to an adjacent domain $j$ ($q_{ij}$ in Figure~\ref{fig:pore-network-DEM}a). The original HM-DEM framework \cite{duan2020initiation} uses an explicit time integration scheme to calculate pipe flow rates, relying on current-timestep pressures ($p_{f,i}^n$ and $p_{f,j}^n$) and converging only within a critical timestep \cite{jain2009preferential,zhang2013coupled}. To eliminate this timestep dependency, we incorporate advanced-timestep pressures ($p_{f,i}^{n+1}$ and $p_{f,j}^{n+1}$) into the integration scheme---\ie~the implicit finite volume approach previously utilized to study fault stability in single-phase hydraulic injection simulations \cite{yang2018two}. The Plane Poiseuille flow equation can thus be expressed as
\begin{equation}
  q_{ij} = \dfrac{a_{ij}^3}{12\eta_{ij}}\dfrac{p_{f,i}^{n+1}-p_{f,j}^{n+1}}{l_{p,ij}},
  \label{eq:plane-pflow}
\end{equation}
where $\eta_{ij}$ is the fluid viscosity, $a_{ij}$ is the aperture width, and $l_{p,ij}$ is the pipe length equal to the sum of contacting grain radii $r_1$ and $r_2$. Depending on whether grain motion is considered, the aperture width in our model can be initialized using either a uniform random distribution or the original contact force field. Fluid mass balance in each domain and flow momentum balance in pipes jointly characterize fluid movement in porous media. Our fluid flow model describes the fluid mass balance in domain $i$ as 
\begin{equation}
  \dfrac{\delta V_i}{\delta t} + \dfrac{V_i}{K_{f,i}}\dfrac{p_{f,i}^{n+1}-p_{f,i}^n}{\delta t} = -\sum_j q_{ij},
  \label{eq:fmass-balance}
\end{equation}
where $\delta V_i$ is the domain volume change, $\delta t$ is the fluid flow timestep, $V_i$ is the domain volume calculated by pore geometry (\ie~polygon area), and $K_{f,i}$ is the fluid bulk modulus. 

In multiphase flow, the fluid bulk modulus, viscosity, and capillary effect should be treated as dynamic variables. We express the fluid bulk modulus and viscosity of the two-phase fluid in weighted-average form, which can be written as
\begin{equation}
  K_{f,i} = S_{\text{inv},i}K_{f,\text{inv}} + (1-S_{\text{inv},i})K_{f,\text{def}}
  \label{eq:mix-kf}
\end{equation}
and
\begin{equation}
  \eta_{ij} = \dfrac{(S_{\text{inv},i}+S_{\text{inv},j})\eta_\text{inv}}{2} + \dfrac{(2-S_{\text{inv},i}-S_{\text{inv},j})\eta_\text{def}}{2},
  \label{eq:mix-eta}
\end{equation}
where $S_{\text{inv},i}$ is the invading fluid volume fraction in a domain (\ie~saturation), and subscripts inv and def denote flow properties of invading and defending fluids, respectively \cite{mavko1998bounds,lenormand1988numerical}. Notably, the capillary effect blocks pipe flow when the pressure difference in the pipe falls below the capillary entry pressure, which occurs in front pipes containing a fluid--fluid interface \cite{lenormand1988numerical,yang2019modeling}. 
The capillary entry pressure ($p_{c,ij}^\text{entry}$) is the threshold capillary pressure in maintaining the fluid--fluid interface in pipes, calculated using the Young-Laplace equation: 
\begin{equation}
  p_{c,ij}^\text{entry} = \dfrac{2\gamma\cos\theta_c}{r_\text{min}},
  \label{eq:young-laplace}
\end{equation}
where $\gamma$ is the interfacial tension, $\theta_c$ is the invading fluid contact angle (180\degree~for drainage flow), and $r_\text{min}=a_{ij}/2$ is the minimum principal curvature radius. 
The flow rate calculation on the flow front, incorporating the capillary entry pressure, can thus be expressed as
\begin{equation}
  q_{ij} = \dfrac{a_{ij}^3}{12\eta_{ij}}\dfrac{\max(p_{f,i}^{n+1}-p_{f,j}^{n+1}+p_{c,ij}^\text{entry},0)}{l_{p,ij}}.
  \label{eq:capi-flow-rate}
\end{equation}

Substituting Eq.~\eqref{eq:capi-flow-rate} into Eq.~\eqref{eq:fmass-balance} yields the finalized formula for the implicit finite volume approach
\begin{equation}
  \dfrac{\delta V_i}{\delta t} + \dfrac{V_i}{K_{f,i}}\dfrac{p_{f,i}^{n+1}-p_{f,i}^n}{\delta t} = -\sum_j \dfrac{a_{ij}^3}{12\eta_{ij}}\dfrac{\max(p_{f,i}^{n+1}-p_{f,j}^{n+1}+p_{c,ij}^\text{entry},0)}{l_{p,ij}}.
  \label{eq:implicit-fva}
\end{equation}
Notably, this implicit approach, owing to its timestep independence, allows for a significantly larger flow timestep (\eg~$10^{-5}$ s) compared with the conventional explicit method, which typically provides a timestep of at most 10$^{-7}$ s, as demonstrated in the slow injection case by Zhang \etal~\cite{zhang2013coupled}. This results in a substantial improvement in simulation efficiency for multiphase flow problems. 

\subsection{Injection-induced grain motion model}
The injection process generates viscous pressures, which act as external forces on grains, leading to grain motion. 
The original model \cite{duan2020initiation} exerts pressures along the line connecting grain centroids, which would, however, overestimate the applied forces when inter-grain distance grows. 
To address this, we consider the pressure-bearing interface at the grain-domain intersecting surface, eliminating the dependence on grain distance, as illustrated in Figure~\ref{fig:pore-network-DEM}b. 
By adopting a local coordinate system ($x'$, $y'$) that bisects the domain corner angle, the applied forces on the grain surface arch can be calculated as
\begin{equation}
  F_p = \int_{-\beta}^\beta p_fr\cos\theta~\dd\theta = 2p_fr\sin\beta,
  \label{eq:applied-force}
\end{equation}
where $\beta$ is the half corner angle of a domain, $p_f$ is the pore pressure, and $r$ is the grain radius \cite{shimizu2011distinct}. 
The force $F_p$ would be decomposed into two lateral components ($F_{px}$ and $F_{py}$) within the global coordinate system ($x$, $y$) and then applied on grains. 
The integrated applied force passes through the grain centroid and does not produce any moment.

The force-displacement law of a linear contact model is employed to capture injection-induced grain motion, resulting in variations in contact forces. The contact system \cite{cundall1979discrete} characterizes linear elastic and frictional behaviors by normal and shear springs as
\begin{equation}
  F_c^n = k_n\delta_n
\end{equation}
and
\begin{equation}
  \Delta F_c^s = -k_s\Delta\delta_s, ~|F_c^s|\leq\mu F_c^n,
\end{equation}
where $F_c^n$ is the normal contact force determined by the grain--grain overlap $\delta_n$ (positive compression), $F_c^s$ is the shear contact force controlled by the slip between grains $\delta_s$, $\mu$ is the friction coefficient, and $k_n$ and $k_s$ represent the normal and shear stiffnesses of the springs. 

Given the set of applied pressures and contact forces (Figure~\ref{fig:pore-network-DEM}c), Newton’s second law governs the translational and rotational motions of each grain. 
The motion equations of grain $i$ can be expressed in vector form as
\begin{equation}
  \sum_k\bm{F}_{c,k} + \sum_j\bm{F}_{p,j} = m_i\Ddot{\bm{x}}_i
  \label{eq:motion-force}
\end{equation}
and
\begin{equation}
  \sum_k\bm{M}_{c,k} = \bm{I}_i\Ddot{\bm{\theta}}_i,
  \label{eq:motion-moment}
\end{equation}
where $\bm{F}_{c,k}$ and $\bm{M}_{c,k}$ are the contact force and moment at contact $k$, $\bm{F}_{p,j}$ is the applied force from an adjacent domain $j$, $m_i$ is the mass, $\bm{I}_i$ is the inertia moment tensor, and $\bm{x}_i$ and $\bm{\theta}_i$ are the position and rotation angle vectors of the grain, respectively. 
It is important to note that the balance of fluid-driven grain motion is a dynamic equilibrium process between the law of motion and the force-displacement law, requiring a simultaneous solution for the entire system.

Fluid injection activity can rearrange pore geometry and alter flow properties in deformable porous media, such as porosity and permeability \cite{bruno1994micromechanics}. Therefore, providing real-time feedback on pore deformation---including domain volume and pipe conductance---is crucial. Up-to-date pore geometry---tracking grain positions and aperture widths---determines domain volume and pipe conductance. The aperture width is characterized by the adjoining grain distance and contact force.
Under compressive normal contact force ($F>0$), it can be calculated as
\begin{equation}
  a = \dfrac{a_0F_0}{F+F_0},
  \label{eq:aperture-width-c}
\end{equation}
where $a_0$ is the residual aperture width when $F=0$, and $F_0$ is the normal contact force when $a=a_0/2$. 
The force-displacement law ceases when contacting grains detach ($F\leq0$). 
We consider this situation fracture initiation in granular media, distinct from the broken bonds in stiff materials (\eg~\cite{li2001simulation,shimizu2011distinct}). The fracturing aperture width can be denoted as
\begin{equation}
  a = a_0 + \lambda(d-r_1-r_2),
  \label{eq:aperture-width-f}
\end{equation}
where $d$ is the distance between adjacent grain centroids, and $\lambda$ is the aperture-width multiplier. 
This multiplier accounts for the substantial permeability of cracks where significant pipe flow can occur \cite{wang2021mechanical}. 
Compared with the merging-fracturing-domain strategy (\eg~\cite{zhang2013coupled,zhou2016numerical}), this approach significantly reduces computational intensity when updating the pore network while preserving the original viscous pressures for advancing pipe flows. Additionally, we update the domain volume and pipe aperture width at each timestep to ensure HM coupling accuracy. 

\section{Numerical implementation}
\label{sec:modeling}

In this section, we implement the fundamental equations of the fully coupled HM-DEM framework using three newly developed numerical components: (i) an implicit pressure solver, (ii) a pressure-volume iteration scheme, and (iii) flow front-advancing criteria. 
These components are developed using the embedded Python programming language in PFC2D. Leveraging a unified platform facilitates data exchange between the two models and expedites simulations. The following subsections describe these custom numerical tools in detail.

\begin{figure}[htbp]
  \centering
  \includegraphics[width=1.0\textwidth]{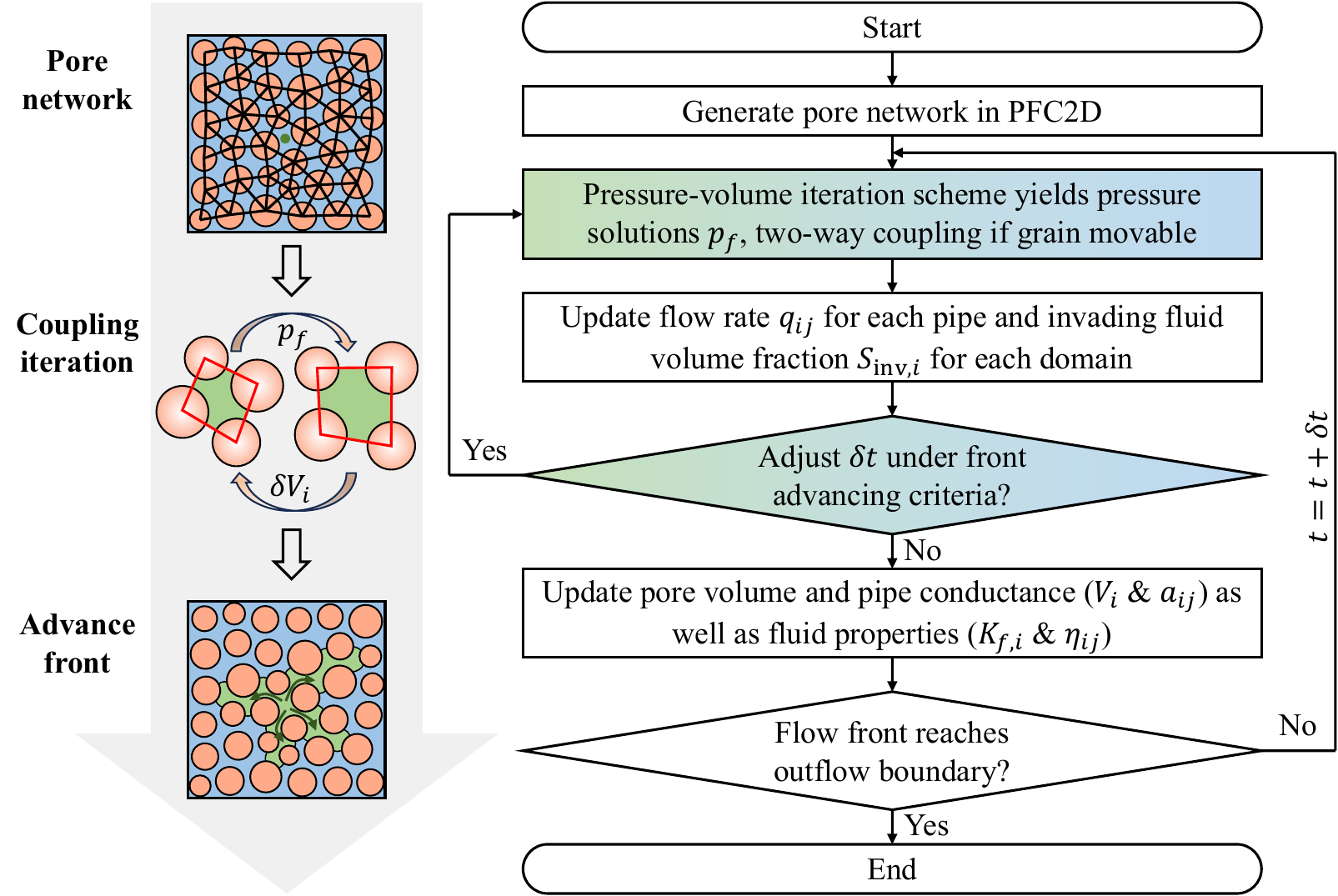}
  \caption{Flowchart of the numerical implementation for the HM-DEM coupled model.}
  \label{fig:implement-flowchart}
\end{figure}

Figure~\ref{fig:implement-flowchart} outlines the complete numerical implementation of the coupled HM-DEM model. 
We begin by discretizing the porous media into a grain-contact interconnected pore network in PFC2D. The implicit pressure solver then provides pressure solutions for moving grains and advancing fluids. Concurrently, the grain motion solver manages the dynamic interactions between grains and contacts in a time-stepping manner, repeatedly applying the laws of motion and force-displacement until force equilibrium is achieved. 
However, the fluid-driven re-arrangement of the pore structure continually challenges pressure solution accuracy. 
We introduce the pressure-volume iteration scheme to synchronize the fluid pressure and grain motion solvers to address this. 
The flow front-advancing criteria enable precise location of the fluid--fluid interface in deformable porous media.
A timestep adjustment occurs when the invading fluid simultaneously advances into two continuous domains or when the capillary effect blocks all potential flow paths of a specific domain. Before initiating the next flow timestep, our model updates the pore volume of each domain ($V_i$), the throat conductance of each pipe ($a_{ij}$), and the fluid bulk modulus ($K_{f,i}$) and viscosity ($\eta_{ij}$) of the front domains. The simulation concludes once the flow front reaches the outflow boundary.

\subsection{Implicit pressure solver}
The implicit pressure solver classifies domains into three types based on pressure conditions: inflow, outflow, and free-pressure domains.
Inflow domains are treated as Neumann boundaries, maintaining a constant flow rate ($q_\text{in}$) as typically employed in Hele-Shaw tests (\eg~\cite{lenormand1988numerical,huang2012granular}). 
Outflow domains are considered Dirichlet boundaries with a constant zero pressure. 
Given these boundary conditions, we can express Eq.~\eqref{eq:implicit-fva} in matrix form for all other domains:
\begin{equation}
  \bm{Ap_f^{n+1}} = \bm{B}.
  \label{eq:general-matrix}
\end{equation}
Entries $\omega$ in coefficient matrix $\bm{A}$ and $o$ in constant matrix $\bm{B}$ can be respectively denoted as
\begin{equation}
  \omega = \begin{cases}
   \dfrac{V_i}{K_{f,i}\delta t} + \displaystyle\sum_{k\in(\mathbb{D}-\mathbb{C})}\dfrac{a_{ik}^3}
   {12\eta_{ik}l_{p,ik}} & \text{if}~i=j \\
   0 & \text{if}~i\not = j \And j \notin (\mathbb{D}-\mathbb{C}) \\
   -\dfrac{a_{ij}^3}{12\eta_{ij}l_{p,ij}} & \text{if}~i\not = j \And j \in (\mathbb{D}-\mathbb{C})
   \end{cases}
\end{equation}
and
\begin{equation}
  o = -\dfrac{\delta V_i}{\delta t} + \dfrac{V_i}{K_{f,i}\delta t}p_{f,i}^n - \displaystyle\sum_{k\in(\mathbb{E}-\mathbb{C})}\dfrac{a_{ik}^3}{12\eta_{ik}l_{p,ik}}p_{c,ik}^\text{entry} + q_{\text{in},i\in\mathbb{F}},
\end{equation}
where $\mathbb{D}$ is the set of surrounding domain indices around domain $i$, $\mathbb{E}$ is the set of domain indices connected to domain $i$ by a pipe containing fluid--fluid interface, $\mathbb{C}$ is the subset of $\mathbb{E}$ where flow is blocked by capillary effects, and $\mathbb{F}$ is the set of inflow domain indices. 
\reviewerone{For simplicity, only the viscous pressure terms include the superscripts $n$ and $n+1$ to distinguish values from the current and advanced timesteps during pressure calculation. All other input parameters, such as $V_i$, $a_{ij}$, $K_{f,i}$, and $\eta_{ij}$, are assumed to represent values at the current timestep unless explicitly stated otherwise.}

We use the conjugate gradient iteration method to solve Eq.~\eqref{eq:general-matrix} and apply the resulting pressure solutions to fluid migration and grain motion calculations. 
The pressure-volume iteration scheme couples the implicit pressure solver with the explicit DEM integration, enabling simultaneous resolution of fluid mass balance, the law of motion, and the force-displacement law. 
Consequently, the fluid flow formulation is effectively implemented within the implicit pressure solver, providing critical quantities such as viscous pore pressures, pipe flow rates, and fluid saturations, which are essential for advancing fluid flow in deformable porous media.

\subsection{Pressure-volume iteration scheme}
The applied forces on grains deform the pore structure, simultaneously altering the initial pressure solutions by incorporating pore volume changes in Eq.~\eqref{eq:implicit-fva}. 
However, the original model for single-phase hydraulic fracturing \cite{duan2020initiation} addresses the coupled problem sequentially: first solving for fluid pressure dissipation, then managing pore deformation due to fluid-driven forces. This approach does not account for pore volume variations in the initial pressure solutions, leading to fluid mass imbalance in deformed domains, such as the injection port. 
Although this error may be negligible for small timesteps, it becomes significant when using a larger timestep. 
To address this, we design the pressure-volume iteration scheme to fully iterate the interaction between fluid pressure and pore deformation, ensuring a two-way hydro-mechanical coupling for pressure solutions.

\begin{figure}[htbp]
  \centering
  \subfloat[]{\includegraphics[width=1.0\textwidth]{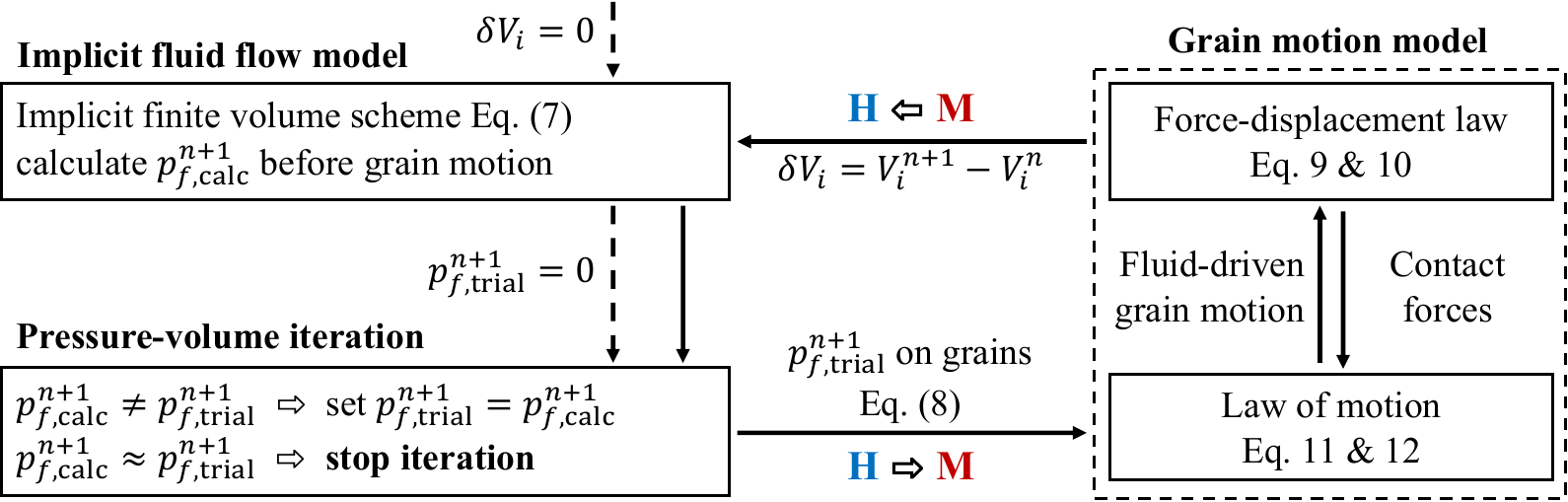}} \\
  \subfloat[]{\includegraphics[width=1.0\textwidth]{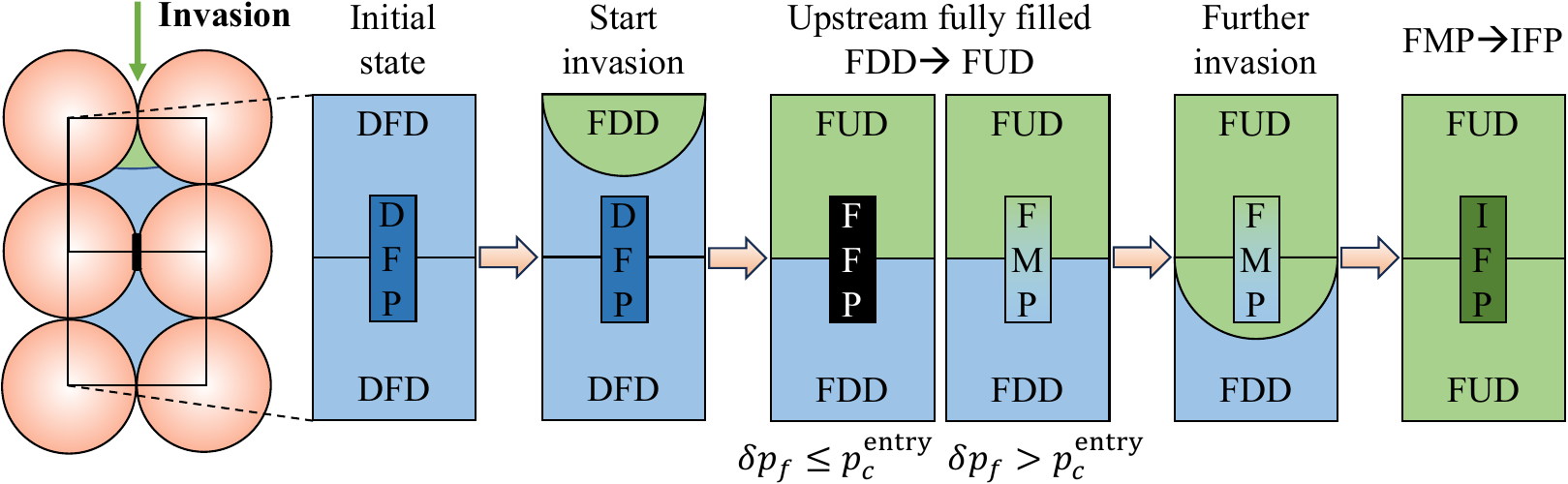}}
  \caption{Key techniques for ensuring numerical accuracy under large timesteps. (a) Pressure-volume iteration scheme and (b) flow front-advancing process between two domains (green indicates invading fluid; blue indicates defending fluid).}
  \label{fig:key-techniques}
\end{figure}

Figure~\ref{fig:key-techniques}a illustrates the detailed pressure-volume iteration process. Central to this process are two advanced-timestep quantities in Eq.~\eqref{eq:implicit-fva}: the domain volume change $\delta V_i$ and the pore pressure $p_{f,i}^{n+1}$. 
The domain volume change $\delta V_i$ is determined by subtracting the current domain volume $V_i^n$ from the advanced one $V_i^{n+1}$. 
Starting from $\delta V_i = 0$, the implicit pressure solver first provides a trial solution $p_{f,\text{trial}}^{n+1}$ (initially zero) for the grain motion model. 
The iteration scheme then updates $p_{f,\text{trial}}^{n+1}$ to match the calculated pore pressure $p_{f,\text{calc}}^{n+1}$. 
Following grain mobilization, the pressure solver recalculates the pressure with the updated pore volume, which is then compared with the previous solution. This iteration process between the implicit pressure solver and grain motion model continues until the difference between $p_{f,\text{trial}}^{n+1}$ and $p_{f,\text{calc}}^{n+1}$ is negligible. 
This convergence ensures that fluid pressure changes harmonize with pore volume variations. 
Although the iteration scheme is applied to all domains, convergence is controlled by the pressure-volume equilibrium at the injection port, where the most significant pore volume changes occur \cite{macminn2015fluid}. 
The fluid flow timestep used for iteration is determined by the fluid flow advancing criteria in the following subsection.
Given that the grain motion model employs an explicit time integration scheme for computational efficiency, we adopt a ratio of the mechanical timestep to the fluid flow timestep of $1/2000$, as recommended by Meng~\etal~\cite{meng2020jamming}, to ensure adequate energy dissipation in grain motion. 

\subsection{Flow front-advancing criteria}
Unlike single-phase hydraulic fracturing problems, implementing fluid flow formulation in multiphase flow problems necessitates continuously updating fluid properties (fluid bulk modulus and viscosity) and monitoring saturations in mixing fluid domains. 
Defining and tracking the dynamic fluid--fluid interface is particularly critical when using remarkable timesteps. To address this challenge, we incorporate the flow front-advancing scheme proposed by Holtzman and Segre~\cite{holtzman2015wettability}, which identifies the flow front as a boundary separating upstream and downstream regions. The front boundary categorizes domains into defending-fluid domain (DFD), front-downstream domain (FDD), and front-upstream domain (FUD). 
Specifically, DFD represents a domain fully saturated with defending fluid, while FUD signifies a domain fully occupied by invading fluid. FDD, situated at the front interface, contains both defending and invading fluids. 
We define four relevant pipe types to determine fluid type during migration: defending-fluid pipe (DFP), invading-fluid pipe (IFP), front-moving pipe (FMP), and front-fixed pipe (FFP). 
The first two pipes carry either defending or invading fluid, while the last two reveal whether the capillary effect is sufficient to impede fluid flow. 

Front advancements invariably lead to transformations in domain and pipe types. Figure~\ref{fig:key-techniques}b demonstrates the key moments of front advancement during a domain-to-domain invasion process. 
When the invading fluid begins to touch the upper domain boundary, the DFD domain promptly converts to an FDD domain. Once fully infiltrated, the domain transitions into an FUD domain. 
Simultaneously, the lower DFD domain shifts to an FDD domain. 
The connecting pipe between these domains is replaced with either an FFP or FMP pipe, depending on whether the pressure difference exceeds the capillary entry pressure. 
As the flow front continues moving forward, the downstream domain and pipe eventually evolve into an FUD domain and an IFP pipe. 
The invasion process is replicated at every front-interface intersecting domain, enabling our model to produce precise fluid--fluid displacement patterns. 

Furthermore, establishing timestep-adjustment criteria is crucial for accurately capturing flow-patterns and achieving convergence in pressure solutions under capillary effects. 
When using a sizeable timestep, if the advanced-timestep invading fluid saturation in any front domain exceeds one ($S_{\text{inv},i}^{n+1} > 1$)---indicating that the flow timestep is too large---we utilize the excess invading fluid volume and the computed flow rate to estimate a refined timestep that precisely fills the domain:
\begin{equation}
  \delta t_{\text{new}} = \dfrac{V_i(1-S_{\text{inv},i}^n)}{\sum_j q_{ij,\text{inv}}},
  \label{eq:new-time-one}
\end{equation}
where $S_{\text{inv},i}^n$ is the original invading fluid saturation and $q_{ij,\text{inv}}$ is the flow rate through adjoining pipes transporting invading fluids. 

In slow drainage scenarios, capillary effects can entirely block surrounding pipes in certain domains. Traditional incompressible-fluid pore-network models (\eg~\cite{lenormand1988numerical,yang2019modeling}) typically require manual selection and opening of a blocked pipe to avoid nil-pressure solutions. 
Although our compressible-fluid model circumvents this issue, estimating an appropriate timestep remains crucial to prevent excessive pore pressure buildup in a flow path-blocked domain, which could otherwise erroneously lead to viscous fingering instead of capillary fingering. 
Considering fluid compressibility---where pressure accumulates in the upstream domain $i$ and dissipates in the downstream domain $j$---the timestep for opening the blocked front pipe with the minimum capillary entry pressure ($p_{c,\text{min}}^\text{entry}$) can be calculated as
\begin{equation}
  \delta t_{\text{new}} = \chi\dfrac{-p_{c,\text{min}}^\text{entry} - p_{f,i}^n + p_{f,j}^n + \dfrac{\delta V_i}{V_i}K_{f,i}-\dfrac{\delta V_j}{V_j}K_{f,j}}{-\dfrac{K_{f,i}}{V_i}\sum_k q_{ik} + \dfrac{K_{f,j}}{V_j}\sum_k q_{jk}},
  \label{eq:new-time-two}
\end{equation}
where $\chi$ is a timestep expanding coefficient (slightly greater than one such as 1.005), ensuring that the pressure difference gradually decreases to the minimum capillary entry pressure, facilitating the opening of the corresponding pipe. 
The model cycles through the HM coupling components until a well-adjusted minimum timestep is obtained using Eq.~\eqref{eq:new-time-one} and Eq.~\eqref{eq:new-time-two}.

\section{Numerical simulations}
\label{sec:parameter-studies}

In this section, we construct both linear and radial flow models to validate the developed HM-DEM framework and explore the underlying mechanisms of multiphase flow in porous media. 
The simulations involve two distinct yet complementary schemes: (i) linear injection tests in fixed-grain porous media to verify the feasibility of producing realistic fluid--fluid displacement patterns and (ii) radial injection tests in movable-grain porous media to demonstrate the capability to replicate complex fluid--grain interactions.

Intrinsic permeability has been proven vital in governing multiphase flow patterns \cite{jain2009preferential,zhang2013coupled,carrillo2021capillary}. 
To harmonize fluid discharge abilities between numerical and experimental samples, we calibrate permeability before injection tests by arbitrarily varying porosity, grain size, and aperture width until the experimental permeability is achieved. 
Two numerical permeabilities are determined for each randomly packed assembly: (1) a geometry-based value derived from the Kozeny-Carman equation \cite{bear2013dynamics} and (2) a value predicted by Darcy flow test from the partial differential equation of diffusion \cite{crank1979mathematics}. 
These two permeabilities mutually verify each other, ensuring that macroscopic fluid conductance aligns with pore-scale behavior. The permeabilities can be expressed as
\begin{equation}
  k_\text{Kozeny-Carman} = \dfrac{1}{180}\dfrac{d^2\phi_{3D}^3}{(1-\phi_{3D})^2}
  \label{eq:kozeny-carman}
\end{equation}
and
\begin{equation}
  k_\text{Darcy-flow} = \dfrac{\eta WQ_\text{in}}{H\Delta P},
  \label{eq:darcy-flow}
\end{equation}
where $d$ is the average grain diameter, $\phi_{3D}$ are the three-dimensional (3D) porosity, $Q_\text{in}$ is the total inflow rate, $\Delta P$ is the steady-state pressure difference across the assembly, and $H$ is the sample height. 
To convert the experimental porosity for 2D applications, we use the linear relationship between 2D and 3D porosities \cite{zhang2013coupled}, expressed as
\begin{equation}
  \dfrac{\phi_{2D} - 0.0931}{0.2146 - 0.0931} = \dfrac{\phi_{3D} - 0.2595}{0.4764 - 0.2595},
  \label{eq:porosity-change}
\end{equation}
where the values represent the porosities of disks (2D) and spheres (3D) packed in the two extreme geometries of regular square (loosest) and hexagonal (densest) assemblies \cite{deresiewicz1958mechanics}. 

To ensure comparability between numerical results and experimental observations, we align injection parameters and hydro-mechanical properties with those used in the experiments. 
We quantify the similarity of injection parameters, such as inflow rate and fluid viscosity, using dimensionless numbers---namely, the viscosity ratio $M$ and the modified capillary number $Ca^*$, given by
\begin{equation}
  M = \dfrac{\eta_\text{inv}}{\eta_\text{def}}
\end{equation}
and
\begin{equation}
  Ca^* = \dfrac{q_\text{in}\eta_\text{inv}}{\sum\gamma\cos\theta_c}\dfrac{L}{\bar{a}},
  \label{eq:ca-number}
\end{equation}
where $\sum$ is the outflow area of the injection port in 2D---equal to the sample height in linear flow or domain perimeter in radial flow---$L$ is the sample length, and $\bar{a}$ is the mean pipe aperture width. 
The modified capillary number $Ca^*$, proposed by Holtzman~\etal~\cite{holtzman2012capillary}, is adopted to eliminate the influence of sample size differences (\ie~$L$ and $\bar{a}$) between numerical and experimental setups, as these differences can alter flow pattern transitions. Specifically, a larger sample length increases the pressure gradient, while a smaller aperture width heightens capillary pressure; these effects must be considered to ensure accurate comparisons.

\reviewertwo{
Notably, our numerical simulations are conducted in a two-dimensional framework, while the experiments are performed in three dimensions. In our 2D model, grains are represented as connected disks with artificial throats for fluid migration. 
This simplification does not fully capture the vertical heterogeneity of pore structures and the complex grain motions present in 3D systems; however, the small thickness-to-length ratio of Hele-Shaw cell---remaining below 0.01 \cite{saffman1958penetration}---ensures that the 3D flow behavior remains comparable to the 2D model. 
By matching permeability and using the same dimensionless parameters ($M$ and $Ca^*$) as in the experiments, we can reproduce the relevant fluid-fluid and fluid-grain displacement patterns.
}

All simulations are conducted on a computer with an 8-Core i7-4790 CPU and 32 GB of RAM. Injection flow simulations take several hours to a full day, depending on the sample dimension, selected timestep, and infiltration scale of the reproduced flow patterns.  
\reviewerone{The robustness and accuracy of the model are validated through supplementary tests, including a resolution convergence study (\ref{appendix_mesh_size}), timestep sensitivity analysis (\ref{appendix_timestep}), pressure solver performance evaluation (\ref{appendix_pressure_solver}), and verification using simple flow scenarios, such as single-phase Darcy flow and two-phase flow in regular packings (\ref{appendix-simple-flow}). Comprehensive details of these validations are provided in the appendices.}

\subsection{Injection tests in fixed-grain porous media}

\subsubsection{Model setup}
\begin{figure}[htbp]
  \centering
  \includegraphics[width=1.0\textwidth]{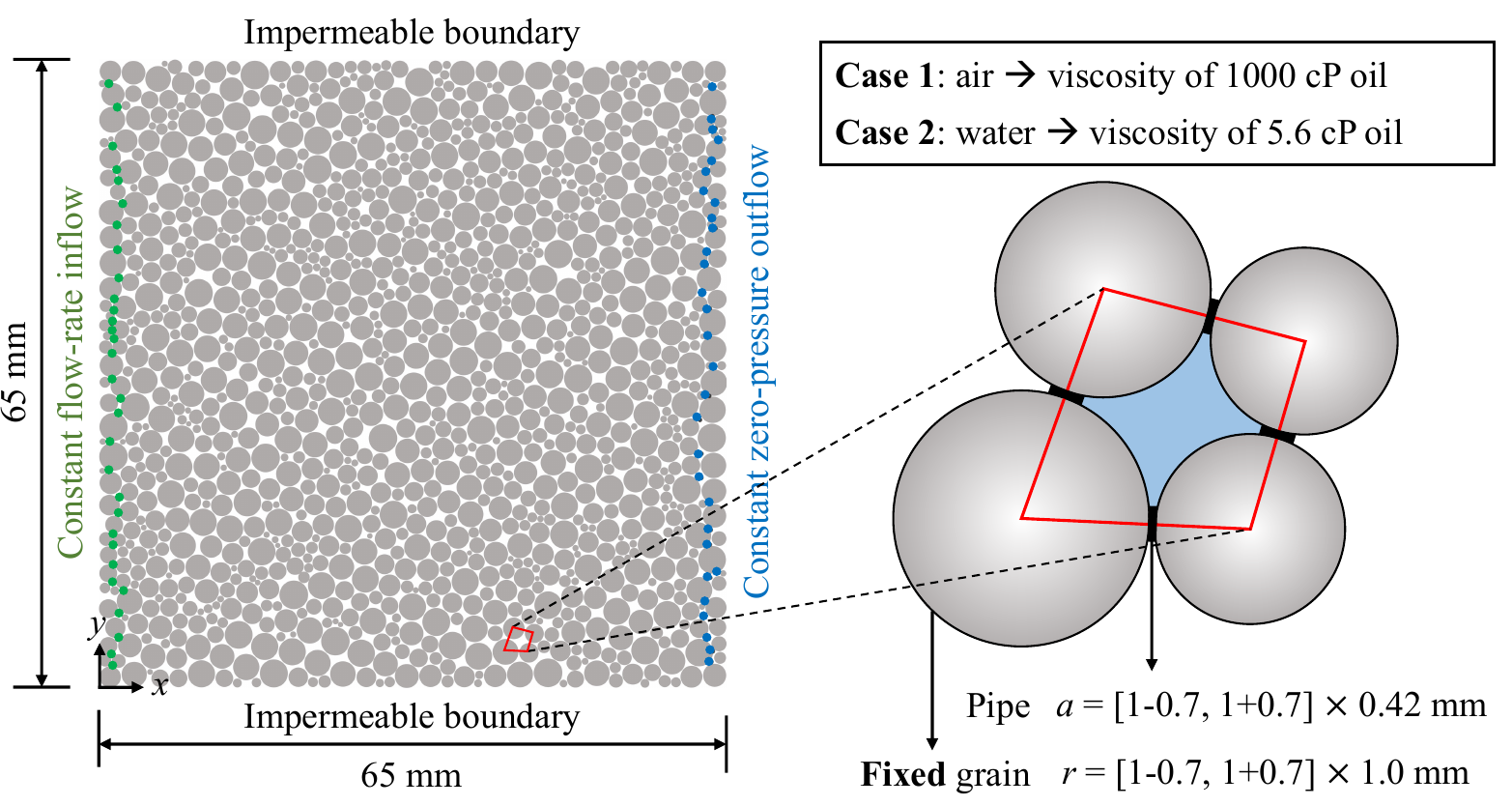}
  \caption{Setup for injection tests in fixed-grain porous media.}
  \label{fig:fixed-setup}
\end{figure}

The fluid injection tests in fixed-grain porous media are based on Lenormand’s experiment \cite{lenormand1988numerical}, a benchmark for fluid--fluid interaction in numerical models (\eg~\cite{primkulov2021wettability}). Lenormand’s experiment used a transparent etched mold with randomly distributed, equally spaced ducts. 
A syringe pump maintained a constant flow rate to inject non-wetting fluids (air or water) laterally into the mold, which was pre-saturated with a wetting fluid (oil). 
As shown in Figure~\ref{fig:fixed-setup}, we construct a grain-fixed linear flow model with drained left-right and impervious top-bottom boundaries. 
The 65 $\times$ 65 mm square assembly conforms to typical granular media, consisting of 1188 uniformly distributed particles with a porosity of 0.117. 
The sample---with an average grain radius of 1.0 mm and an average pipe aperture width of 0.42 mm (distributed uniformly within the interval $[1-0.7, 1+0.7]$)---yields two numerical permeabilities that match with the experimental value (10$^{-9}$ m$^2$): 1.025 $\times$ 10$^{-9}$ m$^2$ using Eq.~\eqref{eq:kozeny-carman} and 1.066 $\times$ 10$^{-9}$ m$^2$ using Eq.~\eqref{eq:darcy-flow}. 
The assembly is further discretized into a pore network comprising 1874 domains and 3061 pipes. 
A constant flow rate is applied to the left inflow boundary, while zero pressure is maintained at the right outflow boundary. 
The remaining domains are set to zero pressure at the initial stage.

To reproduce the observations of Lenormand’s experiment---namely, viscous fingering, capillary fingering, and stable displacement---we conduct two drainage scenarios with viscosity ratios of $1.8\times10^{-5}$ and 101.8. 
Each scenario includes two injection tests with different modified capillary numbers: (i) air displacing very-viscous oil with $Ca^*$ of $1.9\times10^{-4}$ and $3.1\times10^{-7}$, and (ii) glucose solution (mainly water) displacing less-viscous oil with $Ca^*$ of 38.6 and $2.4\times10^{-3}$. 
Table~\ref{tab:fixed-hydraulic} summarizes the flow properties for these drainage cases, ensuring that all hydraulic properties are consistent with Lenormand’s experiment. 
The fluid bulk modulus values are drawn from the literature \cite{klaus1964precise,shimizu2011microscopic}. 
\begin{table}[h!]
    \centering
    \begin{tabular}{l|l|l|l|l}
        \toprule 
        Parameter & Symbol & Unit & Air--oil & Water--oil \\
        \midrule 
        Invading fluid viscosity & $\eta_\text{inv}$ & cP & 0.018 & 570 \\
        Invading fluid modulus & $K_{f,\text{inv}}$ & MPa & 0.14 & 2000 \\
        Defending fluid viscosity & $\eta_\text{def}$ & cP & 1000 & 5.6 \\
        Defending fluid modulus & $K_{f,\text{def}}$ & MPa & 2000 & 2000 \\
        Interfacial tension & $\gamma$ & dyn$/$cm & 20.0 & 14.5 \\
        Contact angle & $\theta_c$ & \degree & 180 & 180 \\
        Aperture width & $a$ & mm & $[0.126,0.714]$ & $[0.126,0.714]$ \\
        \bottomrule
    \end{tabular}
    \caption{Hydraulic properties for injection tests in fixed-grain porous media.}
    \label{tab:fixed-hydraulic}
\end{table}

\subsubsection{Results}
\begin{figure}[htbp]
  \centering
  \includegraphics[width=1.0\textwidth]{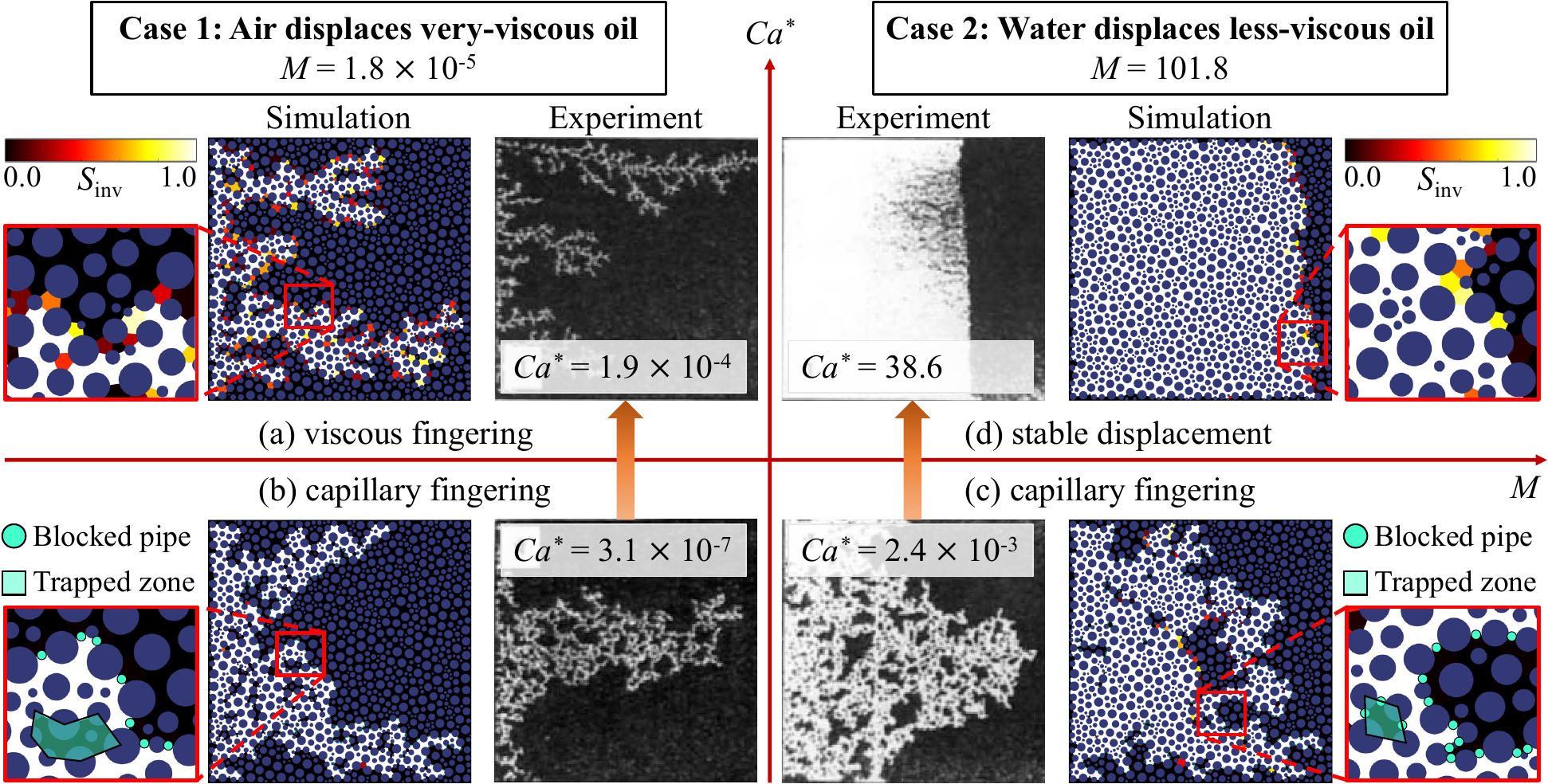}
  \caption{Comparison of fluid--fluid displacement patterns at breakthrough between numerical simulations and observations from Lenormand’s experiment. As the capillary number decreases, the flow regime transitions from (a) viscous fingering to (b) capillary fingering in the case of air displacing very-viscous oil  (Case 1), and from (d) stable displacement to (c) capillary fingering in the case of water displacing less-viscous oil (Case 2).}
  \label{fig:fixed-pattern}
\end{figure}

Figure~\ref{fig:fixed-pattern} qualitatively compares the simulated fluid--fluid displacement patterns with those observed in Lenormand’s experiment. 
The simulated flow patterns closely match the experimental results, effectively capturing the distinct morphological features across various capillary numbers and viscosity ratios. 
Unlike previous simulations of Lenormand's experiment (\eg~\cite{primkulov2021wettability}), which assume incompressible fluids to achieve significant timestep, our implicit pressure solver allows for relatively large timesteps, greatly enhancing simulation efficiency. 
Additionally, our flow front-advancing criteria overcome the limitations of conventional HM-DEM coupled models (\eg~\cite{zhang2013coupled}), enabling accurate representation of diverse fluid-fluid displacement patterns.

Specifically, at a higher $Ca^*$ in Case 1 (Figure~\ref{fig:fixed-pattern}a), the flow pattern exhibits the characteristic tree-like viscous fingering, which spreads toward the outflow boundary and develops sub-branches. 
As $Ca^*$ decreases, Case 1 (Figure~\ref{fig:fixed-pattern}b) displays capillary fingering, characterized by thicker fingers with fewer sub-branches than viscous fingering. 
The capillary effect obstructs the front flow paths when the capillary entry pressure exceeds the viscous pressure drop, visibly marked by blocked pipes---\ie~front-fixed pipes (FFP).  
The invading fluid must then grow vertically or even backward, potentially trapping the defending fluid if it forms a closed loop. 
A similar pattern emerges in Case 2 at a lower $Ca^*$ (Figure~\ref{fig:fixed-pattern}c), where the capillary effect again dominates. 
However, in this case, the main finger branch is significantly thicker than in Figure~\ref{fig:fixed-pattern}b due to the substantial pressure generated by the high-viscous invading fluid.  
Under fast flow, the viscous pressure significantly increases with $Ca^*$ and eventually surpasses the capillary entry pressure, leading to a flat-front stable displacement (Figure~\ref{fig:fixed-pattern}d).

\reviewerone{
Beyond qualitative morphology, the flow regimes can be quantitatively classified by their fractal dimension using the box-counting method \cite{schroeder2009fractals} (see \ref{appendix_fractal_dimension}). 
As $Ca^*$ increases, the fractal dimension decreases in the low $M$ case (Case 1), transitioning from less-branched capillary fingering (1.83) to more-branched viscous fingering (1.63).
In contrast, in the high $M$ case (Case 2), the fractal dimension increases with $Ca^*$, evolving from less-branched capillary fingering (1.84) to flat-front stable displacement (1.96). 
These values align with experimentally determined ranges: viscous fingering [1.62, 1.65] \cite{lovoll2004growth, maaloy1985viscous}, capillary fingering [1.80, 1.83] \cite{lenormand1983mechanisms}, and the typical value of 1.95 for stable displacement \cite{zhang2012pattern}. This demonstrates that our model successfully captures key regime transitions, including the shift from viscous fingering to capillary fingering and the transition from stable displacement to capillary fingering.
}

\begin{figure}[htbp]
  \centering
  \subfloat[]{\includegraphics[width=1.0\textwidth]{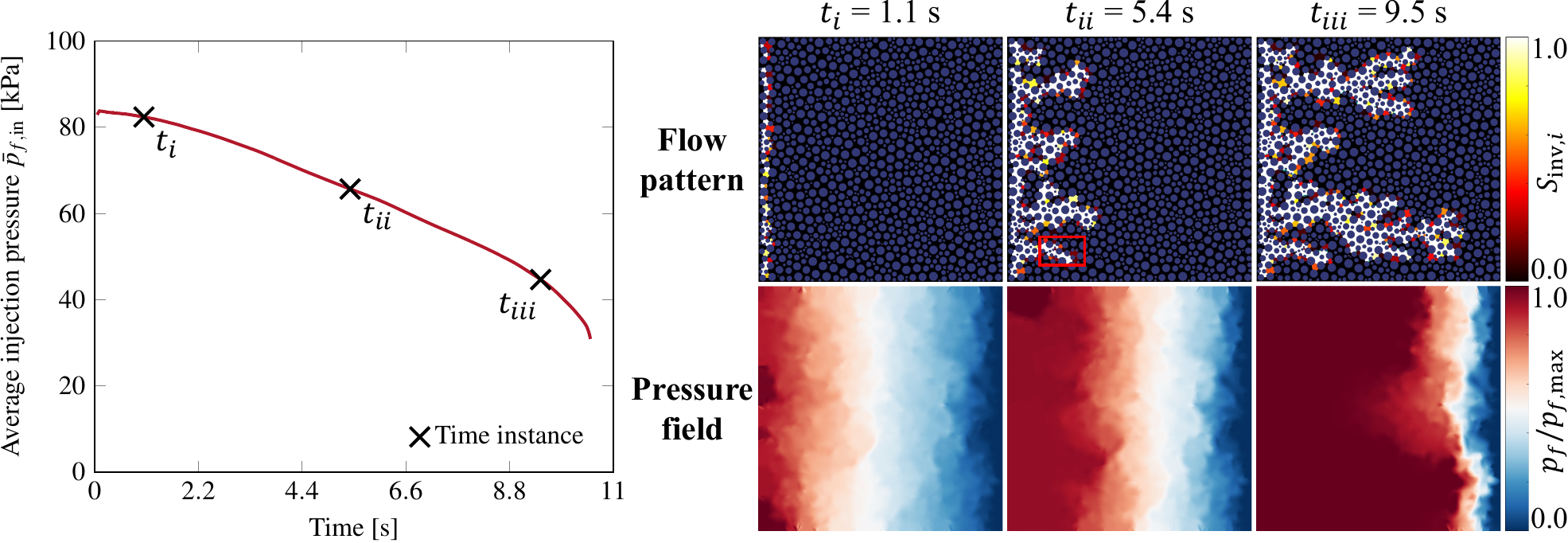}} \\
  \subfloat[]{\includegraphics[width=1.0\textwidth]{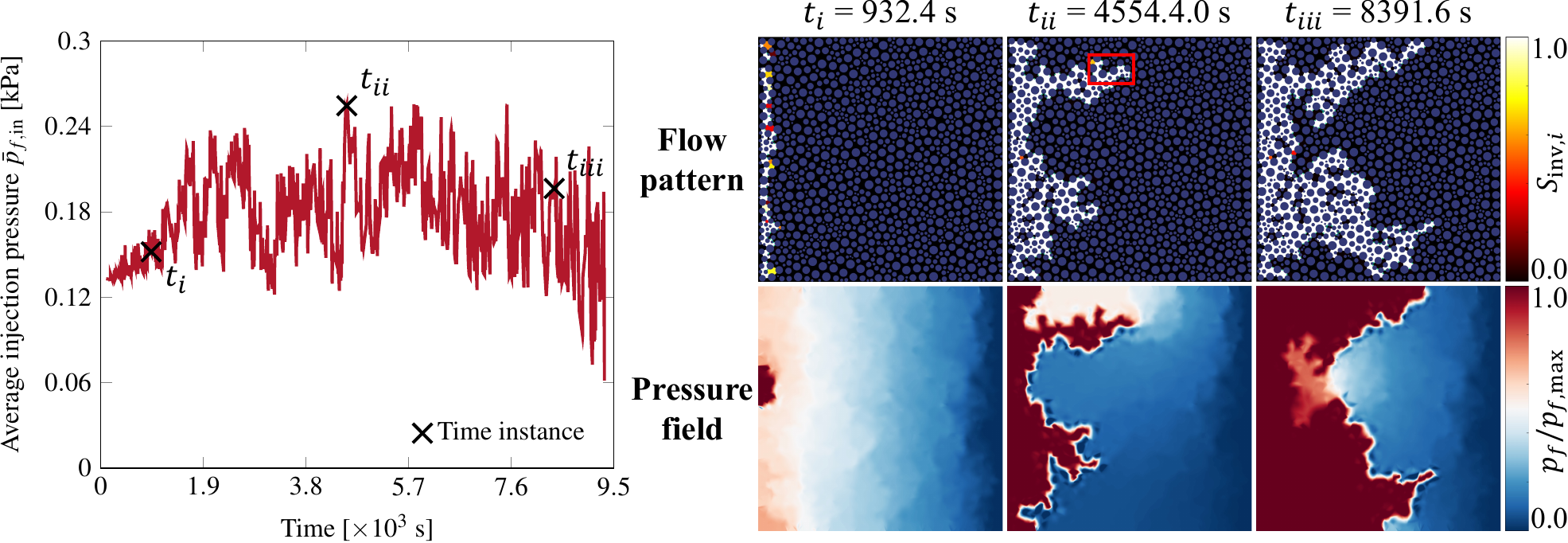}} 
  \caption{Temporal evolution of average injection pressure ($\bar{p}_{f,\text{in}}$) in Case 1 ($M = 1.8\times 10^{-5}$), with selected flow patterns and pressure fields at three time instances. (a) Viscous fingering ($Ca^* = 1.9\times 10^{-4}$) and (b) capillary fingering ($Ca^* = 3.1\times 10^{-7}$).}
  \label{fig:fixed-press-one}
\end{figure}

\begin{figure}[htbp]
  \centering
  \subfloat[]{\includegraphics[width=1.0\textwidth]{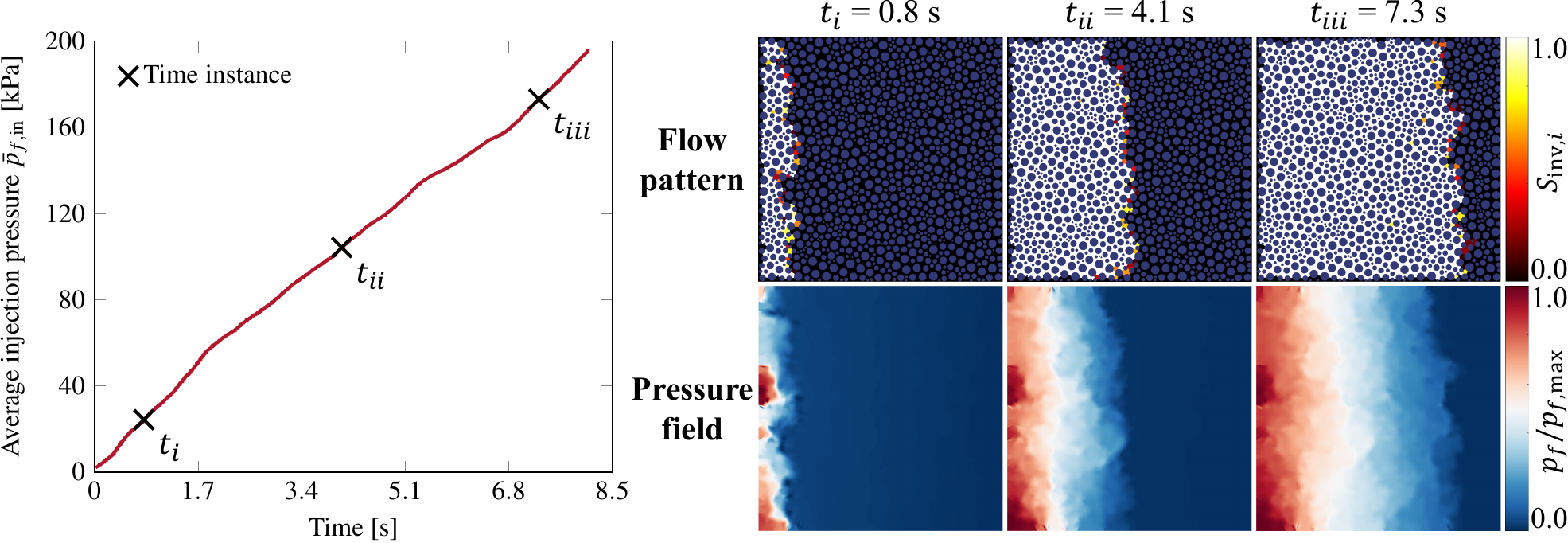}} \\
  \subfloat[]{\includegraphics[width=1.0\textwidth]{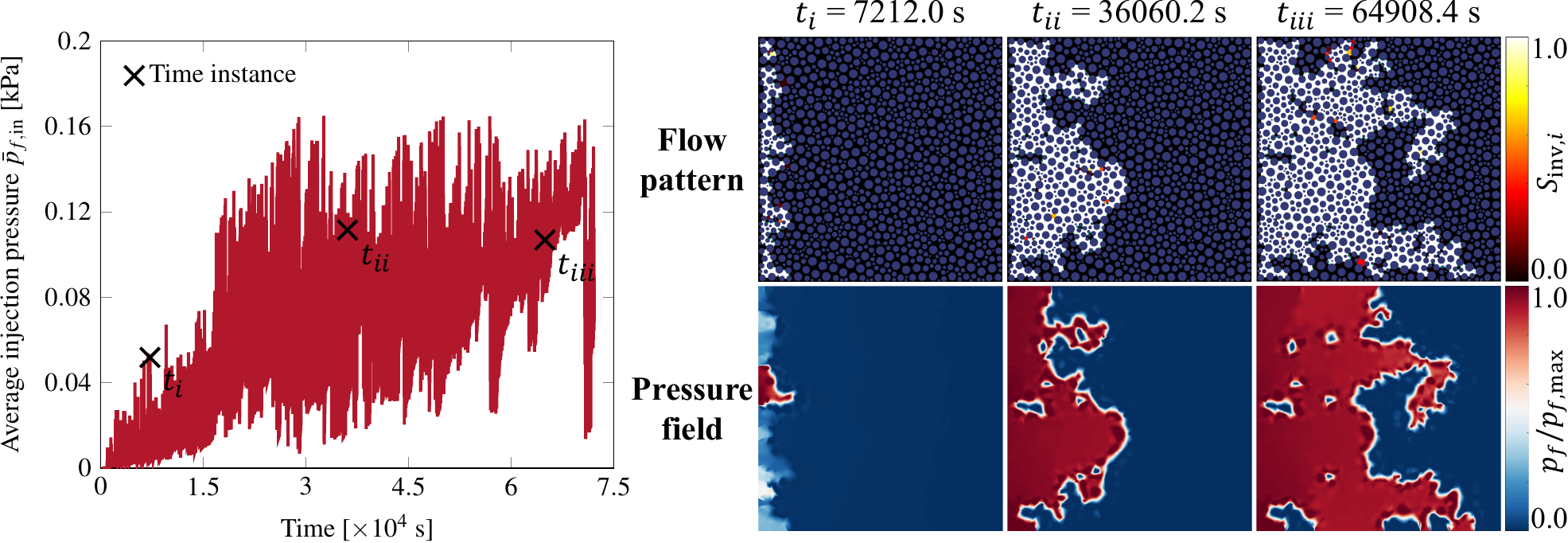}} 
  \caption{Temporal evolution of average injection pressure ($\bar{p}_{f,\text{in}}$) in Case 2 ($M = 101.8$), with selected flow patterns and pressure fields at three time instances. (a) Stable displacement ($Ca^* = 0.25$) and (b) capillary fingering ($Ca^* = 1.6\times 10^{-5}$).}
  \label{fig:fixed-press-two}
\end{figure}

Figure~\ref{fig:fixed-press-one} and Figure~\ref{fig:fixed-press-two} illustrate the fundamental mechanisms driving different flow patterns and their transitions in Case 1 and Case 2, respectively. They depict the temporal evolution of the average injection pressure across all inlet domains ($\bar{p}_{f,\text{in}}$), highlighting three specific time instances ($t/t_\text{max}=$ 0.1, 0.5, and 0.9) to illustrate how the flow pattern and pressure field evolve over time.

In Case 1, with larger $Ca^*$ (Figure~\ref{fig:fixed-press-one}a), the high-viscous defending fluid (oil) initially plays a dominant role in resisting the invading fluid, causing an early injection pressure peak of 83.77 kPa. 
This substantial pressure is attributed to the application of $Ca^*$ in Eq.~\eqref{eq:ca-number}, which significantly increases the inflow rate compared with the experiment, as our simulated sample length (65 mm) is smaller and the aperture width (0.42 mm) is greater than those in the experimental setup (150 mm and 0.35 mm, respectively).
However, as the less-viscous invading fluid (air) gradually occupies the flow path toward the outflow boundary, the pressure steadily decreases to 30.89 kPa. 
The further the finger branches of invading fluid reach, the less impact the defending fluid has on resisting fluid dissipation---reflected by the fading color gradient in the pressure field.
When $Ca^*$ decreases, the initial injection pressure sharply drops to 0.13 kPa (Figure~\ref{fig:fixed-press-one}b), only 0.16$\%$ of the peak pressure observed during viscous fingering and below the capillary entry pressure. 
The capillary effect consequently blocks the flow front, creating two distinct pressure zones, as shown in the pressure field of Figure~\ref{fig:fixed-press-one}b. 
The flow front advances only when the accumulated pressure difference exceeds the capillary entry pressure, leading to intermittent flow and fluctuations in the injection pressure.

\begin{figure}[htbp]
  \centering
  \includegraphics[width=1.0\textwidth]{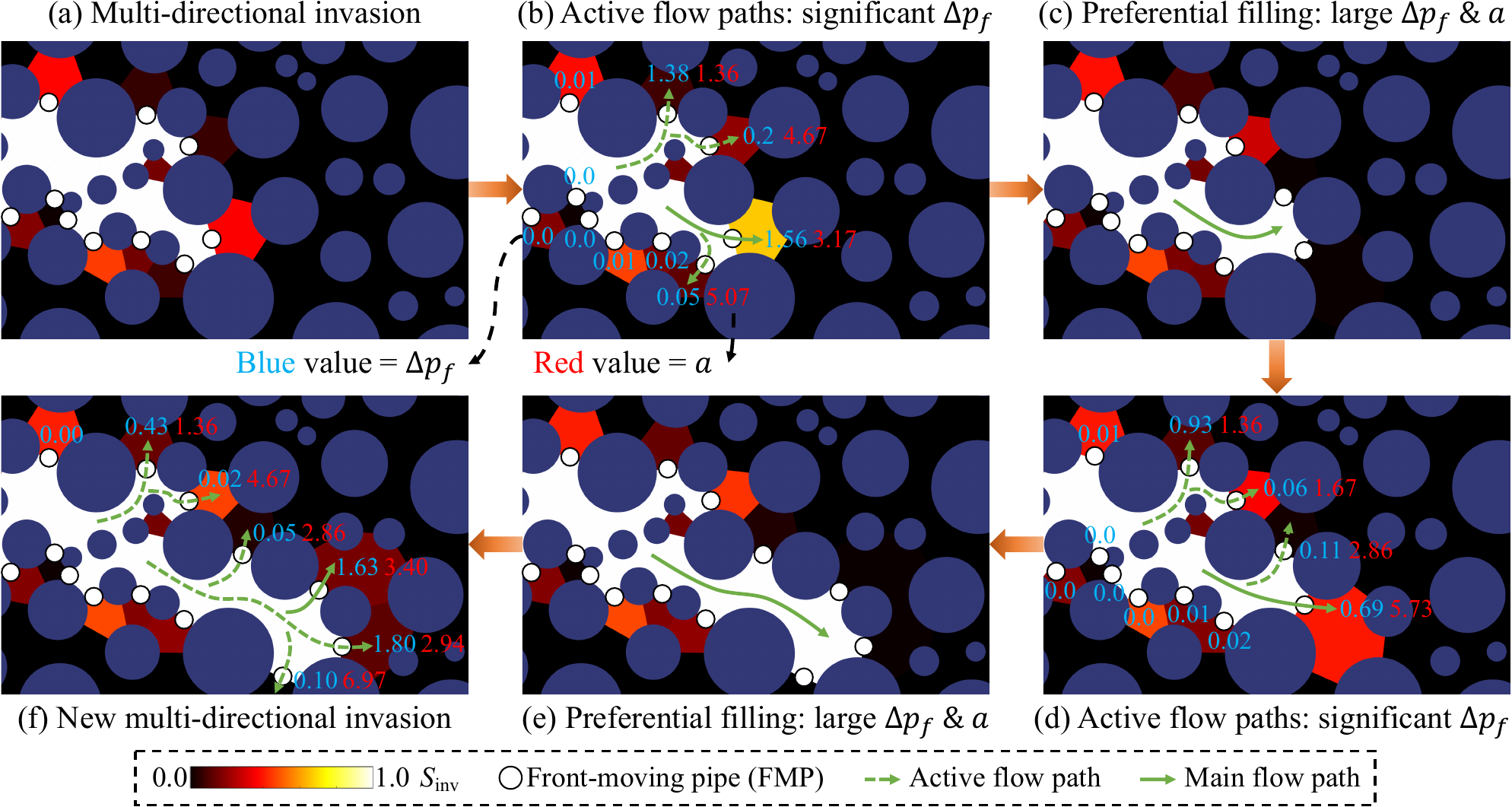}
  \caption{Micro-mechanisms of viscous fingering, with close-ups of the front fingertip in Figure~\ref{fig:fixed-press-one} at sequential timestpes around $t_{ii}$. Blue values represent viscous pressure differences (kPa), red values represent aperture widths ($\times 10^{-4}$ m), and green arrows highlight preferential flow paths.}
  \label{fig:fixed-micro-mechanism-vf}
\end{figure}

\begin{figure}[htbp]
  \centering
  \includegraphics[width=1.0\textwidth]{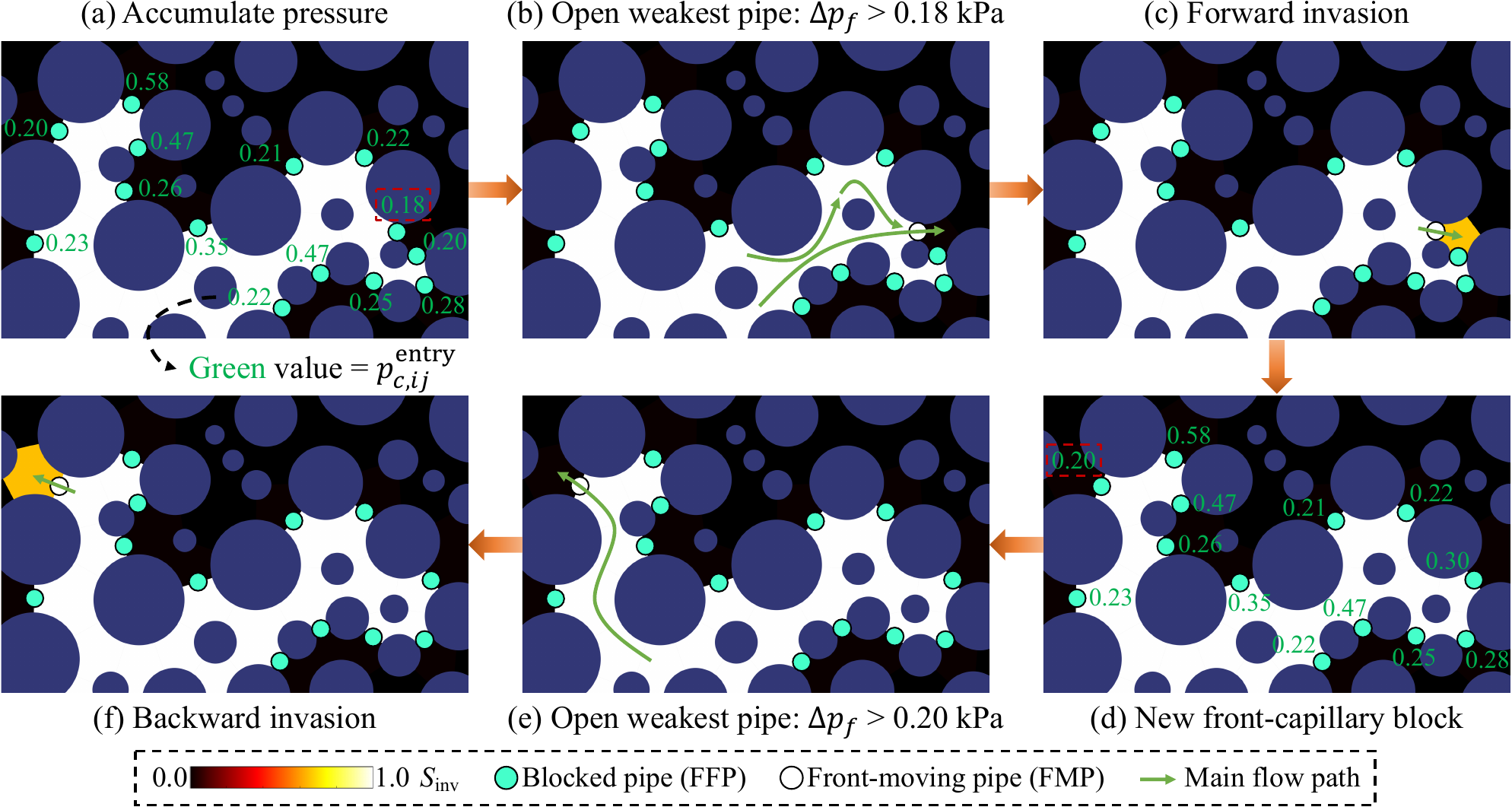}
  \caption{Micro-mechanisms of capillary fingering, with close-ups of the front fingertip in Figure~\ref{fig:fixed-press-one} at sequential timestpes around $t_{ii}$. Green values represent capillary entry pressure (kPa), red dashed boxes indicate the weakest capillary entry-pressure pipe, and green arrows highlight preferential flow paths.}
  \label{fig:fixed-micro-mechanism-cf}
\end{figure}

To explore the micro-mechanisms behind viscous fingering (Figure~\ref{fig:fixed-micro-mechanism-vf}) and capillary fingering (Figure~\ref{fig:fixed-micro-mechanism-cf}), we focus on the evolution of the fingertip during the second time instance ($t_{ii}$) shown in Figure~\ref{fig:fixed-press-one}.
This analysis includes grain-scale viscous pressure differences, aperture widths, and capillary entry pressures.

In Figure~\ref{fig:fixed-micro-mechanism-vf}a, the potential invasion paths at the viscous fingertip are multi-directional. 
As the invading fluid displaces the defending fluid into adjacent domains, viscous pressure increases.
However, the rate of pressure dissipation is primarily influenced by the discharge distance between the defending-fluid domain and the outflow boundary. Shorter distances offer less resistance to dissipation, causing the viscous pressure in the defending-fluid domain at the fingertip front to dissipate more rapidly.
As a result, the front domains develop significant pressure differences, leading to active flow paths, while the domains behind them remain with nearly zero pressure difference (Figure~\ref{fig:fixed-micro-mechanism-vf}b).
Over time, the domain exhibiting a relatively significant pressure difference ($\Delta p_f=1.56$ kPa) and aperture width ($a=3.17 \times 10^{-4}$ m) fills first (Figure~\ref{fig:fixed-micro-mechanism-vf}c). 
The flow then infiltrates the next preferential domain ($\Delta p_f=0.69$ kPa and $a=5.73 \times 10^{-4}$ m) (Figure~\ref{fig:fixed-micro-mechanism-vf}d), which also presents a larger pressure difference due to its short dissipation distance to the outflow boundary.
This domain fills shortly afterward (Figure~\ref{fig:fixed-micro-mechanism-vf}e). 
Given the randomly distributed aperture widths, preferential flow paths are likely to form in vertical directions, potentially developing sub-branches if the aperture widths of the front domain are small.
In Figure~\ref{fig:fixed-micro-mechanism-vf}f, the side pipe with $\Delta p_f=1.63$ kPa and $a=3.40 \times 10^{-4}$ m dominates. Nevertheless, the main branch of viscous fingering continues to advance, as significant pressure differences are concentrated at the leading fingertip.  

Unlike viscous fingering, the flow front paths in capillary fingering (Figure~\ref{fig:fixed-micro-mechanism-cf}a) can be entirely blocked by capillary effects. 
The compressible invading fluid, having no available path, accumulates in upstream domains, causing the invasion pressure to rise until it eventually opens the pipe with the lowest capillary entry pressure (0.18 kPa) (Figure~\ref{fig:fixed-micro-mechanism-cf}b). 
The adjacent domain then fills after the pipe opens (Figure~\ref{fig:fixed-micro-mechanism-cf}c). 
However, as the viscous pressure difference remains lower than the capillary entry pressure, the fingertip is blocked once again (Figure~\ref{fig:fixed-micro-mechanism-cf}d). 
Instead of advancing, the invasion front stays in place until the pressure difference surpasses the lowest capillary entry pressure (0.20 kPa) of a backward pipe (Figure~\ref{fig:fixed-micro-mechanism-cf}e), leading to the invasion of a backward domain (Figure~\ref{fig:fixed-micro-mechanism-cf}f). 
Due to the variability in capillary entry pressures within the randomly distributed aperture-width morphology, opening a pipe requires varying pressure differences, resulting in the injection pressure oscillations observed in Figure~\ref{fig:fixed-press-one}b.

In Case 2, the flow patterns are primarily driven by the viscous pressure gradient of the invading fluid (water), attributed to its high viscosity. 
As the invading fluid gradually displaces the less-viscous defending fluid (oil), it increasingly resists fluid diffusion, causing the injection pressure to build up over time under rapid flow (Figure~\ref{fig:fixed-press-two}a). 
This substantial pressure rise minimizes the impact of localized aperture-width randomness on the selection of preferential flow paths, leading to a flat advancing flow front, unlike the micro-mechanisms of viscous fingering observed in Case 1 (Figure~\ref{fig:fixed-micro-mechanism-vf}). 
Under slow drainage (Figure~\ref{fig:fixed-press-two}b), while there is still an observable upward trend in injection pressure, the maximum injection pressure (0.16 kPa) is 1225 times lower than under fast flow (196.07 kPa). 
As a result, the capillary effect dominates flow front advancement, with varying capillary entry pressures causing significant temporal fluctuations in the injection pressure, following the same micro-mechanism of capillary fingering as seen in Case 1 (Figure~\ref{fig:fixed-micro-mechanism-cf}).

\begin{figure}[htbp]
  \centering
  \includegraphics[width=1.0\textwidth]{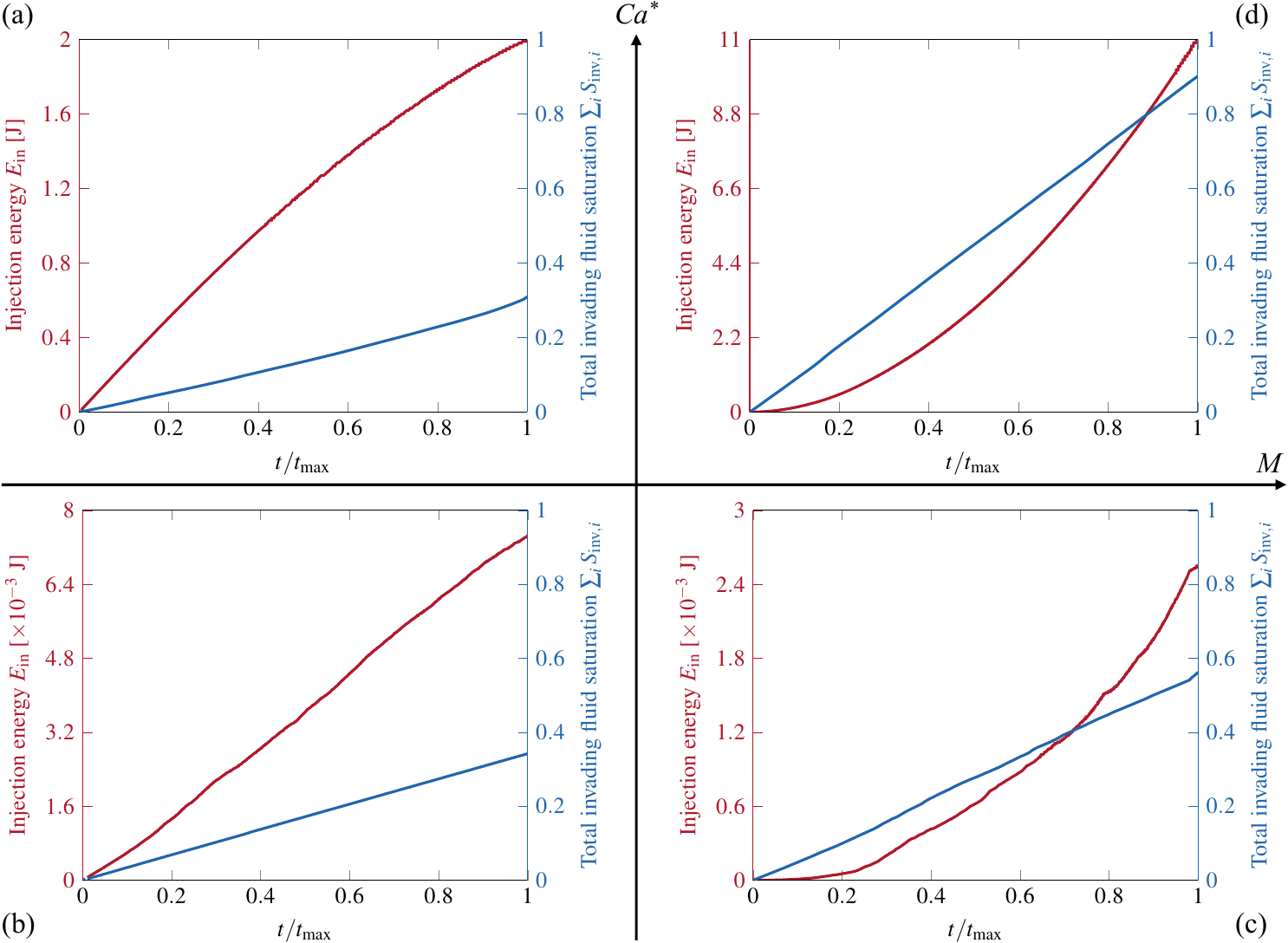}
  \caption{Temporal evolution of injection energy ($E_\text{in}$) and total invading fluid saturation ($\sum_i S_{\text{inv},i}$). (a) Viscous fingering and (b) capillary fingering in the case of air displacing very-viscous oil (Case 1), and (c) capillary fingering and (d) stable displacement in the case of water displacing less-viscous oil (Case 2).}
  \label{fig:fixed-energy-sinv}
\end{figure}

It is noteworthy that, in Figures~\ref{fig:fixed-press-one} and \ref{fig:fixed-press-two}, the capillary fingers take a considerable amount of time---\ie~9324.1 seconds in Case 1 and 72120.4 seconds in Case 2---to reach the outflow boundary, significantly longer than the 10.5 seconds required for viscous fingering and 8.1 seconds for stable displacement.
To further examine the displacement efficiency in both cases, we compare the temporal evolution of injection energy ($E_\text{in}$) and total invading fluid saturation ($\sum_i S_{\text{inv},i}$) as shown in Figure~\ref{fig:fixed-energy-sinv}. 
Injection energy is considered as the cumulative sum of hydraulic power, simplified as the sum of injection energy increments ($\Delta E_\text{in}$) at each timestep ($\delta t_i$). The injection energy can hence be expressed as \cite{goodfellow2015hydraulic}
\begin{equation}
  E_\text{in} = \sum_i(\Delta E_\text{in})_i = \sum_i(\bar{p}_{f,\text{in}}Q_\text{in}\delta t)_i.
  \label{eq:inflow-energy}
\end{equation} 
In Case 1, despite both flow patterns resulting in nearly identical final saturations (Figure~\ref{fig:fixed-energy-sinv}a and b), viscous fingering consumes 2.0 J of injection energy, whereas capillary fingering requires only 7.5 $\times 10^{-3}$ J.  
In Case 2, the more viscous invading fluid displaces a greater amount of defending fluid compared to Case 1, achieving a total invading fluid saturation of 0.56 in capillary fingering (Figure~\ref{fig:fixed-energy-sinv}c) and 0.90 in stable displacement (Figure~\ref{fig:fixed-energy-sinv}d). 
However, stable displacement demands 11.0 J of injection energy, 4297 times higher than required for capillary fingering. 
Therefore, although viscous fingering and stable displacement can achieve the desired results within a short operation period, maintaining a capillary fingering flow pattern is recommended to optimize costs. This insight could guide field operations, such as optimizing injection strategies for geological carbon sequestration or natural gas extraction.

\subsection{Injection tests in movable-grain porous media}

\subsubsection{Model setup}
\begin{figure}[htbp]
  \centering
  \includegraphics[width=1.0\textwidth]{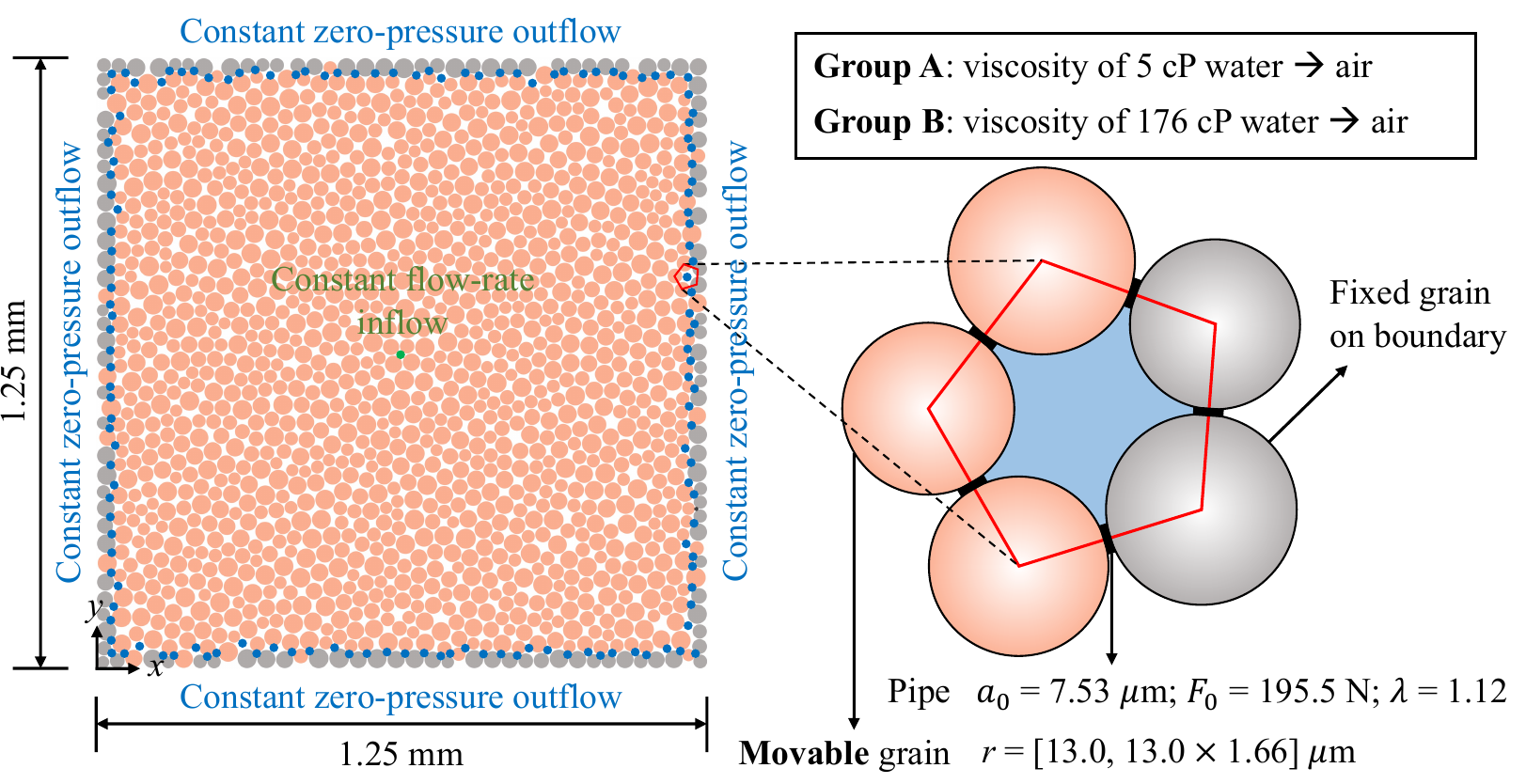}
  \caption{Setup for injection tests in movable-grain porous media.}
  \label{fig:movable-setup}
\end{figure}

The radial flow model reproduces the fluid--grain interactions observed in Huang’s experiment \cite{huang2012granular}. 
The experiment involved injecting a wetting fluid (water) through deformable porous media saturated with a non-wetting fluid (air). The sample was prepared using dry Ottawa F110 sand, with a measured permeability of $7.4 \times 10^{-13}$ m$^2$ under a confining stress of 0.14 MPa. 
We convert the 3D experimental porosity (0.35) using Eq.~\eqref{eq:porosity-change}, yielding a 2D value (0.15) for the numerical assembly packing.
To match the Kozeny-Carman permeability of the numerically confined sample with the experimental permeability, we estimate grain radii ranging from 13 to 21.6 $\mu$m. With the determined porosity and grain size, we initialize a square assembly of size 1.25 $\times$ 1.25 mm, randomly distributing 1395 particles.
The generated pore network comprises 2096 domains and 3490 pipes, with one constant-rate inflow domain at the center and 141 constant-zero-pressure outflow domains on the sides (Figure~\ref{fig:movable-setup}). 
The initial pore pressures of the remaining domains are set to zero. All boundary grains are fixed to imitate the sealed edges of the Hele--Shaw cell, allowing only fluids to pass through.

We select two testing groups from Huang’s experiment---Group A and Group B---with individual invading fluid viscosities of 5 and 176 cP, corresponding to viscosity ratios ($M$) of 277.8 and 9777.8, respectively. 
Each group is simulated with two experimental injection velocities of 8.22 and 165.80 mm/s, corresponding to two modified capillary numbers:
(i) A1 and A3 with $Ca^*$ of 1.6 and 32.3, and (ii) B1 and B3 with $Ca^*$ of 56.3 and 1138.1. 
These simulations cover typical fluid--grain interaction regimes observed in Huang’s experiment: (i) pure fluid infiltration with no fracturing, (ii) infiltration-dominated with fracturing, and (iii) fracturing-dominated, guiding preferential invasion paths.

\subsubsection{Model calibration}
\begin{figure}[htbp]
  \centering
  \subfloat[]{\includegraphics[height=0.38\textwidth]{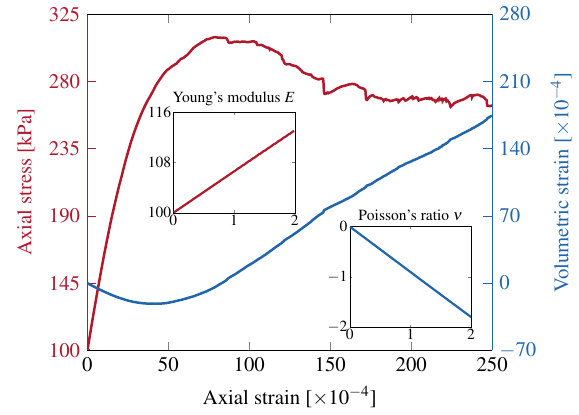}} 
  \subfloat[]{\includegraphics[height=0.38\textwidth]{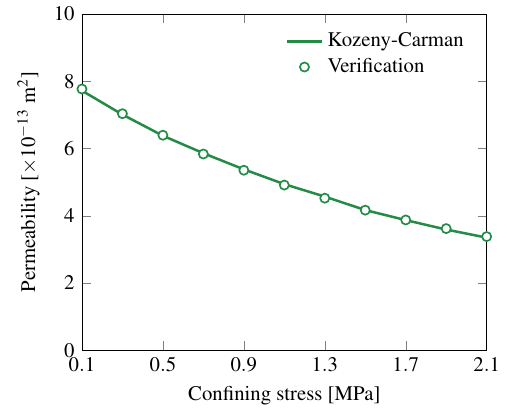}} 
  \caption{Calibration of hydro-mechanical properties. (a) Axial stress and volumetric strain profiles against axial strain under confining stress of 0.1 MPa, and (b) permeability profile against confining stress.}
  \label{fig:calibration-hm-properties}
\end{figure}

We use the biaxial compression test to calibrate the mechanical properties listed in Table~\ref{tab:movable-mechanical-properties}. 
Calibration is achieved by matching the computed Young’s modulus ($E$), Poisson’s ratio ($\nu$), and friction angle ($\varphi$) of the granular assembly with the experimental measurements of Ottawa sand. 
A sample is prepared by isotopically compressing the assembly with a confining stress of 0.1 MPa, consistent with the triaxial compression tests in Huang’s experiment. 
We then apply constant and opposite velocities to the top and bottom boundaries, performing axial compression until the computed vertical strain reaches 2.5$\%$. 
The modeling assembly exhibits typical dense-packing granular media behavior, characterized by a peak in the stress--strain curve and a transition from contraction to dilation in the volumetric deformation curve (Figure~\ref{fig:calibration-hm-properties}a). 

\begin{table}[h!]
    \centering
    \begin{tabular}{l|l|l|l}
        \toprule 
        Parameter & Symbol & Unit & Value \\
        \midrule 
        Grain density & $\rho$ & kg/m$^3$ & 2650 \\
        Grain friction coefficient & $f$ & [-] & 0.6 \\
        Wall friction coefficient & $f_w$ & [-] & 0.0 \\
        Contact normal stiffness & $k_n$ & MPa & 74.9 \\
        Contact shear stiffness & $k_s$ & MPa & 74.9 \\
        Grain radius & $r$ & $\mu$m & $[13.0,21.6]$ \\
        \bottomrule
    \end{tabular}
    \caption{Mechanical properties for injection tests in movable-grain porous media.}
    \label{tab:movable-mechanical-properties}
\end{table}

We employ the elastic segments of these curves to determine the elastic properties of the assembly. The computed Young’s modulus of 65.4 MPa closely matches the measurement from Huang’s experiment (66.2 MPa). The derived Poisson’s ratio of 0.102 is consistent with the laboratory-measured value for gas-saturated Ottawa sand, approximately 0.1, based on the compressional and shear-wave velocities reported by Domenico \cite{domenico1977elastic}. Additionally, applying the Mohr--Coulomb failure criterion $\sigma_1=(1+\sin\varphi)/(1-\sin\varphi)\sigma_3$, where $\sigma_1$ is the peak stress observed in Figure~\ref{fig:calibration-hm-properties}a (309.6 kPa) and $\sigma_3$ is the lateral confining stress, yields a simulated friction angle ($\varphi$) of 30.8\degree.
This value closely corresponds to the macroscopic friction angle of Ottawa F110 sand (31\degree), determined by the simple critical-state test in Santamarina and Cho \cite{santamarina2001determination}. 
Overall, the calibration of mechanical properties maintains an error within 2$\%$, indicating the reliability of our model in predicting the mechanical behavior of Ottawa sand. 

\begin{table}[h!]
    \centering
    \begin{tabular}{l|l|l|l}
        \toprule 
        Parameter & Symbol & Unit & Value \\
        \midrule 
        Invading fluid viscosity & $\eta_\text{inv}$ & cP & $5,176$ \\
        Invading fluid modulus & $K_{f,\text{inv}}$ & MPa & 2000 \\
        Defending fluid viscosity & $\eta_\text{def}$ & cP & 0.018 \\
        Defending fluid modulus & $K_{f,\text{def}}$ & MPa & 0.14 \\
        Interfacial tension & $\gamma$ & dyn/cm & 63.0 \\
        Contact angle & $\theta_c$ & \degree & 0.0 \\
        Aperture width at $F=0$ & $a_0$ & $\mu$m & 7.53 \\
        Normal contact force at $a_0/2$ & $F_0$ & N & 195.5 \\
        Fracturing aperture multiplier & $\lambda$ & [-] & 1.12 \\
        \bottomrule
    \end{tabular}
    \caption{Hydraulic properties for injection tests in movable-grain porous media.}
    \label{tab:movable-hydraulic-properties}
\end{table}

For hydraulic properties listed in Table~\ref{tab:movable-hydraulic-properties}, in addition to properties defined in Huang's experiment, we calibrate the aperture width-related micro-parameters ($a_0$, $F_0$, and $\lambda$) by conducting a series of sample compactions and Darcy flow tests under varying confining stresses from 0.1 to 2.1 MPa. 
\reviewertwo{First, we use the servo mechanism, a feedback control system that dynamically adjusts boundary positions, to isotopically compact the initial assembly under specific confining stress and calculate the Kozeny-Carman permeability using the simulated porosity via Eq.~\eqref{eq:kozeny-carman}.}
Next, we perform Darcy flow tests on the compacted assembly and predict the Darcy-flow permeability using Eq.~\eqref{eq:darcy-flow}. 
Since the effect of hydraulic constants on permeability varies with the number and magnitude of compressive contacts, we arbitrarily test different combinations until finalizing a set of hydraulic constants that results in all Darcy-flow permeabilities matching the corresponding Kozeny-Carman values (with an error within 1$\%$), as shown in Figure~\ref{fig:calibration-hm-properties}b.
In summary, across varying confining stress levels, our DEM-generated sample demonstrates consistent fluid conductance with the experimental sample, indicating similar fluid--grain interactions. 
Following Huang’s experiment, we select a confining stress of 0.14 MPa to prepare the sample for injection simulations, where the predicted Darcy-flow permeability ($7.43\times10^{-13}$ m$^2$) matches the experiment value.

\subsubsection{Results}
\reviewerthr{Table~\ref{tab:movable-simulation-time} presents the fluid flow timestep and total simulation time for movable-grain tests across four distinct combinations of viscosity ratio ($M$) and modified capillary number ($Ca^*$), highlighting our modeling efficiency achieved with reasonable computational time and memory usage.}
The timestep for each $Ca^*$ ensures that our model performs at least three fluid-grain interaction steps to fill one domain.
Compared with other HM-DEM coupled simulations of Huang's experiment, as seen in Zhang \etal~\cite{zhang2013coupled}, our model not only selects more realistic fluid compressibility but also allows for a timestep more than 80 times larger by employing the implicit pressure solver, significantly enhancing simulation efficiency. 
Besides, in their one-way coupling model, the injection pressure under large $Ca^*$ (\ie~B3) causes significant volume changes, leading to pressure fluctuations (sudden drops). 
Our two-way pressure-volume iteration scheme avoids this issue, producing a smoother pressure development curve.

\begin{table}[h!]
    \centering
    \begin{tabular}{l|l|l|l|l|l|l}
        \toprule 
        Test & $M$ & $Ca^*$ & $\delta t_f$ [s] & $t_f$ [s] & $t_c$ [hour] & Memory [MB] \\
        \midrule 
        A1 & 277.8 & 1.6  & $1.0\times 10^{-5}$ & $3.0\times 10^{-2}$ & 21.5 & 270 to 1200 \\
        A3 & 277.8 & 32.3 & $8.0\times 10^{-7}$ & $1.5\times 10^{-3}$ & 25.6 & 270 to 1200 \\
        B1 & 9777.8 & 56.3 & $1.0\times 10^{-5}$ & $2.5\times 10^{-2}$ & 16.2 & 270 to 1200 \\
        B3 & 9777.8 & 1138.1 & $8.0\times 10^{-7}$ & $3.5\times 10^{-4}$ & 2.0 & 270 to 400 \\
        \bottomrule
    \end{tabular}
    \caption{Modeling efficiency of movable-grain tests under various viscosity ratio ($M$) and modified capillary number ($Ca^*$), exhibiting fluid flow timestep ($\delta t_f$), total simulation time ($t_f$), computational time ($t_c$), and memory usage for each case.}
    \label{tab:movable-simulation-time}
\end{table}

\begin{figure}[htbp]
  \centering
  \includegraphics[width=1.0\textwidth]{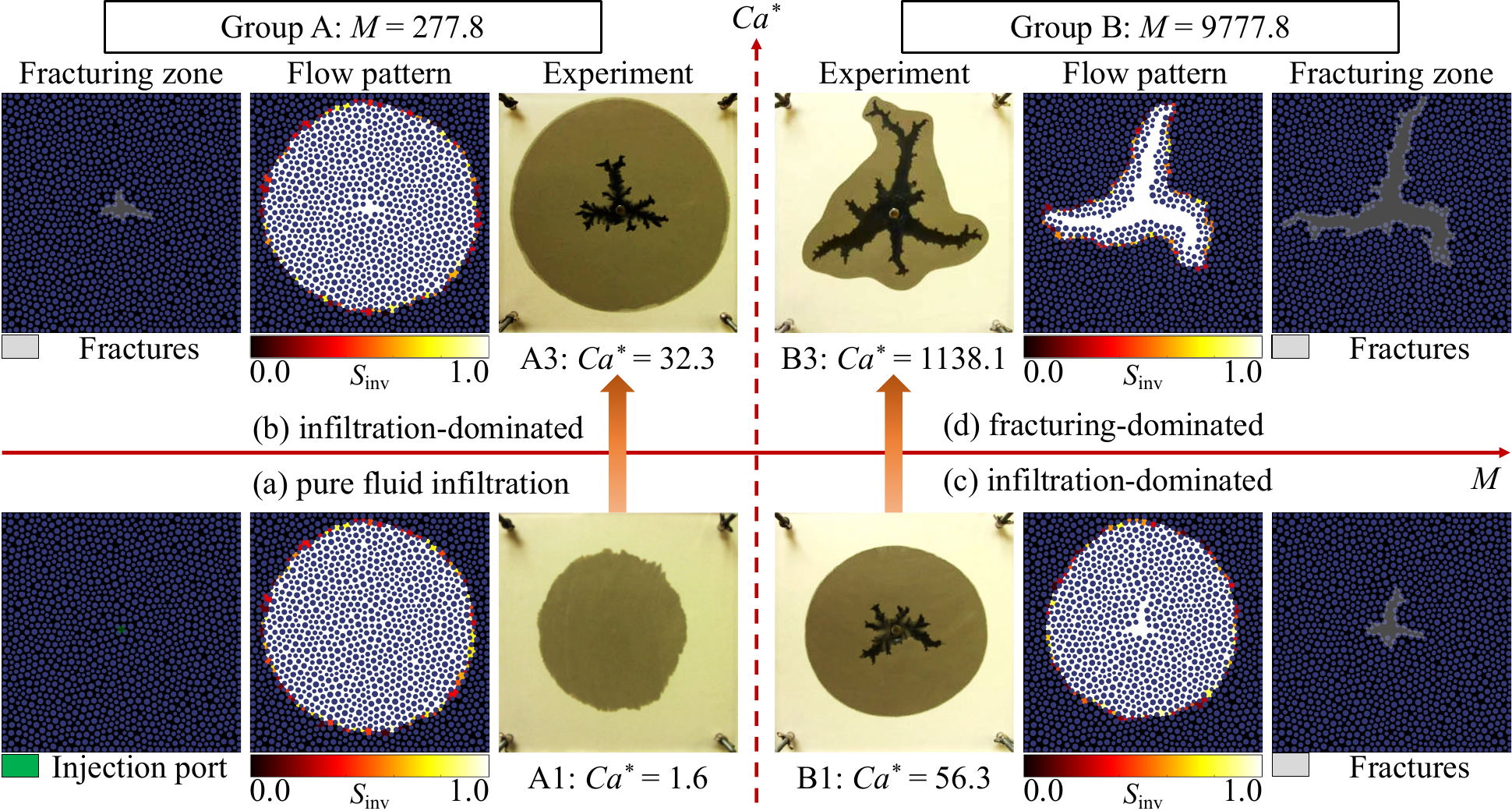}
  \caption{Comparison of fluid--grain interactions at breakthrough between numerical simulations and experimental observations from Huang's experiment. Increasing capillary number results in a transition of dominant roles between fluid infiltration and fracturing: from (a) pure fluid infiltration to (b) infiltration-dominated regime in the 5 cP invading fluid-viscosity case (Group A), and from (c) infiltration-dominated to (d) fracturing-dominated regime in the 176 cP invading fluid-viscosity case (Group B).}
  \label{fig:movable-pattern}
\end{figure}

Figure~\ref{fig:movable-pattern} compares the computational fluid--grain displacement patterns at breakthrough (when the invading fluid reaches the outflow boundary) with the experimental observations from Huang's experiment. 
The simulated displacement patterns closely match the experimental findings. 
As the invading fluid viscosity and injection flow rate (\ie~$Ca^*$) increase, the injection energy intensifies, leading to fractures that ultimately influence the flow pattern.
In A1, the invading fluid viscosity and inflow rate are too low to generate sufficient injection energy to mobilize grains. No fractures are formed, and the invasion front remains circular, characteristic of a stable-displacement flow pattern---\ie~a pure fluid infiltration regime.
As $Ca^*$ increases from A1 to A3 and B1, the injection energy becomes large enough to mobilize grains near the inflow port, initiating fractures. These fractures remain within the fluid infiltration zone, not altering the circular flow front. 
Therefore, A3 and B1 can be classified as infiltration-dominated regimes.
In B3, the high invading fluid viscosity and injection rate generate substantial injection energy, creating fractures significant enough to influence fluid infiltration. The flow front, reflecting fracture propagation, adopts a fingering-like shape, classifying B3 as a fracturing-dominated regime. 

\reviewerone{
Quantitative analysis using the box-counting fractal dimension (see \ref{appendix_fractal_dimension}) shows good agreement with experimental data. 
The fractal dimension of the circular infiltration zones in A1, A3, and B1 matches the experimental value of 1.95 \cite{zhang2012pattern}, while it decreases when fracturing alters the flow front into a fingering-like pattern (B3).
Similarly, the fractal dimension of the fracturing finger decreases as $Ca^*$ increases,  consistent with the experimental range [1.70, 1.82] for enhanced fracture propagation with increasing $Ca^*$ \cite{zhang2012pattern}. 
Overall, our model effectively captures the transition from an infiltration-dominated regime to a fracturing-dominated regime as $Ca^*$ increases, providing a quantitative representation of the evolution of the ramified flow front and fracturing channels.
}

\begin{figure}[htbp]
  \centering
  \subfloat[]{\includegraphics[width=1.0\textwidth]{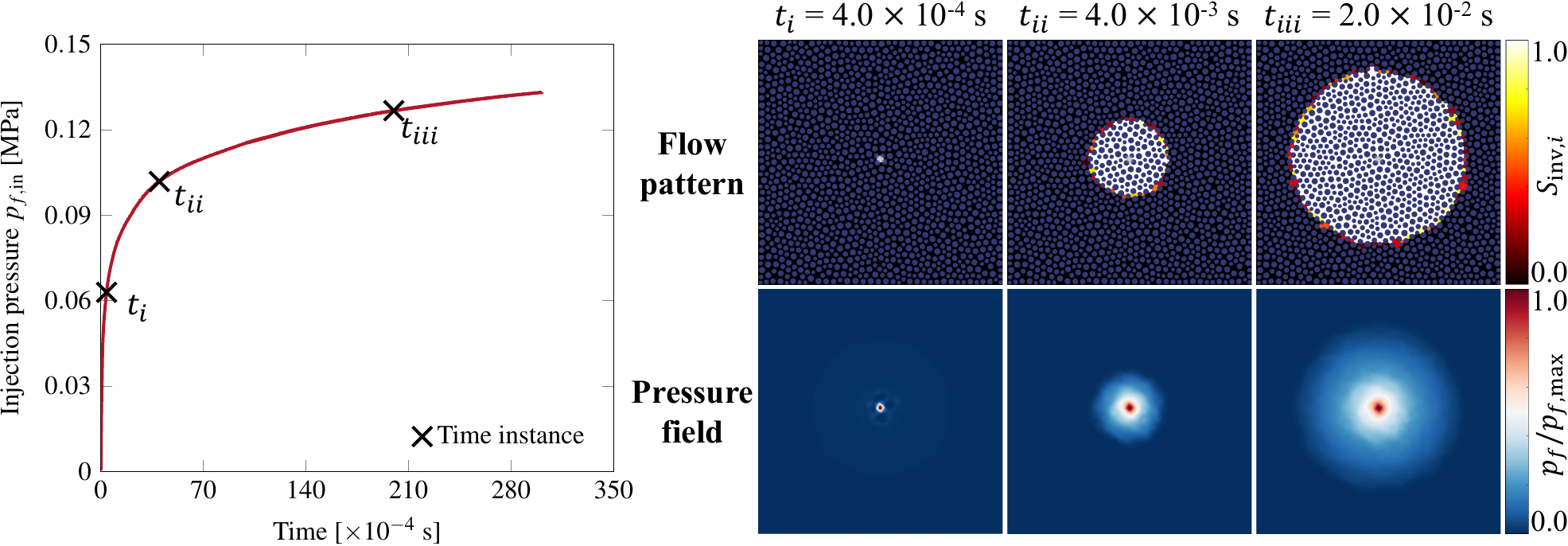}} \\
  \subfloat[]{\includegraphics[width=1.0\textwidth]{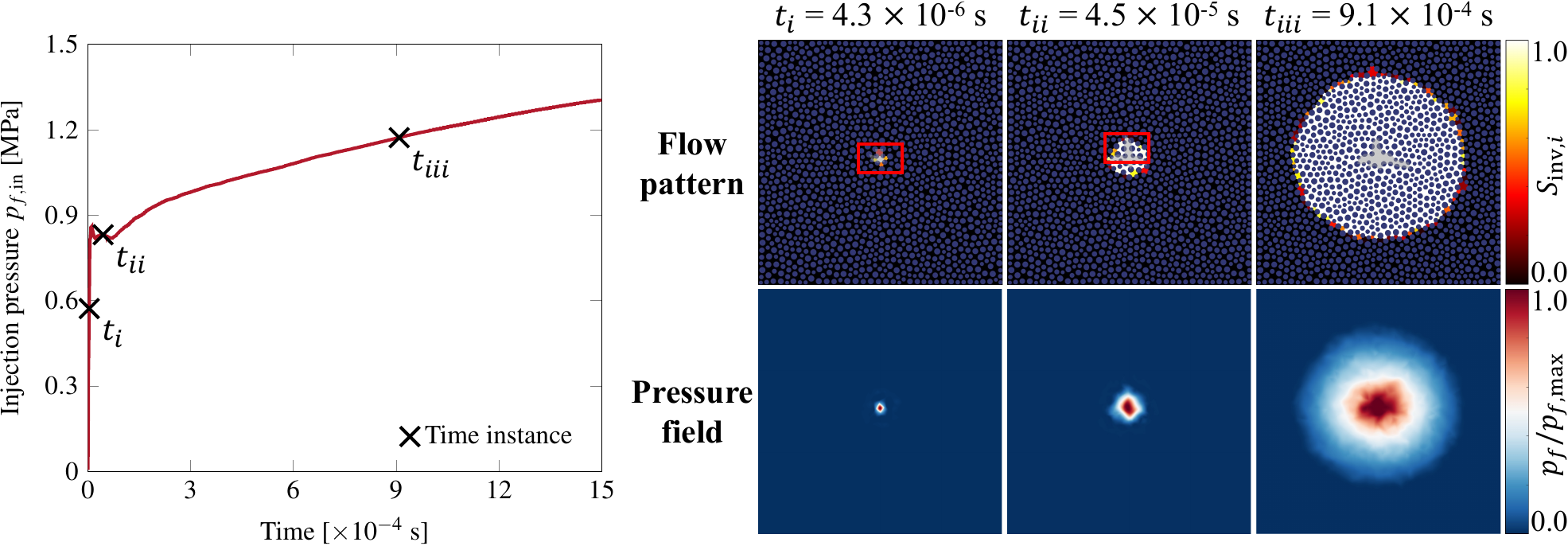}} \\
  \subfloat[]{\includegraphics[width=1.0\textwidth]{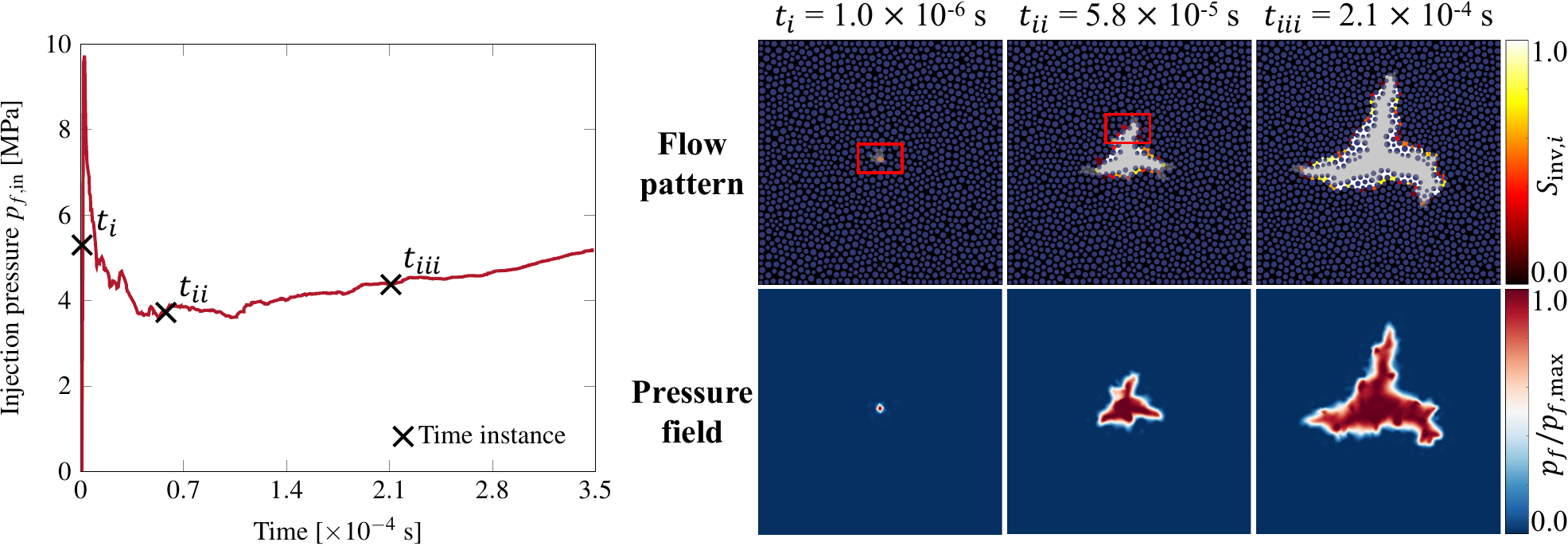}} 
  \caption{Temporal evolution of injection pressure ($p_{f,\text{in}}$) in the movable-grain case, with selected flow patterns and pressure fields at three time instances. (a) Pure fluid infiltration regime (A1: $Ca^* = 1.6$), (b) infiltration-dominated regime (A3: $Ca^* = 32.3$), and (c) fracturing-dominated regime (B3: $Ca^* = 1138.1$).}
  \label{fig:movable-press}
\end{figure}

Figure~\ref{fig:movable-press} explores the underlying mechanisms of fluid-grain interaction regimes (A1, A3, and B3) by examining the development of injection pressure ($p_{f,\text{in}}$) over time, along with flow patterns and pressure fields at selected time instances ($t_i$, $t_{ii}$, and $t_{iii}$). 
For the slow injection case (A1 in Figure~\ref{fig:movable-press}a), the injection pressure gradually increases as the invading fluid occupies the porous media, forming a high-viscous drainage path that causes pressure to accumulate rather than dissipate.
However, even the maximum injection pressure (0.13 MPa) is insufficient to mobilize grains near the injection port, and the flow front remains circular. 

After employing a higher inflow rate (A3 in Figure~\ref{fig:movable-press}b), the injection pressure peaks at 0.86 MPa---6.6 times higher than the maximum pressure in A1---initiating fractures. The fracturing compacts adjacent grains, resulting in larger aperture widths and pore expansion, leading to a concentration of viscous pressure but a drop in injection pressure. The grain compaction strengthens contact forces, resisting further fracturing. Consequently, after $t_{ii}$, fluid infiltration once again controls the rise in injection pressure. 
In B3 (Figure~\ref{fig:movable-press}c), the highest applied invading fluid viscosity and flow rate cause the injection pressure to quickly peak at 9.73 MPa, 11.3 times higher than the early peak in A3. This significant injection pressure creates fractures before fluid infiltration at $t_i$, guiding the fluid to penetrate the fracturing domains at later stages. Along with injection-induced domain expansion, preferential paths enhance pressure dissipation to the fingertips, further decreasing injection pressure. 
When fracturing and infiltration reach a balance, the injection pressure stabilizes at a plateau. 

\begin{figure}[htbp]
  \centering
  \includegraphics[width=1.0\textwidth]{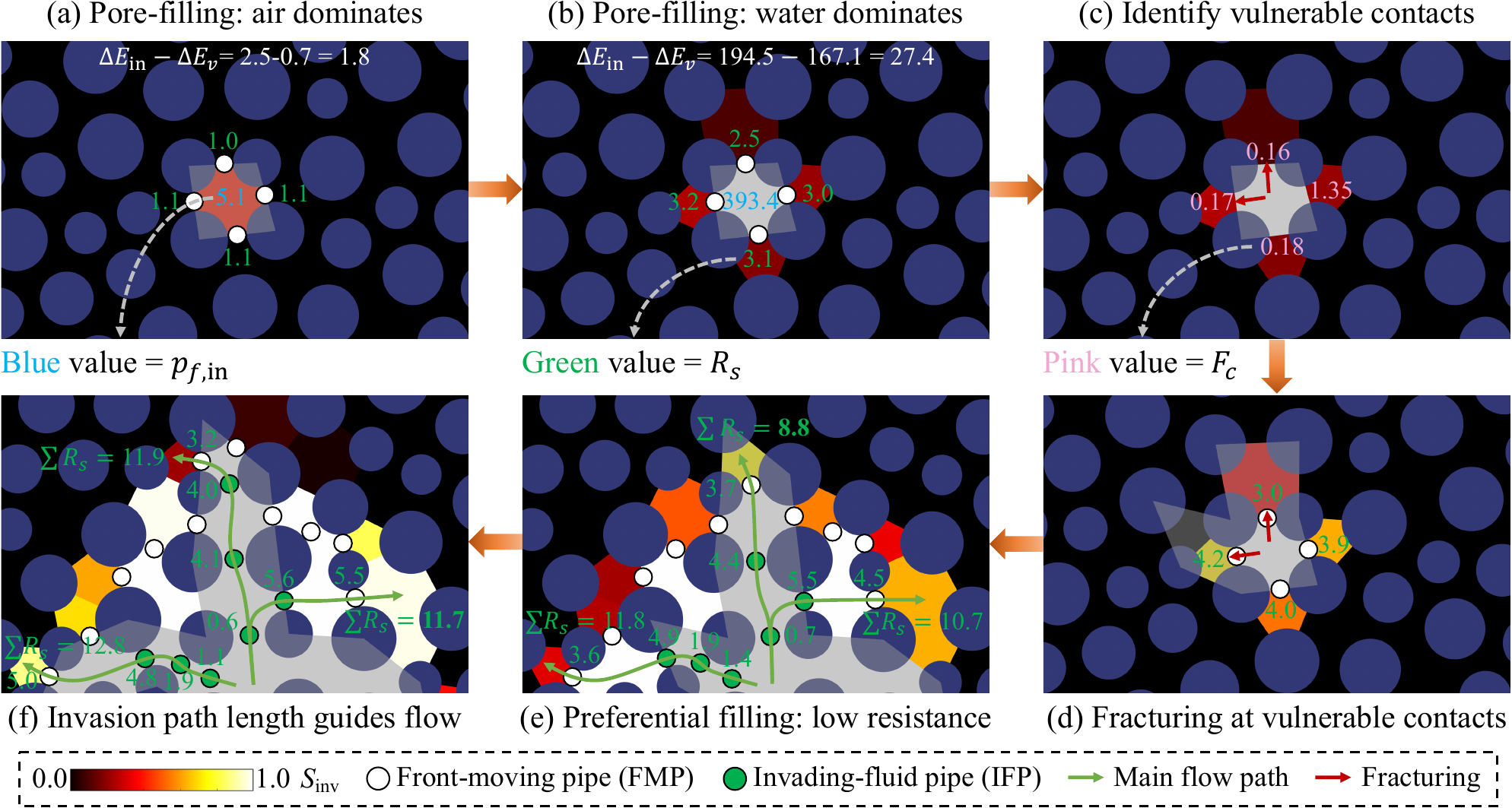}
  \caption{Micro-mechanisms of infiltration-dominated regime (A3), with zoomed-in flow fronts in Figure~\ref{fig:movable-press} at $t_i$ and $t_{ii}$, elaborated with close-ups of selected timesteps from (a) to (f). Blue values represent injection pressure in kPa, green values represent flow resistance in GPa$\cdot$s/m$^2$, and pink values represent contact force in N. All energy values are in $\mu$J.}
  \label{fig:movable-micro-mech-A3}
\end{figure}

\begin{figure}[htbp]
  \centering
  \includegraphics[width=1.0\textwidth]{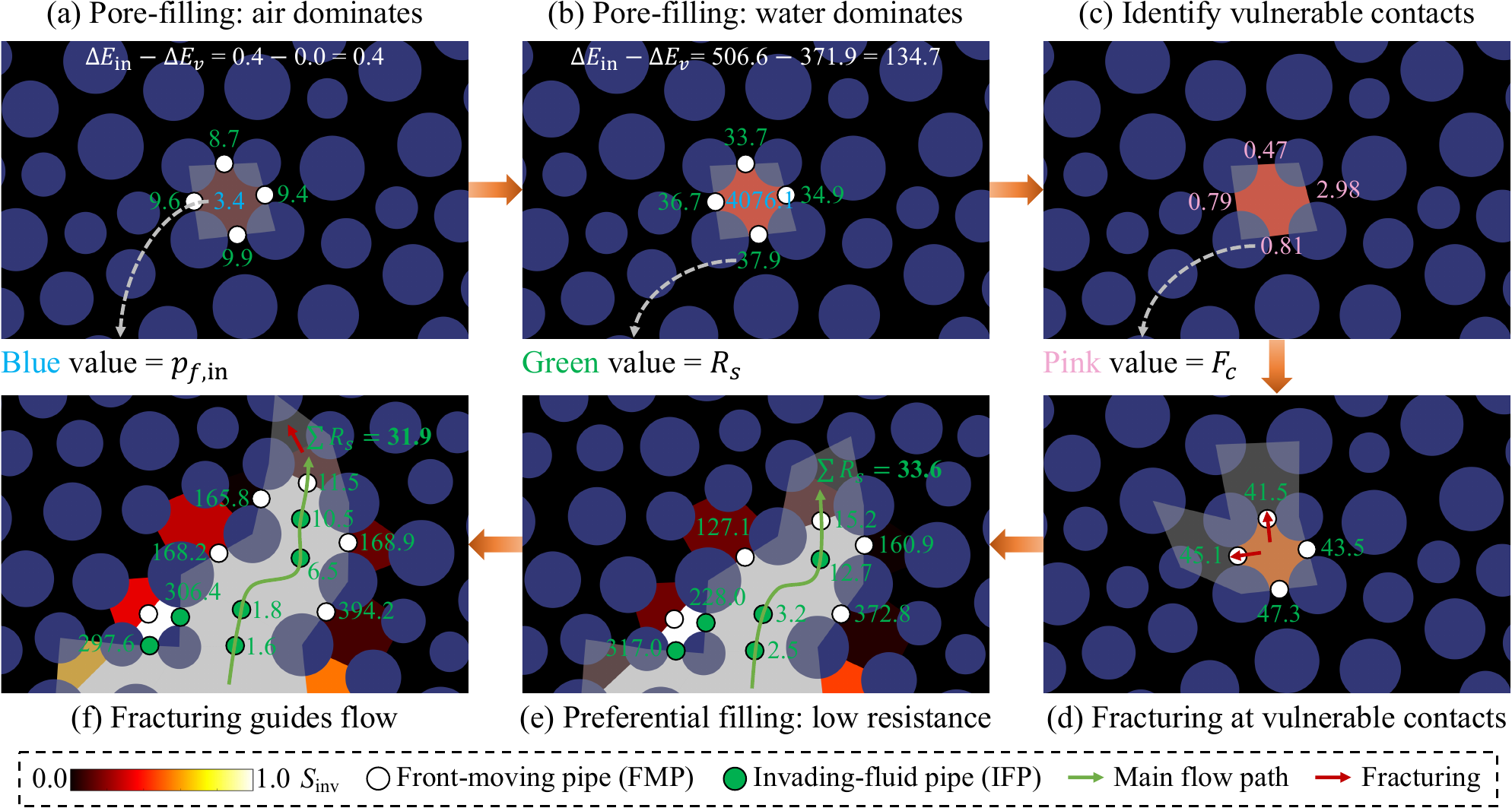}   
  \caption{Micro-mechanisms of fracturing-dominated regime (B3), with zoomed-in flow fronts in Figure~\ref{fig:movable-press} at $t_i$ and $t_{ii}$, elaborated with close-ups of selected timesteps from (a) to (f). Blue values represent injection pressure in kPa, green values represent flow resistance in GPa$\cdot$s/m$^2$, and pink values represent contact force in N. All energy values are in $\mu$J.}
  \label{fig:movable-micro-mech-B3}
\end{figure}

From A3 to B3, under varying degrees of fracturing, fluid infiltration exhibits different patterns. 
To understand the micro-mechanisms behind these fluid–grain interactions, we analyze the grain-scale interplay between fracturing and fluid infiltration from time instance $t_{i}$ to $t_{ii}$ in Figure~\ref{fig:movable-press}, focusing on fracturing energy, flow resistance, and contact force, as shown in Figures~\ref{fig:movable-micro-mech-A3} and \ref{fig:movable-micro-mech-B3}.  
The residual energy for fracturing ($\Delta E_f$) at each timestep is computed as the difference between injection energy ($\Delta E_\text{in}$ in Eq.~\eqref{eq:inflow-energy}) and the total dissipate energy ($\Delta E_v$) of all pipes \cite{zhang2013coupled}, which can be expressed as
\begin{equation}
  \Delta E_f = \Delta E_\text{in}-\Delta E_v = p_{f,\text{in}}Q_\text{in}\delta t-\sum_k q_{ij,k}(p_{f,i}^{n+1}-p_{f,j}^{n+1})_k\delta t.
  \label{eq:viscous-energy}
\end{equation}
Fracturing occurs when the accumulated residual energy is sufficient to generate an applied force ($F_p$) on a grain that exceeds the contact force ($F_c$), initiating a fracture. 
Following fracturing, the aperture width increases, causing variations in fluid infiltration rates through the pipes.
To evaluate how fracturing influences fluid infiltration, we define flow resistance as the ease with which fluid flows through a pipe under a given pressure difference ($\Delta p_{f,ij}$). 
The flow resistance of each pipe is derived from the Plane Poiseuille flow equation (Eq.~\eqref{eq:plane-pflow}), which can be written as
\begin{equation}
  R_s = \cfrac{\Delta p_{f,ij}}{q_{ij}} = \dfrac{12\eta_{ij}l_{p,ij}}{a_{ij}^3}.
  \label{eq:flow-resistance}
\end{equation}
Lower flow resistance indicates a higher flow rate through the pipe, ceteris paribus.

In the light-fracturing case (A3), the injection port is initially saturated with air (Figure~\ref{fig:movable-micro-mech-A3}a), resulting in a low average pipe flow resistance of 1.1 GPa$\cdot$s/m$^2$. 
This low resistance facilitates water injection, producing a low injection pressure of 5.1 kPa. 
Consequently, both injection energy (2.5 $\mu$J) and dissipation energy (0.7 $\mu$J) are low. 
Fluid dissipation consumes only 28$\%$ of the injection energy since the inflow rate exceeds the outflow rate before reaching a steady state.
The residual energy for fracturing ($\Delta E_f$) is 1.8 $\mu$J, insufficient to generate fractures.
Later (Figure~\ref{fig:movable-micro-mech-A3}b), as the injection port fills with water, the average pipe flow resistance increases to 3.0 GPa$\cdot$s/m$^2$, causing the injection pressure to rise to 393.4 kPa. 
More injection energy (194.5 $\mu$J) is required to maintain a constant inflow rate. 
Although fluid dissipation now consumes 86$\%$ of the injection energy due to the displacement of more viscous water instead of air, the residual energy (27.4 $\mu$J) is sufficient to initiate fractures, with vulnerable contacts opening first (with contact forces of 0.16 and 0.17 N in Figure~\ref{fig:movable-micro-mech-A3}c).
After fracturing (Figure~\ref{fig:movable-micro-mech-A3}d), the flow resistance of the fracturing pipes increases to 3.0 and 4.2 GPa$\cdot$s/m$^2$, higher than the initial values of 2.5 and 3.2 GPa$\cdot$s/m$^2$ in Figure~\ref{fig:movable-micro-mech-A3}b, indicating that fracturing in A3 does not significantly alter the aperture width; instead, the two-phase fluid viscosity (Eq.~\eqref{eq:mix-eta}) dominates pipe flow resistance.
As the invading fluid advances (Figure~\ref{fig:movable-micro-mech-A3}e), fluid from the injection center to the flow front must pass through multiple pipes. 
The lowest front-pipe flow resistance (3.6 GPa$\cdot$s/m$^2$) at the bottom left will not be the preferential flow path this time. 
Identifying a preferential flow path requires combining the flow resistance of all pipes along the path and comparing the total flow resistance ($\sum R_s$) with others. 
The preferential filling path is identified as one towards the top with a total flow resistance of 8.8 GPa$\cdot$s/m$^2$. 
After its front domain fills (Figure~\ref{fig:movable-micro-mech-A3}f), the path becomes longer, increasing its total flow resistance. The preferential flow path then shifts to a short-invading-length path with a total flow resistance of 11.7 GPa$\cdot$s/m$^2$ at the bottom right.
Thus, the fluid invasion in A3 can occur in all directions, depending on the invasion path length, eventually causing a circular flow front.

In the heavy-fracturing case (B3), similar to A3, the initial injection pressure is just 3.4 kPa due to the easy dissipation of air (with an average pipe flow resistance of 9.4 GPa$\cdot$s/m$^2$, as shown in Figure~\ref{fig:movable-micro-mech-B3}a). 
The injection energy is minimal at 0.4 $\mu$J, and dissipation energy is negligible.
However, with the invading fluid in B3 having a viscosity of 176 cP---35.2 times greater than in A3 (5 cP)---flow resistance builds up much more rapidly, reaching an average of 35.8 GPa$\cdot$s/m$^2$ (Figure~\ref{fig:movable-micro-mech-B3}b), leading to a high injection pressure of 4076.1 kPa. 
The resulting undissipated energy ($\Delta E_f$) increases to 134.7 $\mu$J, 4.9 times higher than in A3 when fracturing is about to occur, causing fracturing much earlier than fluid infiltration and opening weak contacts with contact forces of 0.47 and 0.79 N (Figure~\ref{fig:movable-micro-mech-B3}c). 
Despite the onset of early fracturing, the flow resistance around the injection port remains relatively high, with an average pipe flow resistance of 44.4 GPa$\cdot$s/m$^2$ (Figure~\ref{fig:movable-micro-mech-B3}d). 
As the fracture propagates, however, significant residual energy compacts grains more in the radial direction than the circumferential direction, resulting in much lower flow resistance inside fractures compared to the sides (Figure~\ref{fig:movable-micro-mech-B3}e).
The lowest pipe flow resistance inside the fracturing zone is only 2.5 GPa$\cdot$s/m$^2$, merely 2$\%$ of the lowest resistance at the sides.
Even though the path is longer, summing all pipe flow resistances along the fractures yields a total of only 33.6 GPa$\cdot$s/m$^2$, causing the domain at the fracturing fingertips to fill first.
After preferential filling, the total flow resistance to the fracturing fingertips remains the lowest, at only 31.9 GPa$\cdot$s/m$^2$. 
Consequently, fracturing guides fluid infiltration in B3, causing a fingering-like front shape.

To sum up, our model successfully simulates fracture initiation and propagation in multiphase flow scenarios, enabling a detailed and visual evaluation of the underlying micro-mechanisms in movable-grain porous media.
Notably, while a higher inflow rate and increased invading fluid viscosity can improve displacement efficiency when grain motion is minimal, these factors may lead to stability issues, such as fracturing, when grain motion significantly dominates fluid--grain interactions.

\section{Conclusions}
\label{sec:conclusion}

This paper presents a comprehensive hydro-mechanical coupled framework for simulating pore-scale multiphase flow in both fixed- and movable-grain porous media. 
Our model represents a significant advancement in multiphase flow simulation, particularly for its efficiency in handling compressible fluids with large timesteps. 
By employing an implicit finite volume approach, we stabilize the pressure solver, enabling unconditionally large timesteps. 
Additionally, we have implemented several techniques to ensure the accuracy of fluid--grain interactions and to preserve the intricate details of multiphase flow patterns, even under substantial timesteps:
\begin{itemize}\itemsep=0pt
  \item The pressure-volume iteration scheme establishes a dynamic feedback loop between the fluid flow and grain motion models. This iterative process continuously updates the fluid pressure acting on grains and the resulting pore structure re-arrangement until convergence, ensuring precise hydro-mechanical coupling, even in scenarios involving significant pore deformation.
  \item The flow front-advancing criteria determine pore-scale front positions and dynamically update mixing-fluid properties, such as bulk modulus and viscosity. The criteria refine the timestep as necessary, ensuring that at most one domain is filled at a time or that pore pressure accumulates sufficiently to open the weakest capillary entry-pressure pipe under slow drainage, especially when dealing with compressible fluids.
\end{itemize}

We have validated the model by successfully reproducing benchmark Hele--Shaw tests. Through rigorous calibration of hydro-mechanical properties (\eg~permeability) and the setup of dimensionless numbers, including the modified capillary number $Ca^*$ and viscosity ratio $M$, our simulations closely align with experimental observations from Lenormand’s fixed-grain experiment and Huang’s movable-grain experiment. This validation confirms that our model is both reliable and efficient in replicating diverse fluid--fluid and fluid--grain displacement patterns under challenging situations (\eg~flow front blockage under capillary effects and large pore volume changes under heaving fracturing).

Moreover, our model offers quantitative insights into the underlying micro-mechanisms governing multiphase flow in porous media, marking the first time that micro-scale interactions have been visually reviewed in terms of viscous and capillary pressures, energies, contact forces, and flow resistances. 
In simulating Lenormand’s experiment, we identify pattern-transition signals from viscous fingering and stable displacement to capillary fingering by comparing throat-scale viscous and capillary entry pressures. 
In simulating Huang’s experiment, we elucidate the governing regimes of fluid--grain interactions by comparing pore-scale residual energies and contact forces for fracturing, and flow resistances for fluid dissipation. 

Notably, while high flow rates and increased invading fluid viscosity in deformable porous media enhance fracturing and influence fluid infiltration paths, this effect contrasts with the positive impact on production efficiency observed in rigid porous media. This discrepancy highlights the need for further systematic parametric studies using our model to fully understand the trade-offs between operational efficiency and system stability in geological processes. 
\reviewerthr{Future work will explore the role of gas compressibility in optimizing gas storage and recovery, leveraging our model's unique capability to account for fluid compressibility---a feature not addressed by most existing models, which often assume incompressible or overly simplified fluids. Additionally, we aim to enhance the model by replacing fixed boundaries with confined ones, enabling more realistic simulations of multiphase flow patterns and displacement efficiency under stress-dependent conditions.}

\section*{Acknowledgments}

This work was supported by the HKU Postgraduate Scholarship. Kand Duan acknowledges funding from the National Key R$\&$D Program of China (No. 2023YFB2390300). We thank Dr. Yue (Olivia) Meng (Purdue University) and Dr. Zhibing Yang (Wuhan University) for insightful discussions that contributed to this study.

\section*{CRediT author statement} 
\label{sec:credit}

\textbf{Quanwei Dai}: Conceptualization, Methodology, Software, Validation, Formal Analysis, Investigation, Data Curation, Writing - Original Draft, Visualization.
\textbf{Kang Duan}: Conceptualization, Methodology, Formal Analysis, Investigation, Writing - Review \& Editing, Visualization, Supervision.
\textbf{Chung-Yee Kwok}: Conceptualization, Methodology, Investigation, Resources, Writing - Review \& Editing, Visualization, Supervision, Project Administration, Funding Acquisition.

\section*{Data Availability Statement} 
\label{sec:data-availability} 

The data that support the findings of this study are available from the corresponding author upon reasonable request.

\appendix

\reviewerone{
\section{Resolution convergence}
\label{appendix_mesh_size}
We here describe additional tests conducted to ensure that our validation samples have sufficient resolution, with enough particles to produce consistent results. The key criterion for resolution convergence is that flow patterns and temporal pressure evolution remain consistent across different resolutions, achieved by varying grain sizes while maintaining constant permeability and operational parameters ($Ca^*$ and $M$). This ensures that viscous pressure and capillary entry pressure balance at similar levels within the displacement regime.

We use Lenormand's fixed-grain case, where air displaces a very viscous oil (Case 1), as an example. 
Specifically, we vary the original average grain radius ($r_\text{ave}$) by decreasing it by 30$\%$ ($r_\text{ave}=0.7$ mm) and increasing it by 30$\%$ ($r_\text{ave}=1.3$ mm), resulting in particle counts of 2305 (refined) and 724 (coarser), respectively.
The sample size, random control interval ($[1-0.7, 1+0.7]$), and boundary conditions are kept the same as in Figure~\ref{fig:fixed-setup}.
The corresponding refined and coarser pore structures are shown in Figure~\ref{fig:mesh-independent-pattern}.
The original pipe network, consisting of 1874 domains and 3061 pipes, is refined to 3403 domains and 5707 pipes for the smaller grain size and coarsened to 1160 domains and 1883 pipes for the larger grain size. To ensure that permeability remains consistent across all samples, the aperture width range is adjusted proportionally. For the refined sample, the average aperture width decreases to 0.35 mm, while for the coarser sample, it increases to 0.47 mm, compared to 0.42 mm in the original case. 
These adjustments ensure that all samples achieve the experimental permeability of 10$^{-9}$ m$^2$.

As shown in Figure~\ref{fig:mesh-independent-pattern}, the simulated flow patterns exhibit key features of viscous fingering (tree-like morphology) and capillary fingering (thicker fingers with blocked front pipes) for both the reduced- and increased-grain-size samples. These observations are consistent with the original $r_\text{ave}=1.0$ mm case, as shown in Figures~\ref{fig:fixed-pattern}a and \ref{fig:fixed-pattern}b.
We also compare the injection pressure development between the original sample and the grain-size-modified samples for the viscous fingering regime. As illustrated in Figure~\ref{fig:mesh-independent-pressure}, the pressure curves exhibit excellent agreement in both magnitude and temporal trends, confirming resolution convergence and demonstrating the robustness of our simulation framework across variations in pore structure.

\begin{figure}[htbp]
  \centering
  \subfloat[]{\includegraphics[width=1.0\textwidth]{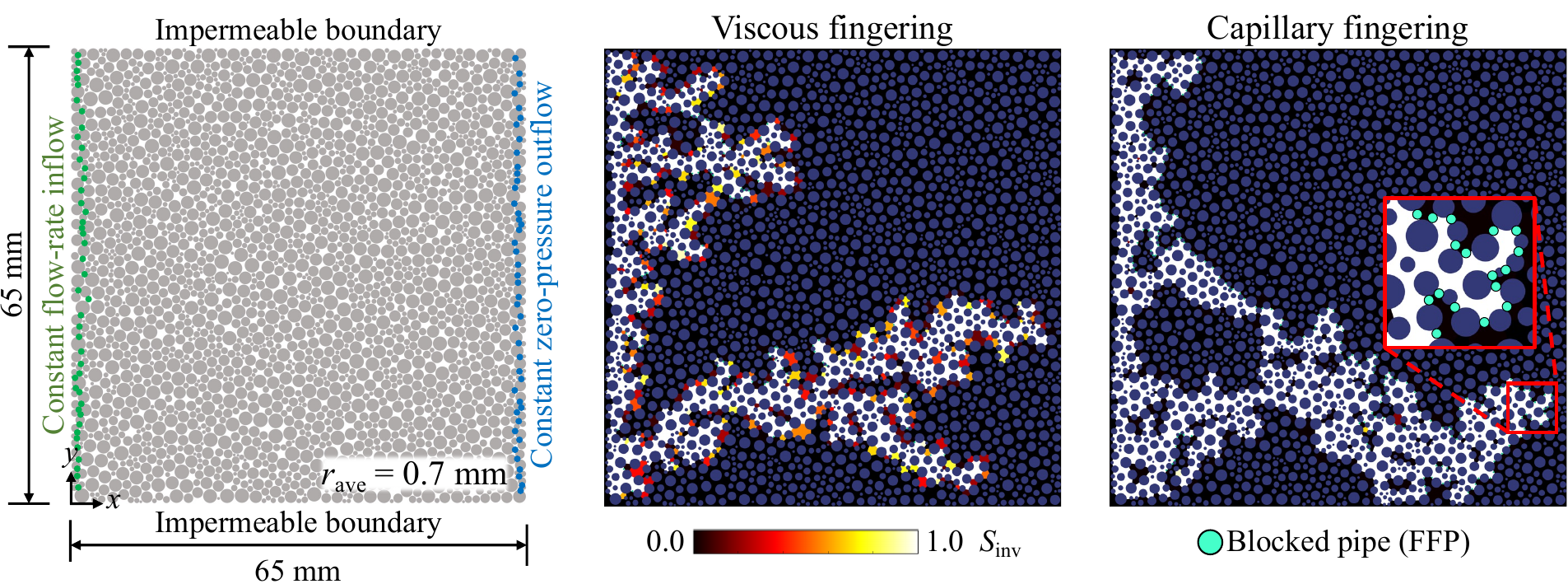}} \\
  \subfloat[]{\includegraphics[width=1.0\textwidth]{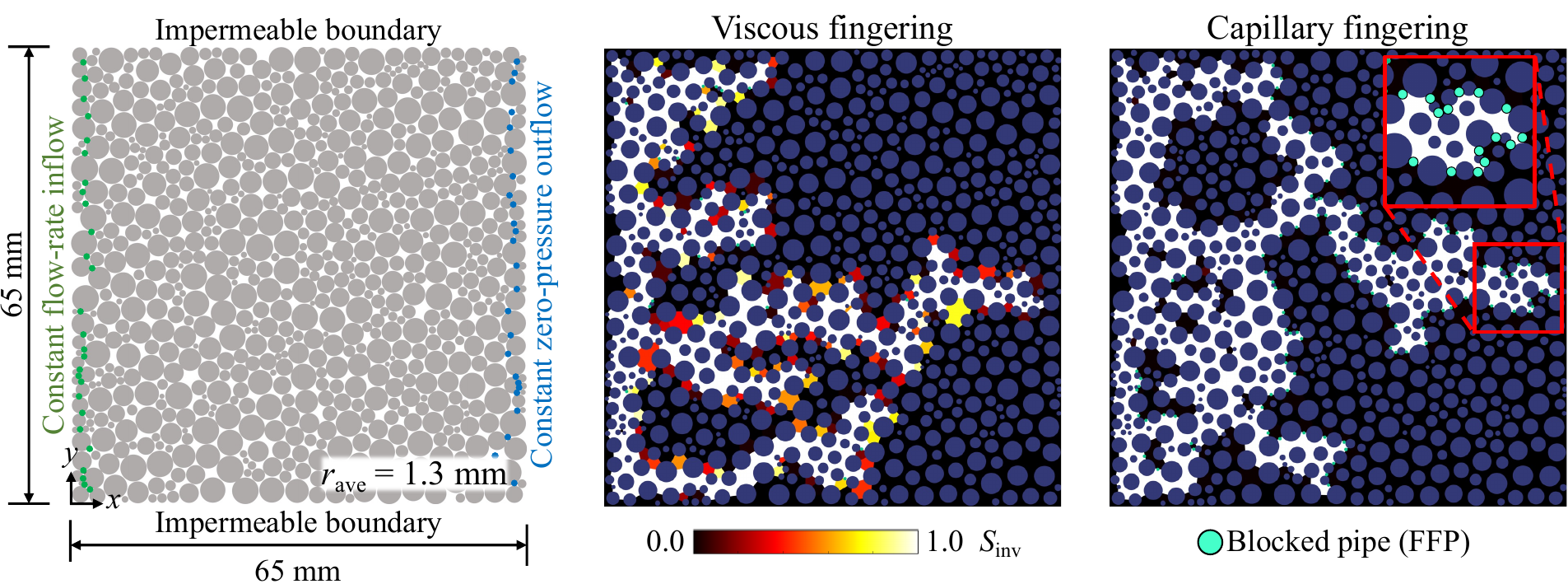}} 
  \caption{Effects of grain size on viscous fingering ($Ca^*=1.9\times10^{-4}$) and capillary fingering ($Ca^*=3.1\times10^{-7}$) in Lenormand's experiment, under same sample permeability and operational parameters ($Ca^*$ and $M$). Model setup and flow patterns are shown for (a) refined and (b) coarser samples.}
  \label{fig:mesh-independent-pattern}
\end{figure}

\begin{figure}[htbp]
  \centering
  \subfloat[]{\includegraphics[height=0.38\textwidth]{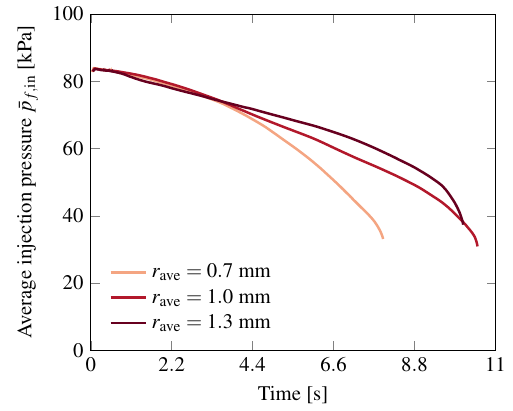}} 
  \subfloat[]{\includegraphics[height=0.38\textwidth]{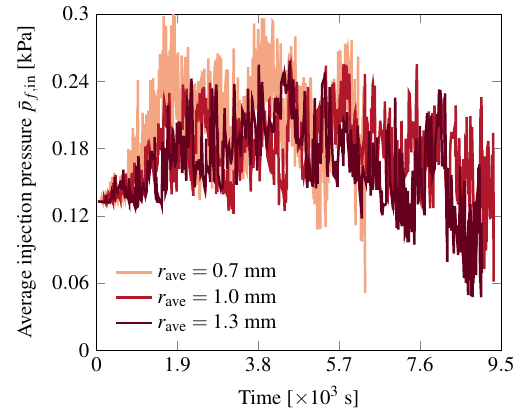}} 
  \caption{Temporal evolution of injection pressure during (a) viscous fingering ($Ca^*=1.9\times10^{-4}$) and (b) capillary fingering ($Ca^*=3.1\times10^{-7}$) in Lenormand’s experiment under varying grain sizes.}
  \label{fig:mesh-independent-pressure}
\end{figure}
}

\reviewerone{
\section{Timestep sensitivity}
\label{appendix_timestep}
We here investigate the sensitivity of our numerical validations of Lenormand's and Huang's experiments to timestep variations.
Specifically, we take viscous fingering (VF: $Ca^*=1.9\times10^{-4}$), capillary fingering (CF: $Ca^*=3.1\times10^{-7}$), fluid infiltration-dominated stable displacement (B1: $Ca^*=56.3$), and fracturing-dominated stable displacement (B3: $Ca^*=1138.1$) as representative examples.
For these cases, we vary the timesteps by increasing or decreasing them by 50$\%$. For instance, in CF of Lenormand's experiment, where the original timestep is 100 s, additional simulations are conducted with $\delta t=50$ s and $\delta t=200$ s. 
The model setups remain consistent with those shown in Figure~\ref{fig:fixed-setup} for fixed-grain injection cases (Lenormand's experiment) and Figure~\ref{fig:movable-setup} for movable-grain injection cases (Huang's experiment).

As shown in Figures~\ref{fig:timestep-effect-lenormand} and \ref{fig:timestep-effect-huang}, both the fluid-fluid displacement patterns (\ie~viscous fingering and capillary fingering) and fluid-grain displacement patterns (\ie~fluid infiltration-dominated and fracturing-dominated regimes) closely match the original validation results of Lenormand's (Figure~\ref{fig:fixed-pattern}) and Huang's experiments (Figure~\ref{fig:movable-pattern}). The temporal evolution of injection pressure also agrees well between the additional tests and the original simulations, in terms of pressure magnitude and development trends.
The only notable variation occurs in the peak injection pressure of B3, which features a high inflow rate and invading fluid viscosity. When the timestep is increased from $4.0 \times 10^{-7}$ to $1.6 \times 10^{-6}$ s, the peak injection pressure changes from 9.0 to 11.1 MPa. 
This variation is attributed to differences in the frequency of updating the mixing water-air fluid bulk modulus in the injection domain under varying timesteps. While this change may influence the early stages of fracture initiation and propagation, it does not impact the final fluid-grain displacement pattern.

The use of larger timesteps in our fully coupled HM-DEM framework is inherently not problematic. Our default flow front-advancing criteria (Eqs.~\eqref{eq:new-time-one} and \eqref{eq:new-time-two}) automatically refine the timestep to ensure stability when any front domain becomes overfilled or when flow is impeded due to blocked front pipes caused by slow drainage. However, excessively small timesteps reduce simulation efficiency, particularly for fluid infiltration-dominated stable displacement (B1), as shown in Table~\ref{tab:timestep-effect-time}.
In summary, all our validation results demonstrate both accuracy and robustness with respect to timestep variations.

\begin{figure}[htbp]
  \centering
  \subfloat[]{\includegraphics[width=1.0\textwidth]{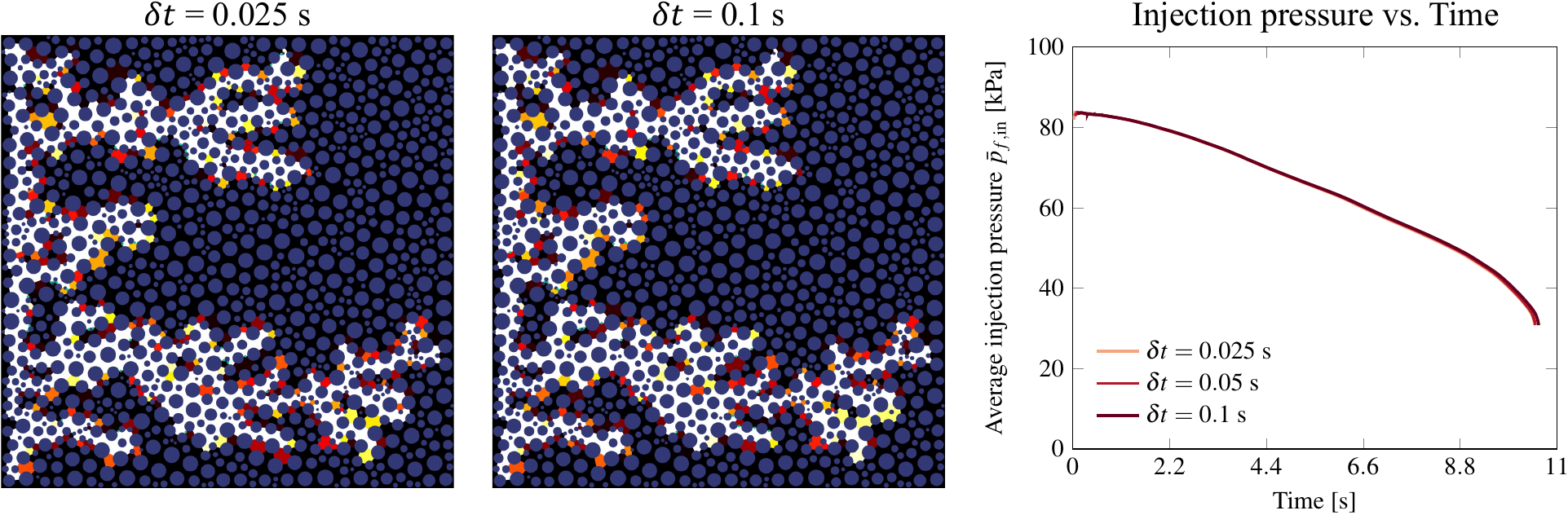}} \\
  \subfloat[]{\includegraphics[width=1.0\textwidth]{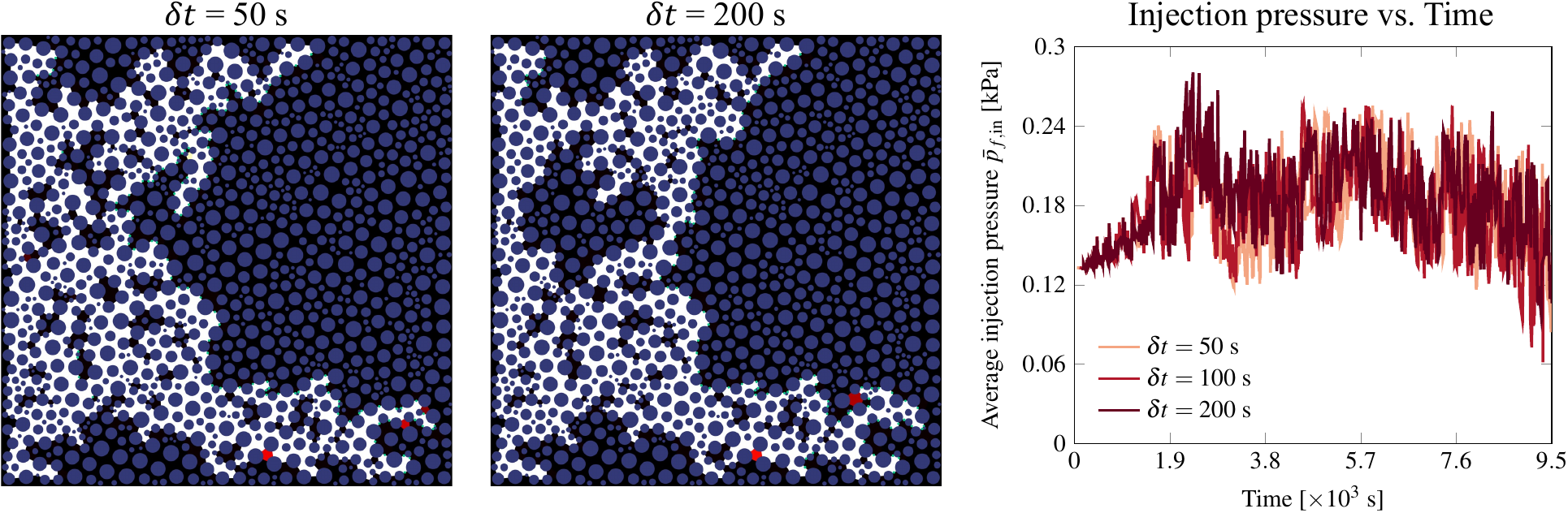}} 
  \caption{Effects of timestep on (a) viscous fingering ($Ca^*=1.9\times10^{-4}$) and (b) capillary fingering ($Ca^*=3.1\times10^{-7}$) in Lenormand's experiment, shown through flow patterns and the temporal evolution of injection pressure under different timesteps.}
  \label{fig:timestep-effect-lenormand}
\end{figure}

\begin{figure}[htbp]
  \centering
  \subfloat[]{\includegraphics[width=1.0\textwidth]{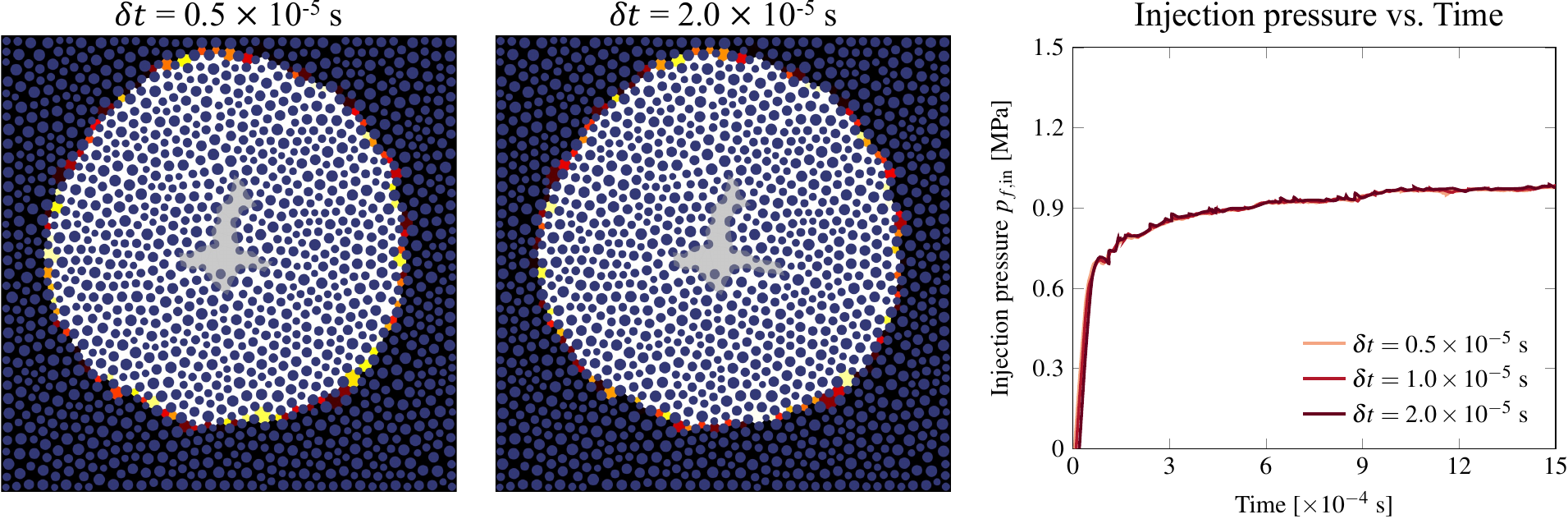}} \\
  \subfloat[]{\includegraphics[width=1.0\textwidth]{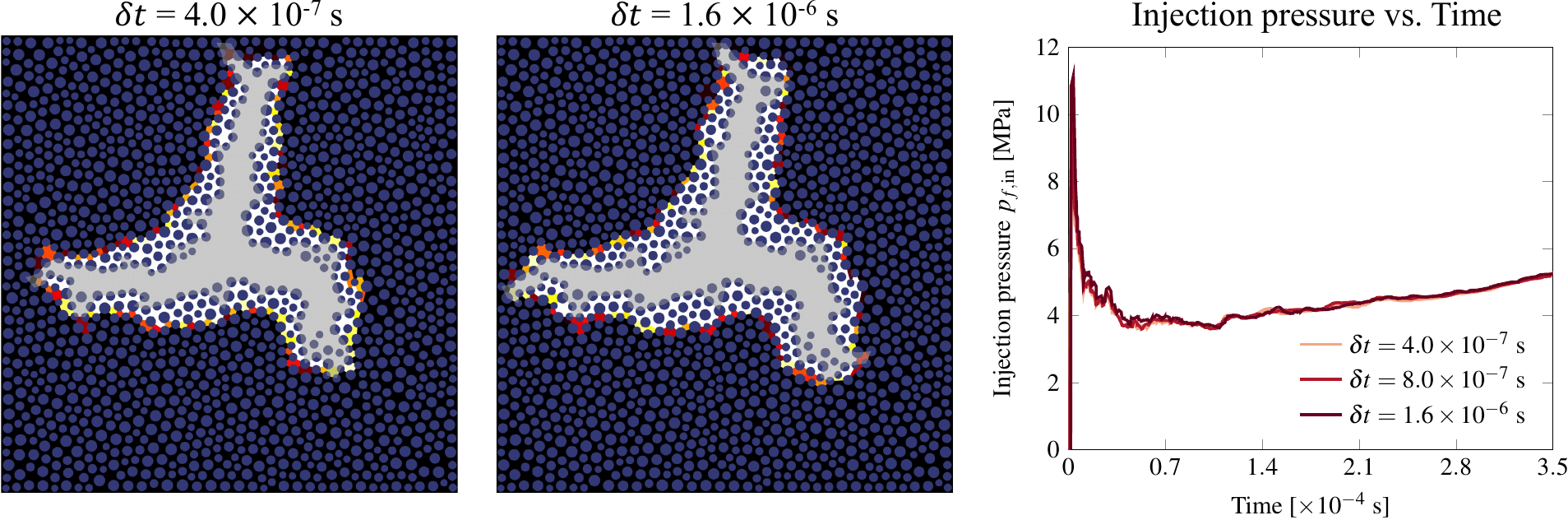}} 
  \caption{Effects of timestep on (a) infiltration-dominated regime (B1: $Ca^* = 56.3$) and (b) fracturing-dominated regime (B3: $Ca^* = 1138.1$) in Huang's experiment, shown through flow patterns and the temporal evolution of injection pressure under different timesteps.}
  \label{fig:timestep-effect-huang}
\end{figure}

\begin{table}[h!]
    \centering
    \begin{tabular}{l|l|l|l}
        \toprule 
        Lenormand-VF & $\delta t=0.025$ s & $\delta t=0.05$ s  & $\delta t=0.1$ s \\
        \midrule 
        $t_c$ [hour] & 1.2 & 1.0 & 1.0 \\
        \bottomrule 
        \toprule 
        Lenormand-CF & $\delta t=50$ s & $\delta t=100$ s  & $\delta t=200$ s \\
        \midrule 
        $t_c$ [hour] & 7.0 & 7.5 & 8.3 \\
        \bottomrule 
        \toprule 
        Huang-B1 & $\delta t=0.5\times10^{-5}$ s & $\delta t=1.0\times10^{-5}$ s  & $\delta t=2.0\times10^{-5}$ s \\
        \midrule 
        $t_c$ [hour] & 45.5 & 16.2 & 7.2 \\
        \bottomrule 
        \toprule 
        Huang-B3 & $\delta t=4.0\times10^{-7}$ s & $\delta t=8.0\times10^{-7}$ s  & $\delta t=1.6\times10^{-6}$ s \\
        \midrule 
        $t_c$ [hour] & 7.5 & 2.0 & 1.5 \\
        \bottomrule
    \end{tabular}
    \caption{Effects of timestep on simulation efficiency for Lenormand's and Huang's experiments, comparing the computational time ($t_c$) for each simulation.}
    \label{tab:timestep-effect-time}
\end{table}
}

\reviewertwo{
\section{Pressure solver performance}
\label{appendix_pressure_solver}
In this study, we apply an implicit pressure solver to overcome the timestep limitations inherent in explicit integration schemes commonly used in pore network models (\eg~\cite{duan2021modeling, jain2009preferential}). To demonstrate its performance improvement, we conduct additional simulations using an explicit pressure solver and compare the results with those obtained using our implicit solver. Representative cases from Lenormand’s and Huang’s experiments---including viscous fingering (VF: $Ca^*=1.9\times10^{-4}$), capillary fingering (CF: $Ca^*=3.1\times10^{-7}$), fluid infiltration-dominated stable displacement (B1: $Ca^*=56.3$), and fracturing-dominated stable displacement (B3: $Ca^*=1138.1$)---are used for this comparison.

Using an explicit time integration scheme to solve the fluid mass balance (Eq.~\eqref{eq:fmass-balance}) necessitates a very small timestep for stability \cite{jain2009preferential}. Ignoring domain volume changes, the characteristic timestep for fluid dynamics within a domain is given by
\begin{equation}
  \delta t \backsim \dfrac{V_i}{K_{f,i}}\dfrac{\delta p}{\sum_j -q_{ij}},
  \label{eq:explicit-fmass-balance}
\end{equation}
where $\delta p=p_{f,i}^{n+1}-p_{f,i}^n$ is the pressure variation per timestep. The explicit form of the Plane Poiseuille flow equation (Eq.~\eqref{eq:plane-pflow}) is approximated as
\begin{equation}
  q_{ij} = \dfrac{a_{ij}^3}{12\eta_{ij}}\dfrac{p_{f,i}^{n}-p_{f,j}^{n}}{l_{p,ij}} \backsim \dfrac{a_{ij}^3}{12\eta_{ij}}\dfrac{-\delta p}{l_{p,ij}}.
  \label{eq:explicit-plane-pflow}
\end{equation}

Substituting this expression into the fluid mass balance and applying the scalings $V_i \backsim r_\text{ave}^2$, $a_{ij} \backsim r_\text{ave}$, and $l_{p,ij} \backsim 2r_\text{ave}$ (with $r_\text{ave}$ the average grain radius), the critical timestep becomes
\begin{equation}
  \delta t_\text{crit} \backsim \dfrac{V_i\delta p}{K_{f,i}}/{\sum_j \dfrac{a_{ij}^3}{12\eta_{ij}}\dfrac{\delta p}{l_{p,ij}}} \backsim \dfrac{24\eta_{ij}}{N K_{f,ij}} \backsim \dfrac{24}{N}\min(\dfrac{\eta_{\text{inv}}}{K_{f,\text{inv}}},\dfrac{\eta_{\text{def}}}{K_{f,\text{def}}}),
  \label{eq:explicit-timestep}
\end{equation}
where $N$ is the number of pipes per domain (assumed to be 5 to represent the critical situation).
Given the extremely small critical timestep, the mechanical timestep in explicit simulations is set equal to the fluid flow timestep.

Table~\ref{tab:timestep-pressure-solver} summarizes the performance comparison between implicit and explicit pressure solvers, including fluid flow timestep ($\delta t_f$), flow-to-mechanical timestep ratio ($\delta t_f:\delta t_m$), total simulation time ($t_f$), computational time ($t_c$), number of run steps, and solver efficiency (dimensionless time: $t_f/t_c$). 
The number of steps also accounts for refined timesteps required to manage overfilled domains and blocked front pipes.
Although the explicit solver executes more steps within the same computational time---largely due to fewer mechanical steps and the reduced computational load of solving the pressure matrix---this apparent efficiency is misleading. Closer examination reveals that the dimensionless time for the explicit solver is orders of magnitude smaller than that of the implicit solver. 
Specifically, the implicit solver outperforms the explicit solver by roughly nine orders of magnitude in CF ($Ca^*=3.1\times10^{-7}$) and two orders of magnitude in B3 ($Ca^*=1138.1$).
This disparity arises because the explicit solver’s extremely small timesteps severely limit the progress of fluid dynamics per step, ultimately rendering it computationally inefficient.

Flow pattern comparisons (Figure~\ref{fig:explicit-flow-pattern}) further emphasize this point. Despite executing far more steps, the explicit pressure solver barely advances the invading fluid within the injection domain. In contrast, the implicit solver, with far fewer steps, allows the invading fluid to reach the outflow boundary within the same computational time. This demonstrates the implicit solver’s ability to utilize much larger timesteps, offering a clear computational advantage.

In summary, while the explicit solver executes more steps, the implicit solver’s ability to handle larger timesteps and generate realistic multiphase flow patterns makes it a significantly more efficient and practical choice for simulating multiphase flow in deformable porous media.

\begin{table}[h!]
    \centering
    \begin{tabular}{l|l|l|l|l|l|l}
        \toprule 
        Test          & $\delta t_f$ [s]       & $\delta t_f:\delta t_m$ & $t_f$ [s]      & $t_c$ [hour] & Steps & $t_f$/$t_c$ \\
        \midrule 
        VF-implicit   & $5.0\times10^{-2}$     & [-]       & 10.5                  & 1     & 703   & $2.9\times10^{-3}$  \\
        VF-explicit   & $6.0\times10^{-10}$    & [-]       & $2.4\times10^{-6}$    & 1     & 4010  & $6.7\times10^{-10}$ \\
        CF-implicit   & $100.0$                & [-]       & 9324.1                & 7.5   & 1783  & $3.5\times10^{-1}$  \\
        CF-explicit   & $6.0\times10^{-10}$    & [-]       & $1.7\times10^{-5}$    & 7.5   & 27555 & $6.1\times10^{-10}$ \\
        \midrule
        B1-implicit   & $1.0\times10^{-5}$     & $2000:1$  & $2.5\times10^{-2}$    & 16.2  & 2910  & $4.3\times10^{-7}$ \\
        B1-explicit   & $4.0\times10^{-10}$    & $1:1$     & $5.0\times10^{-6}$    & 16.2  & 12415 & $8.5\times10^{-11}$ \\
        B3-implicit   & $8.0\times10^{-7}$     & $2000:1$  & $3.5\times10^{-4}$    & 2     & 612   & $4.8\times10^{-8}$ \\
        B3-explicit   & $4.0\times10^{-10}$    & $1:1$     & $8.8\times10^{-7}$    & 2     & 2211  & $1.2\times10^{-10}$ \\
        \bottomrule
    \end{tabular}
    \caption{Performance comparison between implicit and explicit pressure solvers, including fluid flow timestep ($\delta t_f$), flow-to-mechanical timestep ratio ($\delta t_f:\delta t_m$), total simulation time ($t_f$), computational time ($t_c$), number of run steps, and solver efficiency (dimensionless time: $t_f/t_c$).}
    \label{tab:timestep-pressure-solver}
\end{table}

\begin{figure}[htbp]
  \centering
  \captionsetup[subfigure]{labelformat=empty}
  \subfloat[(a)]{\includegraphics[width=0.5\textwidth]{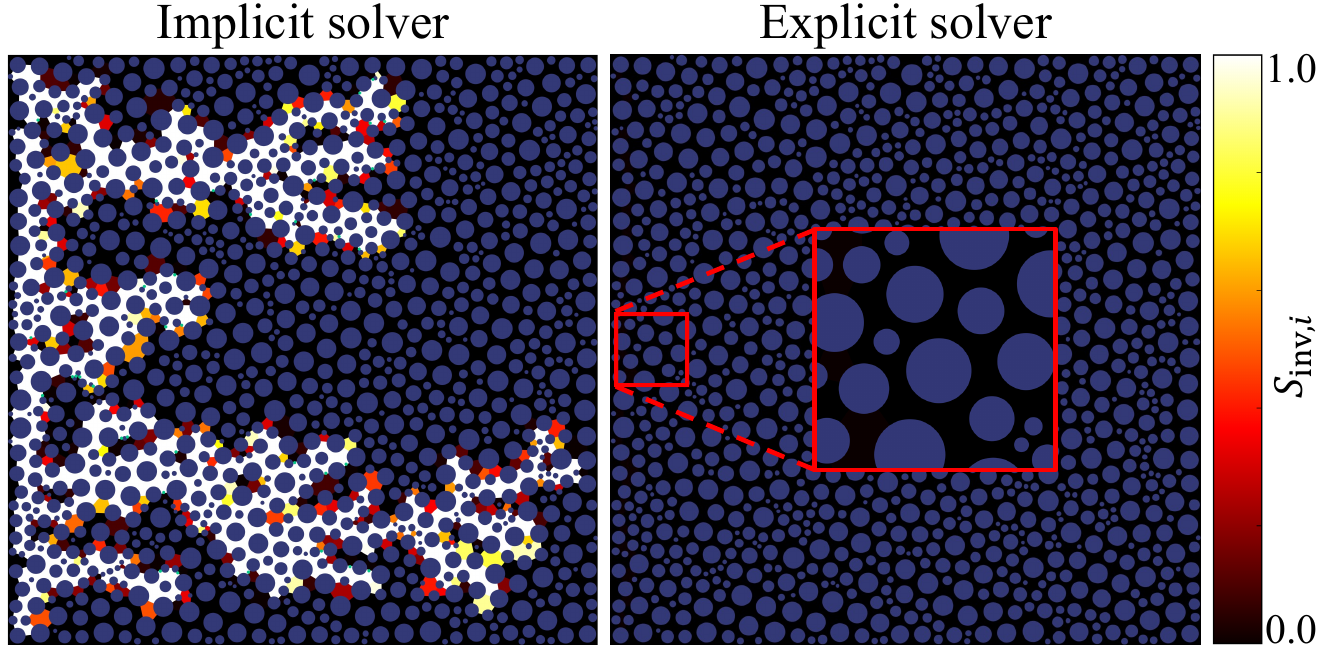}}
  \subfloat[(c)]{\includegraphics[width=0.5\textwidth]{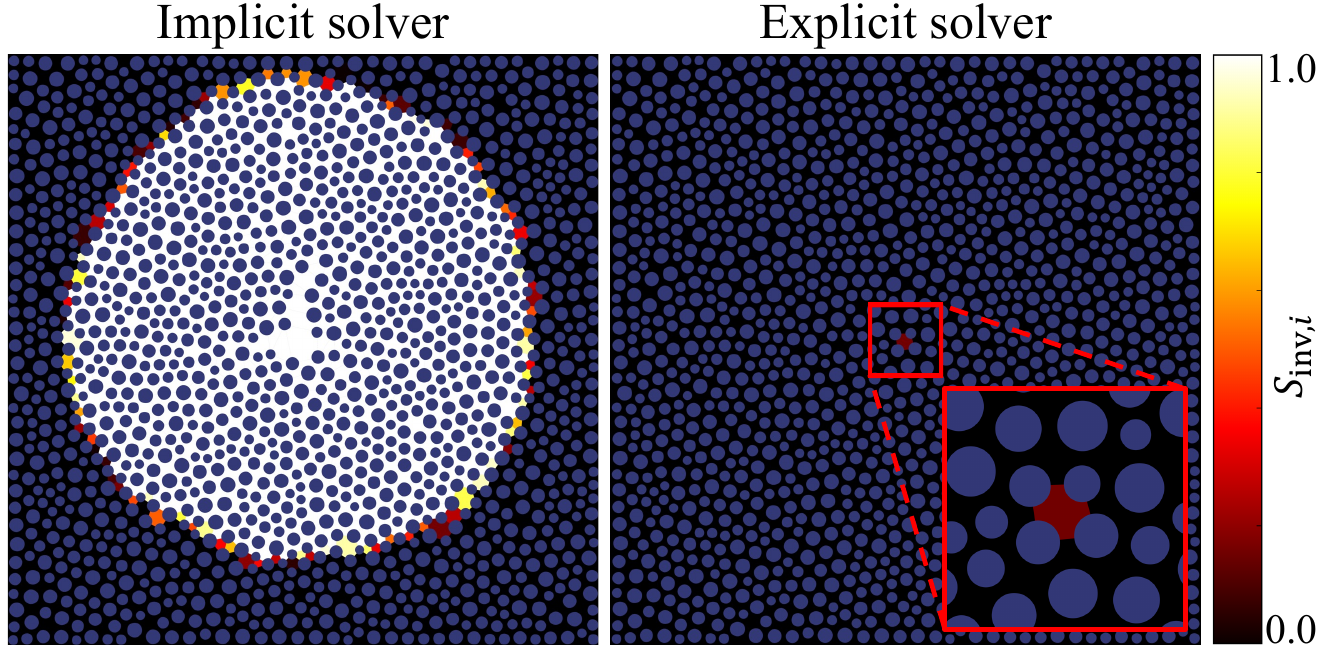}} \\
  \subfloat[(b)]{\includegraphics[width=0.5\textwidth]{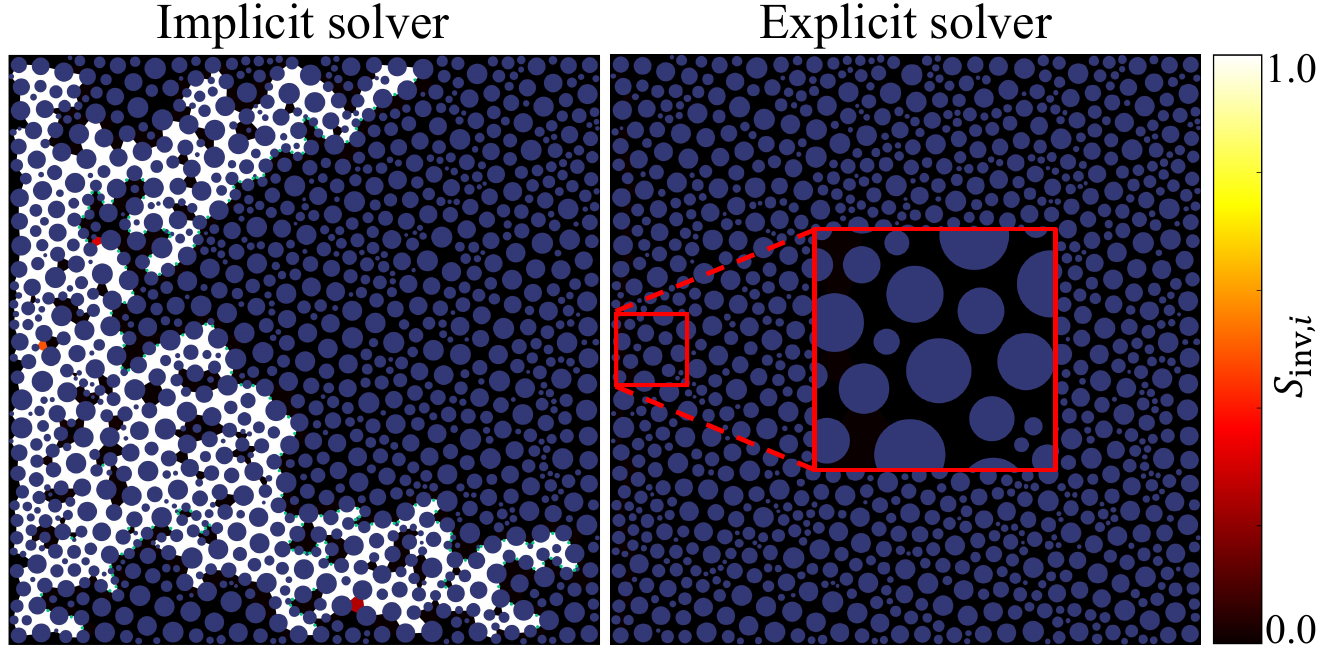}} 
  \subfloat[(d)]{\includegraphics[width=0.5\textwidth]{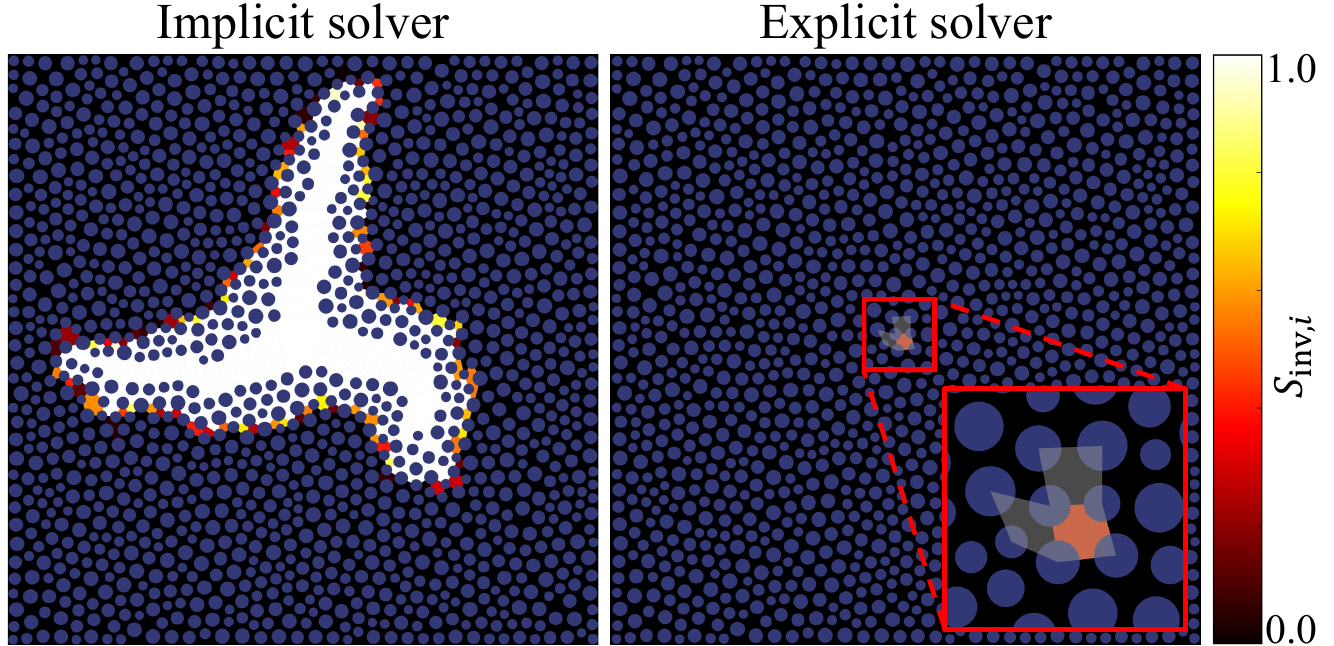}} 
  \caption{Comparison of flow patterns obtained using implicit and explicit pressure solvers within the same computational time. (a) Viscous fingering (VF, $Ca^*=1.9\times10^{-4}$), (b) capillary fingering (CF, $Ca^*=3.1\times10^{-7}$), (c) infiltration-dominated stable displacement (B1: $Ca^*=56.3$), and (d) fracturing-dominated stable displacement (B3: $Ca^*=1138.1$).}
  \label{fig:explicit-flow-pattern}
\end{figure}
}

\reviewertwo{

\section{Simple flow tests}
\label{appendix-simple-flow}
While our validation results using randomly packed samples align well with experimental observations in both Lenormand’s and Huang’s experiments (Figures~\ref{fig:fixed-pattern} and \ref{fig:movable-pattern}), we further evaluate the model’s accuracy through two simple flow tests: single-phase Darcy flow in randomly packed media and two-phase flow in regularly packed media. Unlike multiphase fluid injection in highly randomized samples, Darcy flow can be validated against analytical solutions, and two-phase flow in regular packing allows precise reproduction of experimental results, such as the number and direction of fingers.

\subsection{Single-phase Darcy flow test}
We simulate one-dimensional (1D) single-phase linear flow by injecting highly viscous oil (same as Case 1 in Table~\ref{tab:fixed-hydraulic}) into oil-saturated, fixed-grain porous media. The model setup is the same as in Lenormand’s experiment (Figure~\ref{fig:fixed-setup}), with an injection rate of $31\times10^{-5}$ m$^3$/s applied at the left boundary. The simulation continues until inflow and outflow reach steady state.

The simulated results are validated against analytical solutions for 1D Darcy-flow problem, governed by the diffusion equation
\begin{equation}
    \dfrac{\partial^{2}p_{D}}{\partial x_{D}^{2}}=\dfrac{\partial p_{D}}{\partial t_{D}},
    \label{eq:diffusion-equation}
\end{equation}
with the dimensionless quantities for pressure $p$, $x$-coordinate distance from the inflow boundary $x$, and time $t$ defined, respectively, by
\begin{equation}
    p_{D}=-\dfrac{kWH}{\eta L Q_{\text{in}}}(p-p_{0}),
\end{equation}
\begin{equation}
    x_{D}=\dfrac{x}{L},
\end{equation}
\begin{equation}
    t_{D}=\dfrac{kK_{f}}{c_{v}\eta L^{2}}t.
\end{equation}
Here, $k$ is the intrinsic permeability, $p_0$ is the initial pressure, $c_v$ is the consolidation coefficient, $\eta$ is fluid viscosity, $H$ and $L$ are the height and length of the porous media, $W$ is the out-of-plane thickness (1 m for 2D problems), and $Q_\text{in}$ is the total inflow rate. 

The initial and boundary conditions are
\begin{equation}
    p_{D}(x_D,0)=0, \quad 0 \leq x_D \leq 1,
\end{equation}
and
\begin{equation}
    \dfrac{\partial p_{D}}{\partial t_{D}}(0,t_D)=1, \quad p_D(1,t_D)=0, \quad t_D>0.
\end{equation}

We derive analytical solutions for fluid pressure and flow rate by the method of separation \cite{crank1979mathematics}, which can be expressed as
\begin{equation}
    p_{D}(x_D,t_D)=-1+x_{D}+\dfrac{8}{\pi^{2}}\sum_{n=0}^{\infty}\dfrac{\cos\tfrac{2n+1}{2}\pi x_{D}}{(2n+1)^2}e^{-\tfrac{(2n+1)^2}{4}\pi^2t_{D}}
\end{equation}
and
\begin{equation}
    q_{D}(x_D,t_D)=1-\dfrac{4}{\pi}\sum_{n=0}^{\infty}\dfrac{\sin\tfrac{2n+1}{2}\pi x_{D}}{2n+1}e^{-\tfrac{(2n+1)^{2}}{4}\pi^{2}t_{D}}.
\end{equation}
Theoretical inflow and outflow rates are evaluated at $x_D=0$ and $x_D=1$, respectively.
Permeability $k$ is determined from the steady-state pressure gradient and flow rate using Eq.~\eqref{eq:darcy-flow}, while $c_v$ is calibrated by matching simulated inflow and outflow curves at steady state.

Figure~\ref{fig:appendix-darcy-flow} compares simulation results with analytical solutions. The simulated inflow and outflow rates match closely with analytical solutions when $c_v=0.945$ (Figure~\ref{fig:appendix-darcy-flow}a), confirming the consistency between theory and simulation during transient and steady-state conditions. Similarly, the simulated pressure profiles align with analytical predictions at all selected time stages (Figure~\ref{fig:appendix-darcy-flow}b). Despite minor randomness introduced by variations in domain and pipe sizes, the results demonstrate excellent agreement.

These findings validate the model’s ability to accurately simulate fluid migration in permeable media, confirming its reliability and robustness in Darcy-flow analysis.

\begin{figure}[htbp]
  \centering
  \subfloat[]{\includegraphics[height=0.38\textwidth]{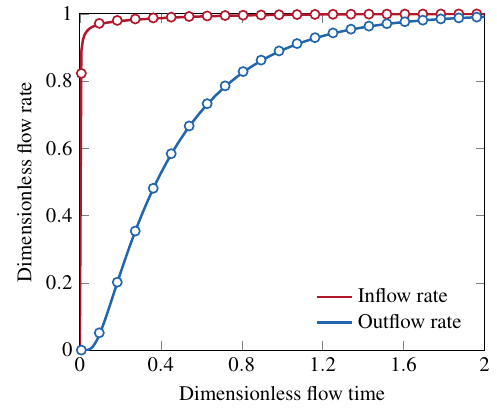}} 
  \subfloat[]{\includegraphics[height=0.38\textwidth]{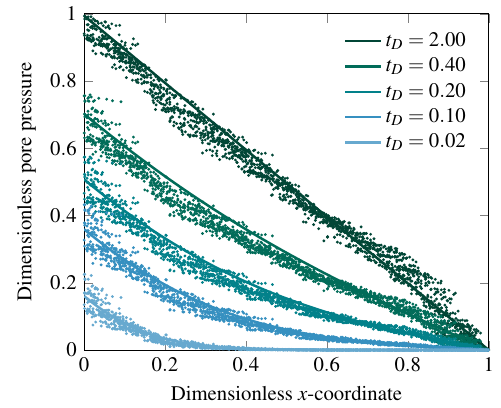}} 
  \caption{Comparison of Darcy flow behaviors between numerical simulations (dots) and analytical solutions (solid lines). (a) Inflow and outflow rates as a function of dimensionless flow time and (b) dimensionless pore pressure profiles as a function of dimensionless $x$-coordinate.}
  \label{fig:appendix-darcy-flow}
\end{figure}

\subsection{Two-phase flow in regular packing}
In random packing, two-phase flow patterns are typically classified based on morphological features (\eg~tree-like structures or blocked front pipes) or fractal dimensions. In contrast, regular packing, while overly simple for representing real geological formations, allows flow patterns such as the number and direction of fingers to be reproduced exactly as observed in experiments.

To validate this, we replicate a viscous fingering experiment conducted by Chen and Wilkinson \cite{chen1985pore}, where non-wetting oil ($\eta_\text{inv}=1$ cP) is injected into wetting glycerine ($\eta_\text{def}=1200$ cP) within a square network of equally spaced, interconnected channels etched into glass. 
While the original experiment did not measure permeability, we set the grain radius ($r$) and aperture width ($a$) to 0.25 mm, matching experimental values.

Our model constructs a 39 $\times$ 39 square network of channels, each 0.5 mm in length, consisting of 1521 domains and 3120 pipes (Figure~\ref{fig:regular-flow-pattern}). 
The inflow domain is located at the center, while 72 outflow domains along the edges are maintained at a constant zero pressure. All initial domain pressures are set to zero.
The fluid bulk modulus is 2000 MPa for both oil and glycerine, with interfacial tension of 20 dyn/cm and a contact angle of 180$\degree$.

We simulate drainage with a modified capillary number $Ca^*=0.03$, corresponding to the experimental inflow rate of $Q_\text{in}=1.4\times10^{-10}$ m$^3$/s, replicating the experiment’s viscous fingering behavior. As shown in Figure~\ref{fig:regular-flow-pattern}, the simulation reproduces the observed pattern, forming four fingers along the $xy$-axes, identical in number and direction to the experiment.

This accuracy arises from the uniformity of pore and aperture sizes, which ensures consistent pressure distribution favoring flow along the $xy$-axes, resulting in the formation of four fingers. This successful reproduction underscores the accuracy of our HM-DEM framework in simulating two-phase flow in regular media, providing a reliable benchmark for validating its application to more complex systems such as Lenormand’s and Huang’s experiments.

\begin{figure}[htbp]
  \centering
  \includegraphics[width=1.0\textwidth]{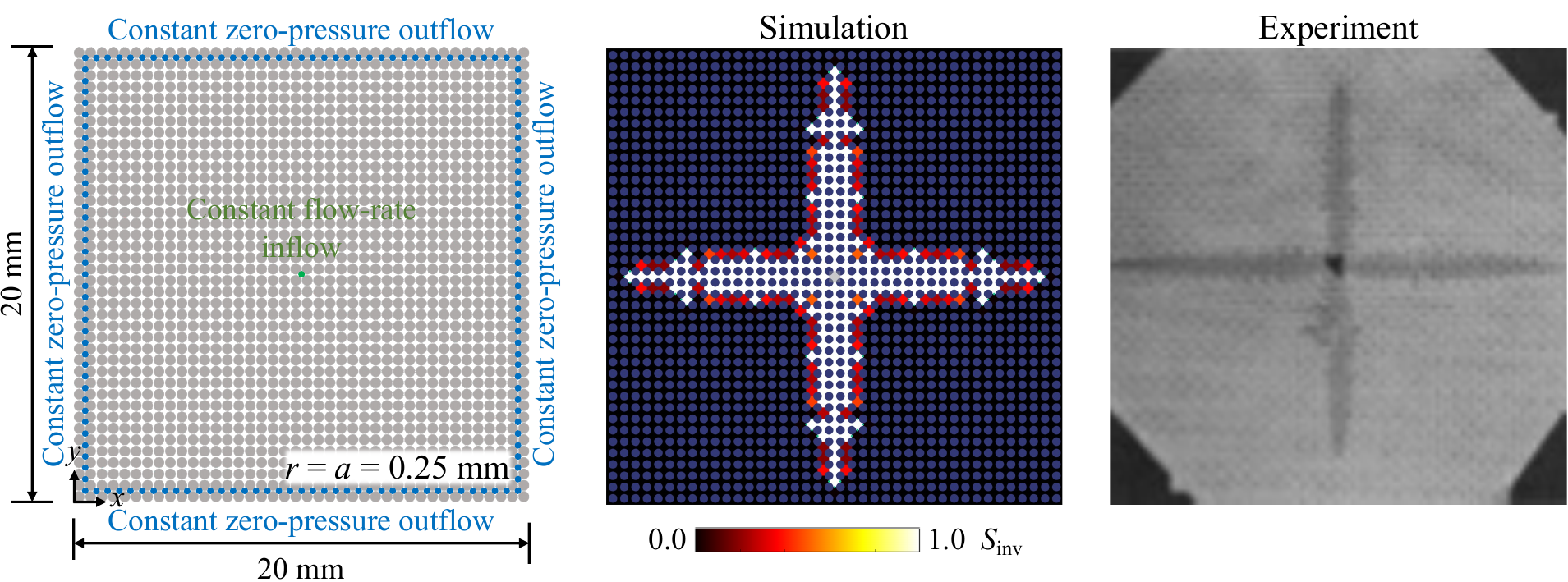}
  \caption{Flow pattern in regularly packed media: numerical simulation reproduces the four-finger pattern observed in Chen’s experiment.}
  \label{fig:regular-flow-pattern}
\end{figure}

}

\reviewerone{

\section{Fractal dimension calculation}
\label{appendix_fractal_dimension}
The classification of various fluid-fluid and fluid-grain displacement patterns is often based on visual inspection of their characteristic morphological features, such as the highly branched viscous fingering or the less-branched capillary fingering. While such qualitative approach can be useful, we also adopt a quantitative method to classify displacement patterns based on their fractal dimension, $D_f$, calculated via the box-counting method \cite{schroeder2009fractals}---a widely used technique for quantifying the complexity of irregular shapes or flow patterns. 
In this method, the fractal domain is discretized into a grid of square boxes with side length $s$, and the minimum number of boxes, $N(s)$, required to cover the domain is counted. 
The process is repeated for a range of box sizes, and the fractal dimension is determined as the slope of the linear region in a log–log plot of $N(s)$ versus $s$:
\begin{equation}
  D_f= \cfrac{d\log N(s)}{d\log s}.
  \label{eq:box-counting-df}
\end{equation}

In practice, the resolution of the image or the pore structure in a pipe network model (\eg~grain and pore sizes) can greatly influence the accuracy of fractal dimension estimation. To address this issue, we employ a “local fractal dimension” approach. Instead of fitting a single slope across all scales, we calculate the slope over smaller intervals in the log-log space and then average these local slopes over the scales most relevant to the observed patterns. This approach enhances robustness by accounting for variations in resolution and focusing on the box sizes that best capture the dominant morphological features, providing a more reliable estimate of the fractal dimension for fluid-fluid and fluid-grain displacement patterns.

Before applying the box-counting method, the flow pattern image is first binarized to isolate the infiltration or fracturing regions. Subsequently, we plot $\log N(s)$ versus $\log s$ as well as the local $D_f$ versus $\log s$ to determine the fractal dimension of the specific flow pattern while accounting for pore structure resolution.
As a demonstration, we calculate the fractal dimension for three representative cases: viscous fingering ($Ca^*=1.9\times10^{-4}$) and capillary fingering ($Ca^*=3.1\times10^{-7}$) from Lenormand's experiment, and the fracturing-dominated stable displacement case (B3) from Huang's experiment. The results are shown in Figures~\ref{fig:fractal-dimension-lenormand} and \ref{fig:fractal-dimension-huang}. 
A summary of the calculated fractal dimensions for all simulated displacement patterns is provided in Figure~\ref{fig:all-fractal-dimension}.

\begin{figure}[htbp]
  \centering
  \subfloat[]{\includegraphics[width=1.0\textwidth]{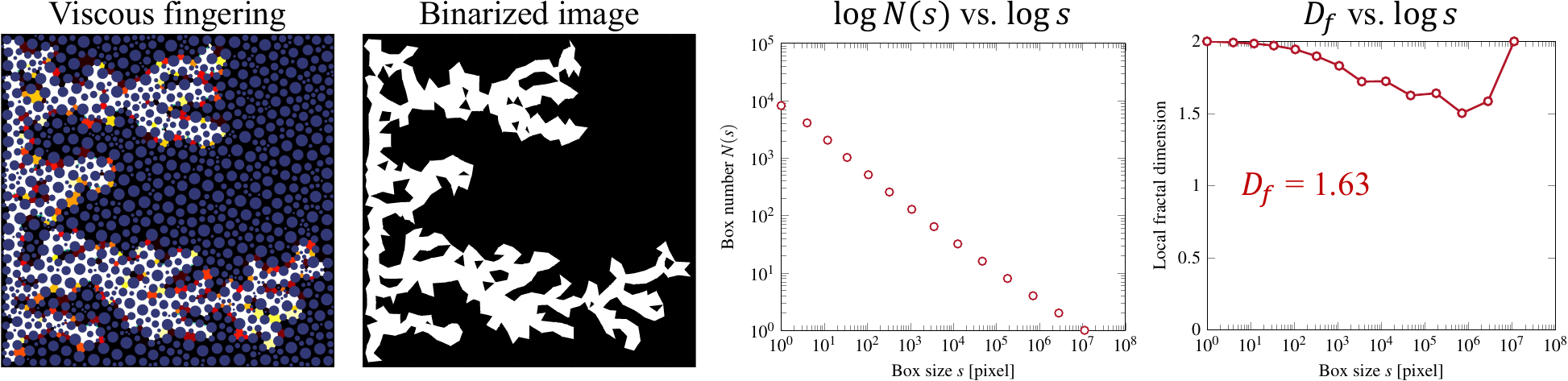}} \\
  \subfloat[]{\includegraphics[width=1.0\textwidth]{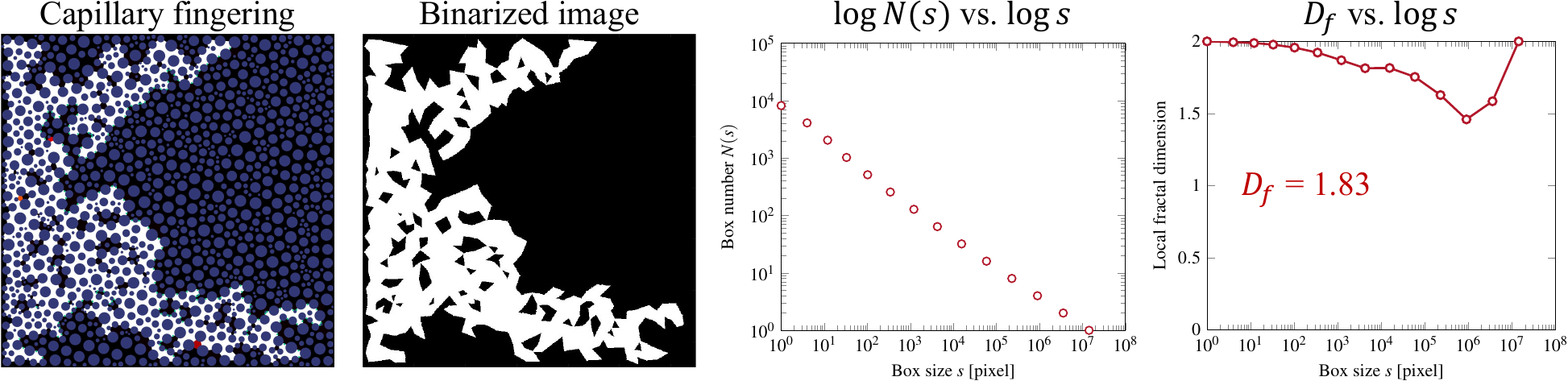}}   
  \caption{Fractal dimension estimation of (a) viscous fingering ($Ca^*=1.9\times10^{-4}$) and (b) capillary fingering ($Ca^*=3.1\times10^{-7}$) in Lenormand's experiment. Shown are the binarized displacement patterns, the relationship between box count ($N(s)$) and box size ($s$), and the local fractal dimension ($D_f$) as a function of $s$.}
  \label{fig:fractal-dimension-lenormand}
\end{figure}

\begin{figure}[htbp]
  \centering
  \subfloat[]{\includegraphics[width=1.0\textwidth]{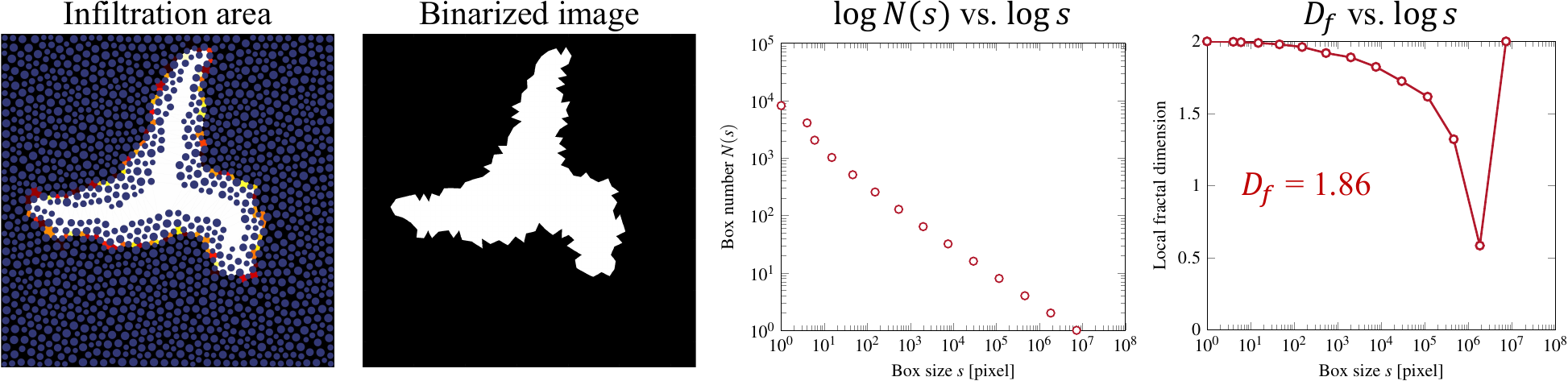}} \\
  \subfloat[]{\includegraphics[width=1.0\textwidth]{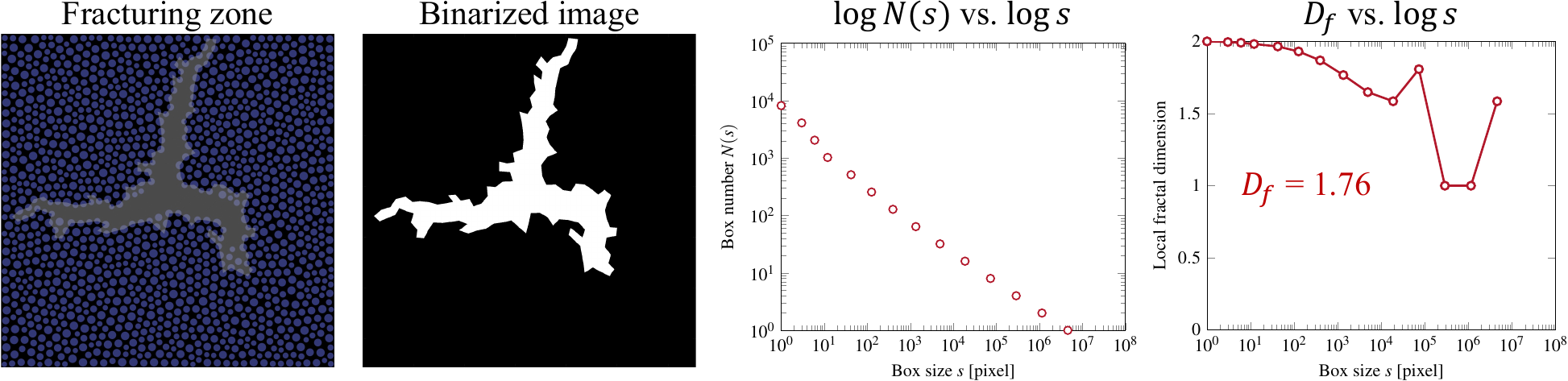}}   
  \caption{Fractal dimension estimation of (a) fluid infiltration area and (b) fracturing zone in B3 from Huang's experiment (fracturing-dominated stable displacement). Shown are the binarized displacement patterns, the relationship between box count ($N(s)$) and box size ($s$), and the local fractal dimension ($D_f$) as a function of $s$.}
  \label{fig:fractal-dimension-huang}
\end{figure}

\begin{figure}[htbp]
  \centering
  \includegraphics[width=0.5\textwidth]{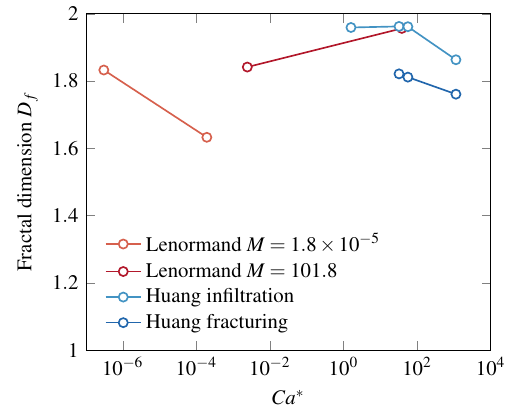}
  \caption{Fractal dimension for each fluid-fluid and fluid-grain displacement pattern in Lenormand's and Huang's experiments.}
  \label{fig:all-fractal-dimension}
\end{figure}

}

\bibliography{references}

\end{document}